\documentclass[aps,prd,10pt,notitlepage,nofootinbib,superscriptaddress,showkeys,showpacs]{revtex4-1}

\usepackage{amstext,amsmath,amssymb,amsfonts,bbm}
\usepackage[latin1]{inputenc}
\usepackage{epsfig}
\usepackage{hyperref}
\usepackage{amsthm}
\usepackage{subfigure}
\usepackage{color}
\usepackage{multirow}

\usepackage{graphicx}

\usepackage{latexsym}

\usepackage{cancel}

\newcommand{\bea}{\begin{eqnarray}}	
\newcommand{\eea}{\end{eqnarray}}
\newcommand{\beq}{\begin{equation}}	
\newcommand{\eeq}{\end{equation}}

\newcommand{\cB}{{\mathcal B}}
\newcommand{\cG}{{\mathcal G}}
\newcommand{\cS}{{\mathcal S}}
\newcommand{\cF}{{\mathcal F}}
\newcommand{\bG}{{\partial\mathcal G}}
\newcommand{\tJ}{{\widetilde{J}}}

\newcommand{\cL}{{\mathcal L}}
\newcommand{\cV}{{\mathcal V}}

\newcommand{\cexG}{\mathcal G_{\rm{color}}}
\newcommand{\bJ}{ J_{\partial} }

\newcommand{\ren}{\text{ren\,}}

\newcommand{\kin}{\rm{kin\,}}
\newcommand{\inter}{\rm{int\,}}
\newcommand{\ext}{\rm{ext\,}}

\newcommand{\col}{ \rm{color} } 
\newcommand{\uncol}{ \rm{uncolor} }

\newcommand{\dee}{{\mathbf d}}
\newcommand{\bee}{{\mathbf b}}

\newcommand{\bmu}{\boldsymbol{\mu}}
\newcommand{\N}{{\mathbb N}}
\newcommand{\Z}{{\mathbb Z}}

\newtheorem{lemma}{Lemma}

\newtheorem{theorem}{Theorem}
\newtheorem{proposition}{Proposition}

\newtheorem{remark}{Remark}

\newcommand{\bra}{ \langle} 
\newcommand{\ket}{  \rangle } 

\begin{document}

\title{Renormalizable  Models in Rank $d\geq 2$ Tensorial Group Field Theory}
\author{\large Joseph Ben Geloun}\email{jbengeloun@perimeterinstitute.ca}
\affiliation{Perimeter Institute for Theoretical Physics, 31 Caroline
St, Waterloo, ON, Canada \\
International Chair in Mathematical Physics and Applications, 
ICMPA-UNESCO Chair, 072BP50, Cotonou, Rep. of Benin}

\date{\small\today}

\begin{abstract}
 Classes of renormalizable models in the Tensorial Group Field Theory framework are investigated. The rank $d$ tensor fields are defined over $d$ copies of a group manifold $G_D=U(1)^D$ or $G_D= SU(2)^D$ with no symmetry and no gauge invariance assumed on the fields. In particular, we explore the space of 
renormalizable models endowed with a kinetic term corresponding to a sum of momenta of the form $p^{2a}$, $a\in ]0,1]$. This study is tailored for models equipped with Laplacian dynamics on $G_D$ (case $a=1$) but also for more exotic nonlocal models in quantum topology (case $0<a<1$). A generic model can be written $(_{\dim G_D}\Phi^{k}_{d}, a)$, where $k$ is the maximal valence of its interactions. Using a multi-scale analysis for the generic situation, we identify several classes of renormalizable actions including matrix model actions. In this specific instance, we find a tower of renormalizable matrix models parametrized by $k\geq 4$. 
In a second part of this work, we focus on the UV behavior of the models
up to maximal valence of interaction $k =6$. All rank $d\geq 3$  tensor models proved renormalizable are asymptotically free in the UV.  All matrix models 
with $k=4$ have a vanishing $\beta$-function at one-loop and, very likely, reproduce the same feature of the Grosse-Wulkenhaar model [Commun. Math. Phys. {\bf 256}, 305 (2004)].

\medskip
\noindent Pacs numbers: 11.10.Gh, 04.60.-m, 02.10.Ox
pi-qg-329 and ICMPA/MPA/2013/004 \\

\end{abstract}

\maketitle

\tableofcontents

\section{Introduction}

In attempts to generalize in higher dimensions matrix model results on 2D quantum gravity (QG)  \cite{Di Francesco:1993nw}, tensor models have been examined since the early 90's \cite{ambj3dqg,mmgravity,sasa1}.  These models stem from the idea that the classical geometry of some manifold could emerge from the statistical
sum of random geometries associated with triangulations of this background.
They might also pertain to a broader proposal that gravity originates from more fundamental (quantum) objects and laws \cite{Konopka:2006hu}. 
The special case of matrices provides one of the most compelling results in that direction. Indeed, the Feynman integral of matrix models generates ribbon graphs organized in a $1/N$ (or genus) expansion \cite{'tHooft:1973jz} so that this statistical sum is well controlled through only analytical tools. 
The real beauty of these models reveals itself after a phase transition \cite{Kazakov:1985ds}\cite{David:1985nj}: the resulting
model maps to a 2D theory of gravity coupled with Liouville conformal field \cite{kpz,David:1988hj,conf1,conf2}.
From this framework and the tools developed within, a wealth of 
implications on integrability and statistical mechanics 
followed and then led to the renown of matrix models \cite{Di Francesco:1993nw}. The framework of random matrices still attracts a lot of attention in both physicist and mathematician  communities \cite{Duplantier:2011sy,Duplantier:2010yw,m1,m2,m3,m4}.  

For higher rank tensor models, the story turns out to be a far greater challenge \cite{ambj3dqg,mmgravity}. The crucial $1/N$ expansion tool leading to  
the understanding and control of the partition function in the case of matrix models
was missing in the tensor case then. A way to understand analytically the partition function of tensor models was quite abandoned and, consequently, computations in theories implementing a discrete version of QG in higher dimension strongly rests on numerics up to today.  With somehow a different perspective and still in the same period, Boulatov showed that the amplitudes of a simplicial theory of 3D complexes made of tensors equipped with a particular invariance reproduce several features of amplitudes of a lattice gauge theory \cite{Boul,oog}. The type
of invariance of the Boulatov model will turn out to be interesting on its
own and leads to several connections with other QG approaches \cite{oriti,Oriti:2006ar}. 
 
Concerning analytical calculations, the interest in tensor models could have been certainly and significantly improved if these were provided with an appropriate notion of $1/N$-expansion. 
This was indeed what happened after the spotless discovery by Gurau of a genuine  notion of large $N$-expansion for a particular class of random tensor models \cite{Gur3,GurRiv,Gur4}. This particular class refers to the colored tensor models which prove to be associated with triangulations of simplicial pseudo-manifolds in any dimension \cite{color,Gurau:2009tz,Gurau:2010nd,Gurau:2011xp}.
The critical behavior for this class of tensor models has been investigated. They are found to undergo a phase transition towards the so-called branched polymer phase \cite{Bonzom:2011zz, Gurau:2013cbh,Gurau:2013pca}.   
More results provide answers on longstanding questions on statistical mechanics on random lattices \cite{Bonzom:2011ev,Benedetti:2011nn} and mathematical physics \cite{Gurau:2011sk,Gurau:2011tj,Gurau:2012ix}. Another  profound result is that there exists an extension of the universality and Wigner-Dyson law valid for tensors \cite{Gurau:2011kk} (for more results in a short review, see either \cite{Gurau:2012vu} or \cite{Gurau:2012vk}). 

It was reasonable to expect that more will be unraveled from 
such developments. Indeed, the $1/N$-expansion revealed a 
basis of unitary trace invariants for (unsymmetrized) tensors \cite{Gurau:2011tj,Gurau:2012ix,Bonzom:2012hw}. In simple words, 
these extend the unitary trace invariants ${\rm tr}[(M^\dag M)^p]$, for $p\geq 0$,
built from matrices $M$ themselves generalizing the unique unitary invariant
built from vectors $|\,\vec \phi\,|^2$. Unitary tensor invariants have been studied long ago by mathematicians \cite{gord}. The point of the previous works comes from the fact that all these unitary tensor  invariants were captured from a path integral field theory formalism defined on tensor colored models and simply encoded in a colored graph representation. 

Considered as basic observables and interactions, the same trace invariants were at the basis of the uncovering of the first renormalizable tensor models of rank greater than or equal to 3 \cite{BenGeloun:2011rc,Geloun:2012fq}. 
These models were investigated quite recently 
(see the reviews \cite{Rivasseau:2011xg,Rivasseau:2011hm,Rivasseau:2012yp})
as extensions of the so-called Grosse-Wulkenhaar (GW) model \cite{Grosse:2004yu},
a renormalizable matrix model issued from noncommutative geometry \cite{Rivasseau:2007ab}. The renormalizability of tensor models promotes
the latter to the rank of well-behaved quantum field theories.
We will refer these models to as Tensorial Group Field Theory (TGFT).\footnote{ 
The appearance of the name ``Group'' comes from the fact that the tensors
considered in these models are nothing but the Fourier components of some
class of functions or fields defined on an abstract group $G$. }
Why renormalizability is important for tensor models?
Renormalizability for any quantum field theory is a very desirable feature because
it mainly ensures one that the theory will survive after several energy scales.  
All known interactions of the standard model are renormalizable. This feature
gives a sense to a system dealing with several types of infinities (infinitely many degrees of freedom, divergence occurring in their physical quantities). Quantum field theory predictions rely 
on the fact that, from the Wilsonian or Renormalization Group (RG) point of view, these infinities should be not hidden or ignored but should  locally  (from one scale to the other) reflect a change in the form of the theory \cite{Rivasseau:1991ub}. 
In particular, if  tensor models intend to describe at low energy any physical reality like our spacetime, and since generically they possess divergent 
correlation functions, one must explain these divergences. 
The renormalization program is built for that purpose
and the RG offers a natural mechanism to flow from a certain model at some scale to another at another scale while  dealing consistently with these infinities.

Before reviewing the main results obtained in TGFTs, let us give now some precisions and basic terminology about tensor
models. Consider a model defined via a tensor field of rank $d$. This field represents a $(d-1)$ simplex. The interaction consists in a $d$ simplex
obtained by the gluing these fields or $(d-1)$ basic simplexes along their boundary. 
The path integral of a such model generates $d$ dimensional simplicial complexes
from the gluing of the interaction terms along their boundary. Hence, the rank $d$ of the tensor field and the dimension $d$ of the simplicial complexes generated are merely the same. 
The renormalization program for TGFT has achieved many results in 
the last four years \cite{BenGeloun:2011rc,Geloun:2012fq}
\cite{Geloun:2009pe, Geloun:2010nw, Geloun:2010vj, Geloun:2011cy,
BenGeloun:2012pu, Geloun:2012bz, BenGeloun:2012yk, Carrozza:2012uv,
Samary:2012bw, Carrozza:2013wda, Samary:2013xla}. So far, one identifies two types of renormalizable TGFTs.
One of them implements the gauge invariance by Boulatov \cite{oriti}. Referring to this particular type of TGFT, we shall use the terminology gi-TGFT. Discussing models without gauge invariance we will 
sometimes use ``simple'' TGFTs but the most
of time we will simply say TGFTs when the context does not lead
to any confusion. 

Table \ref{table:mod} collects the different features of both super-renormalizable and just-renormalizable models. It surprisingly happens that most of the just-renormalizable models discovered so far (gauge invariant or not) turns out to be asymptotically free\footnote{The model by Carrozza et al. \cite{Carrozza:2013wda} is presently under analysis.}.  We are led to the important question: Is asymptotic freedom a generic feature in TGFTs ? In general, a model is called UV asymptotically free if
it makes sense at arbitrary high energy scales and possesses a trivial UV fixed point defined by the free theory. QCD or the theory of strong interactions is a typical example of this kind. From the UV going in the IR direction, the renormalized coupling constant
grows up to some critical value for which one reaches a new phase
described in terms of new degrees of freedom (quark confinement in QCD). If tensor models are generically asymptotically free, this could be a nice feature because it would mean that, (1) in the case that these models actually describe a theory of gravity, this theory would be sensible at arbitrary small distances and, (2) in the IR, the models likely experience a phase transition after which, hopefully, the final degrees of freedom may encode more geometrical data than the initial ones (which are totally topological) may lead to a notion of invariance under coordinate change in the new action.

\begin{table}
\begin{center}
\begin{tabular}{lccccccccc|cc|}
\hline
TGFT (type)&& Group && $\Phi^{k_{\max}}$  &&  $d$ && Renormalizability &&   Asymptotic freedom \\
\hline
 && $U(1)$ && $\Phi^{4}$ && 4 &&\cite{BenGeloun:2011rc} Just- &&  $\sqrt{}$ \cite{BenGeloun:2012yk}\\
 && $U(1)$ && $\Phi^{3}$  && 3 &&\cite{BenGeloun:2012pu} Just- && $\sqrt{}$\cite{BenGeloun:2012pu} \\
\hline 
 gi-&& $U(1)$ && $\Phi^{2k}$  && 4&& \cite{Carrozza:2012uv} Super- &&  -\\
 gi-&&  $U(1)$ && $\Phi^{4}$  && 6 &&\cite{Samary:2012bw} Just- &&  $\sqrt{}$\cite{Samary:2013xla} \\
 gi-&&  $U(1)$ && $\Phi^{6}$  && 5 &&\cite{Samary:2012bw} Just- && $\sqrt{}$\cite{Samary:2013xla}\\
 gi-&&  $U(1)$ && $\Phi^{4}$  && 5 && \cite{Samary:2012bw} Super- &&-\\
 gi- && 
$SU(2)^3$ && $\Phi^{6}$  && 3 && \cite{Carrozza:2013wda} Just- && ?\\
\hline
\end{tabular}\end{center}
\caption{List of renormalizable models and their features
($\sqrt{}\equiv$ asymptotic freedom proved; $k_{\max}$
is the maximal valence of the vertex).}
\label{table:mod}
\end{table}

For the next discussion, it is instructive to provide details on
these renormalizable tensor models. The type of interactions which triggers renormalizable models are of the form of the unitary tensor invariants. Concerning the dynamics,
it was a unexpected fact that, starting with a rank 3 gi-TGFT with trivial 
dynamics in the form of a mass term and expanding
the two-point function, one was able to generate diverging corrections of a Laplacian form \cite{Geloun:2011cy}. Thus, this suggested that one needs to introduce a Laplacian dynamics in order to make sense of a renormalization program in gi-TGFTs. After this stage, 
introducing a group Laplacian in the kinetic term played a major role in the proof that several models were just-renormalizable  indeed
\cite{BenGeloun:2011rc,Samary:2012bw,Carrozza:2013wda}. 
However, this type of kinetic term is not the only possible which might lead to just-renormalizable theories. For instance, the rank $d=3$ model in \cite{BenGeloun:2012pu} has a kinetic term written in momentum space as $(\sum_{s=1}^3 |p_s| + \mu)$,
$p_s\in \Z$ representing the momentum associated with the direct space coordinate
$\theta$ parametrizing the circle $S^1 \sim U(1)$. There is, at this point,
no direct space formulation of this model. A way to think about such 
a formulation would be to introduce  anti-commuting fields $\psi$\footnote{Note
that the first color model \cite{color} was defined with anti-commuting
Grassmann variables.} and deal with a Dirac field formalism. This leads to another question about the statistics
of the tensors and the representation of the Lorentz group associated with
it. Nonetheless, at this QG energy level, there is no reason to enforce that
 Lorentz invariance should hold and that our fields should be some Dirac spinors. More just-renormalizable classes of tensor models of this kind have been found and several of them are related with classes of matrix models \cite{Geloun:2012bz}. 
This urges us to think about some physical selection criteria for tensor models. Since QG is very ``special'', we should adopt an inclusive attitude and will gain certainly by scrutinizing the space of all possible models with at least some particular features among which just-renormalizability.  According to some minimal physical axioms, the present work establishes that the space of just-renormalizable tensor models is not that huge that one might think and, in fact, several rank $d\geq 3$
 models in this space are asymptotically free in the UV. 

In this paper, we consider complex and arbitrary rank $d$ TGFT models (without gauge invariance, this is the simplest class of tensor model) written in the momentum space 
of $G_D = U(1)^D$ or $G_D =SU(2)^D$, and by introducing a free parameter $a \in ]0,1]$ as the power of momenta $p^{2a}$ in the kinetic term, we explore the space of models in order to find renormalizable theories characterized by the maximal valence of the interaction term $k_{\max}$. Any of these models can be written 
\beq
(_{\dim G_D}\Phi^{k_{\max}}_d, a), \quad a\in ]0,1], \quad D\in \N\setminus\{0\},
\quad k_{\max}\in 2\N\setminus\{0,2\}\, \quad \text{ and } \quad d\in \N\setminus\{0,1\}\,.
\eeq  
Note that our study includes the case of matrix models recovered for $d=2$. 

This work reports the following new results: 

$\bullet$ A multi-scale analysis and power-counting theorem (Theorem \ref{powcont}) for any theory of rank $d$ with kinetic term with at most quadratic momenta and with field background space $SU(2)^D$ and $U(1)^D$. Note that in all previous works discussing renormalization 
of tensor models except two (\cite{BenGeloun:2012pu,Geloun:2012bz}), the authors perform their analysis by restricting to a unique group $SU(2)$ and $U(1)$ with exactly Laplacian dynamics. 
The present analysis allows to address in a row several other possible renormalizable tensor models. 

$\bullet$ The tensor models $(_1\Phi^6_{3}, a=\frac23)$ over $G=U(1)$, 
$(_2\Phi^4_{3}, a=1)$ over $G=U(1)^2$, 
$(_1\Phi^4_{4}, a=\frac34)$ over $G=U(1)$, 
$(_1\Phi^4_{5}, a=1)$ over $G=U(1)$, are all just-renormalizable 
(Theorem \ref{theo:rentensor}). 
These models, in addition to 
$(_1\Phi^{6}_{4}, a=1)$ over $U(1)$ \cite{BenGeloun:2011rc}
and $(_1\Phi^{4}_{3}, a=\frac12)$  over $U(1)$ \cite{BenGeloun:2012pu},
are all are asymptotically free in the UV (Section \ref{subsect:tensbeta}). 
 
$\bullet$ There is a family parametrized by $k \in \N \setminus\{0,1\}$ of rank 3 tensor models $(_1\Phi^{2k}_{3}, a=1-\frac1k)$ over $G=U(1)$, which are all potentially just-renormalizable.  In the following, we refer such a family of models parametrized by the maximal valence of the interactions to as a tower of models. The family described above is 
clearly missed if
one restricts the analysis to models endowed with Laplacian dynamics. 

$\bullet$ There are two towers of matrix models $(_1\Phi^{2k}_{2}, a=\frac12(1-\frac{1}{k}))$ over $G=U(1)$ and 
$(_2\Phi^{2k}_{2}, a=1-\frac{1}{k})$ over $G=U(1)^2$,
supplemented by the models 
$(_3\Phi^6_{2}, a=1)$ over $G=U(1)^3$ or $G=SU(2)$,
$(_3\Phi^4_{2}, a=\frac34)$ over $G=U(1)^3$ or $G=SU(2)$,
$(_4\Phi^4_{2}, a=1)$ over $G=U(1)^4$
which are just-renormalizable (Theorem \ref{theo:renormat}). 

$\bullet$ The matrix models $(_1\Phi^4_{2}, a)$ over
their corresponding group have all a vanishing $\beta$-function at one-loop (Section \ref{subsect:matbeta}). Very likely, they will be  
all asymptotically safe at all orders like the GW model \cite{Grosse:2004yu,Grosse:2012uv}. 
The models $(_1\Phi^6_{2}, a)$ over their corresponding group
have all a Landau ghost \cite{Rivasseau:1991ub}. 

$\bullet$ For $k \geq 2$, the towers of matrix models,  
$(_1\Phi^{2k}_{2}, a=\frac12)$ over $G=U(1)$,  and
$ (_2\Phi^{2k}_{2}, a=1)$  over $G=U(1)^2$, and the
tower of rank 3 tensor models $(_1\Phi^{2k}_{3}, a=1)$ over $G=U(1)$, 
define all super-renormalizable models (Theorem \ref{theo:superren}).

\medskip 

The plan of this paper is as follows. Section \ref{sect:graphs} defines
the combinatorial ingredients and the topological content of the category of graphs which will support the perturbative expansion of the models
discussed in this work. Section \ref{sect:justren} is devoted to the construction of models and the ensuing multi-scale analysis leading to the general power counting theorem for a large class of models. 
We determine specific criteria for seeking super- and just-renormalizable tensor models. Section \ref{sect:rank3} achieves the proof of 
the renormalizability of some rank $d\geq 3$ tensor models
up to maximal valence of the interaction of order 6. 
Section \ref{sect:matrix} provides a similar analysis and the proof of 
the renormalizability of several matrix models
up to a finite but arbitrary maximal valence of the interaction. 
Super-renormalizable models are discussed in Section \ref{sect:Supren}
whereas Section \ref{sect:beta} undertakes the computation of the first order of $\beta$-functions of all just-renormalizable models find in this work up to maximal valence of the interaction of order 6. 
Section \ref{concl} wraps up our results and some consequences. 
An appendix provides the proof of some claims in the
main text together with interesting illustrations and peculiar features of  the real matrix model case.

\section{Colored and uncolored tensor graphs}
\label{sect:graphs}

Colored tensor models \cite{color, Gurau:2011xp} expand in perturbation theory as colored Feynman graphs which have a rich stranded structure. From these colored tensor graphs, one builds another type of graphs called uncolored \cite{Gurau:2011tj,Bonzom:2012hw} which will be the useful category of graphs  we will be dealing with. In this section, for the sake of self-containedness of this work, we  quickly review the basic definitions of  these combinatorial objects and concepts in above references. 

\subsection{Combinatorial and topological structures on colored tensor graphs}
\label{subsect:colgra}

\begin{figure}[h]
\begin{center}
\includegraphics[width=3cm, height=1cm]{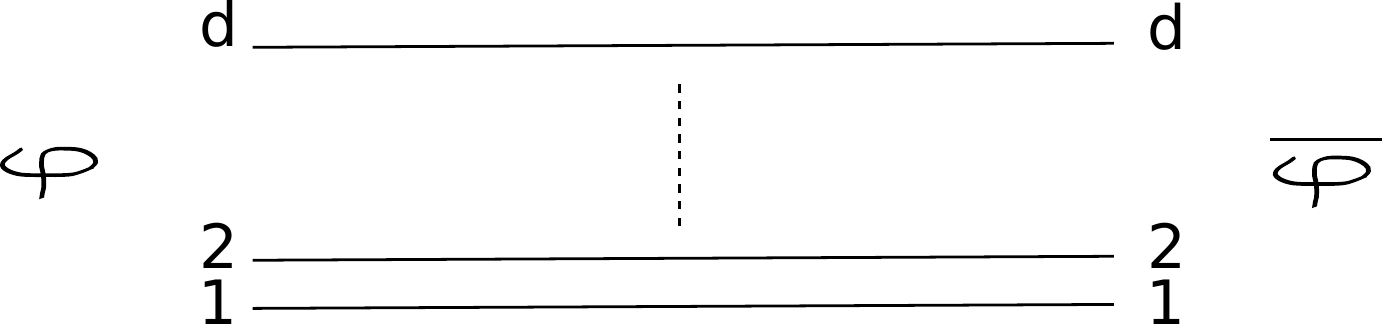}  
\caption{{\small The propagator or line in a rank $d$ tensor model
decomposes in $d$ labeled strands. }}
\label{fig:prop}
\end{center}
\end{figure}

\noindent{\bf Colored tensor graphs.} 
In a rank $d$ colored tensor model, a graph is a collection of edges or lines and vertices glued together according to quantum field theory rules
 dictated by the field measure.  In such theory, a graph (or tensor graph) has a stranded structure because its main ingredients obey the following properties \cite{avohou}: 

- each edge corresponds to a propagator and is represented by a line with $d$ strands, see Figure \ref{fig:prop}
(fields $\varphi$ are half-lines with the same structure);

- there exists a $(d+1)$ edge or line coloring;

- each vertex has coordination or valence $d+1$ with the complete graph $K_{d+1}$-type connection between its legs, namely each leg connects all half-lines hooked to the vertex. Due to the stranded structure at the vertex and the existence of an edge coloring, one defines a strand bi-coloring which associates to each strand leaving a leg of color $a$ and joining a leg of color $b$, $a\neq b$, in the vertex
the couple of colors $(ab)$;

- there are two-types of vertices, black and white and one enforces
that the graph is bipartite. This also provides an orientation to all lines,
say each line leaves a black vertex to a white vertex. 

\begin{figure}[h]
 \centering
     \begin{minipage}{.7\textwidth}
\includegraphics[angle=0, width=10cm, height=3cm]{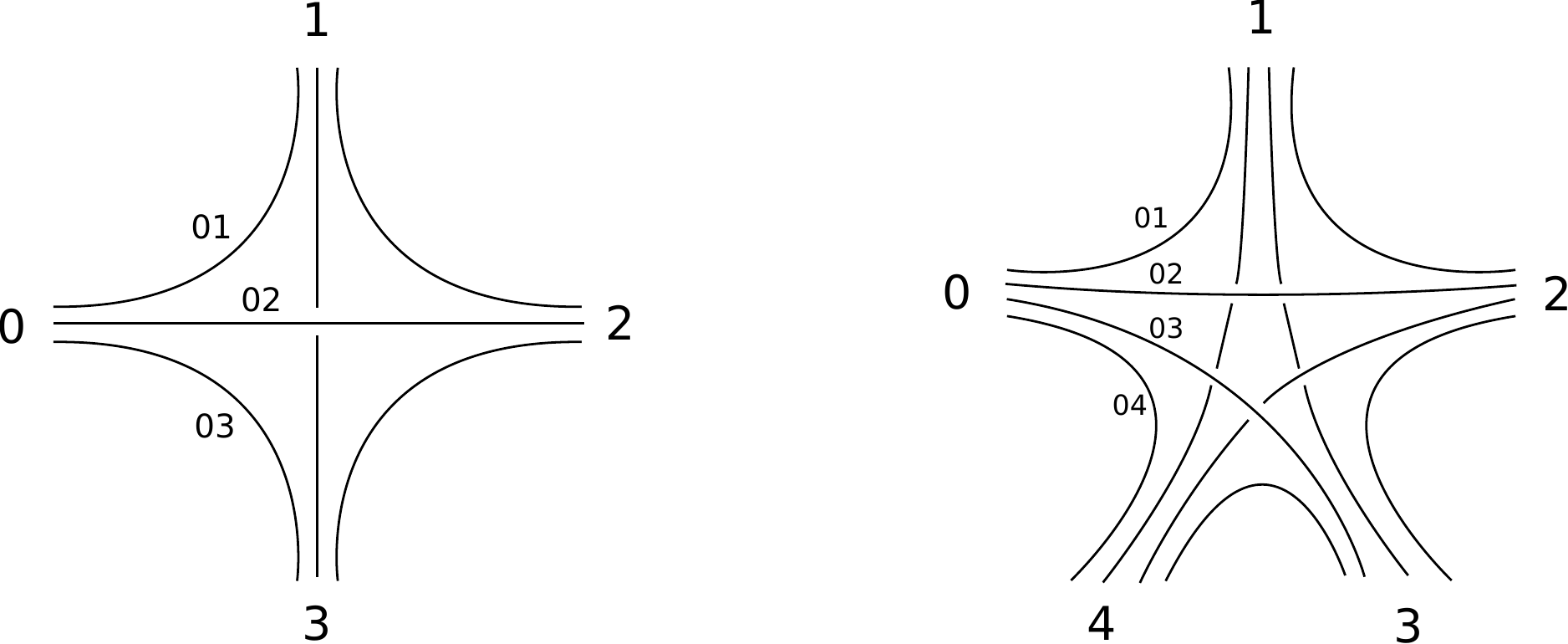} \\
\caption{ {\small Two vertices in rank $d=3$ (left) and $d=4$ (right)
colored models: they connect like the $K_{d+1}$-graph. Each leg of the vertex points towards a different colored line providing a bi-coloring on  strands. }}\label{fig:cvert}
\end{minipage}
\end{figure}

\begin{figure}[h]
 \centering
     \begin{minipage}{.7\textwidth}
\includegraphics[angle=0, width=6cm, height=2.5cm]{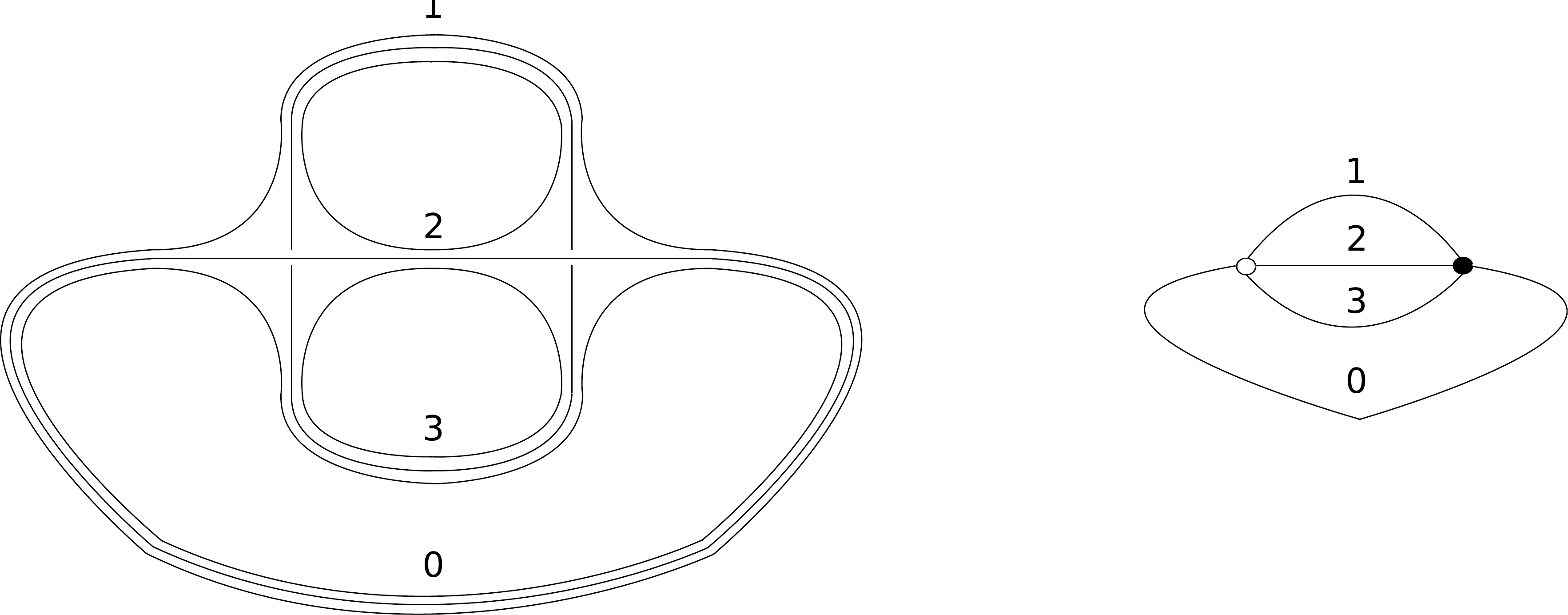} \\
\caption{ {\small A rank 3 colored tensor graph (left) and its compact colored bi-partite representation (right). }}\label{fig:com}
\end{minipage}
\end{figure}

Illustrations on rank $d=3,4$ white vertices are provided in Figure \ref{fig:cvert}. Black vertices have a very similar structure 
but with labels denoted counterclock-wise. The labels of the black 
vertices possess a bar. An example of a rank 3 colored tensor graph is given by Figure \ref{fig:com} (left). 
We will also use simplified diagrams and collapse all
the stranded structure into a simple colored graph capturing all the information of the former (see Figure \ref{fig:com}). 
The result of a collapse procedure is called simplified, compact or collapsed colored graph.

In \cite{Gurau:2010nd}, Gurau proves that rank $d$ colored tensor graphs are
dual to simplicial pseudo-manifolds in dimension $d$. This property might deeply matter 
if one would expect to achieve that the type of topological spaces
generated by some effective action of the model describes or includes
manifolds with a nice topology and smooth geometry like the one of our spacetime.
The color prescription drastically reduces the type of simplicial 
complexes possibly spanned by the partition function.

\medskip 

\begin{figure}[h]
 \centering
     \begin{minipage}[t]{.8\textwidth}
\includegraphics[angle=0, width=6cm, height=2.5cm]{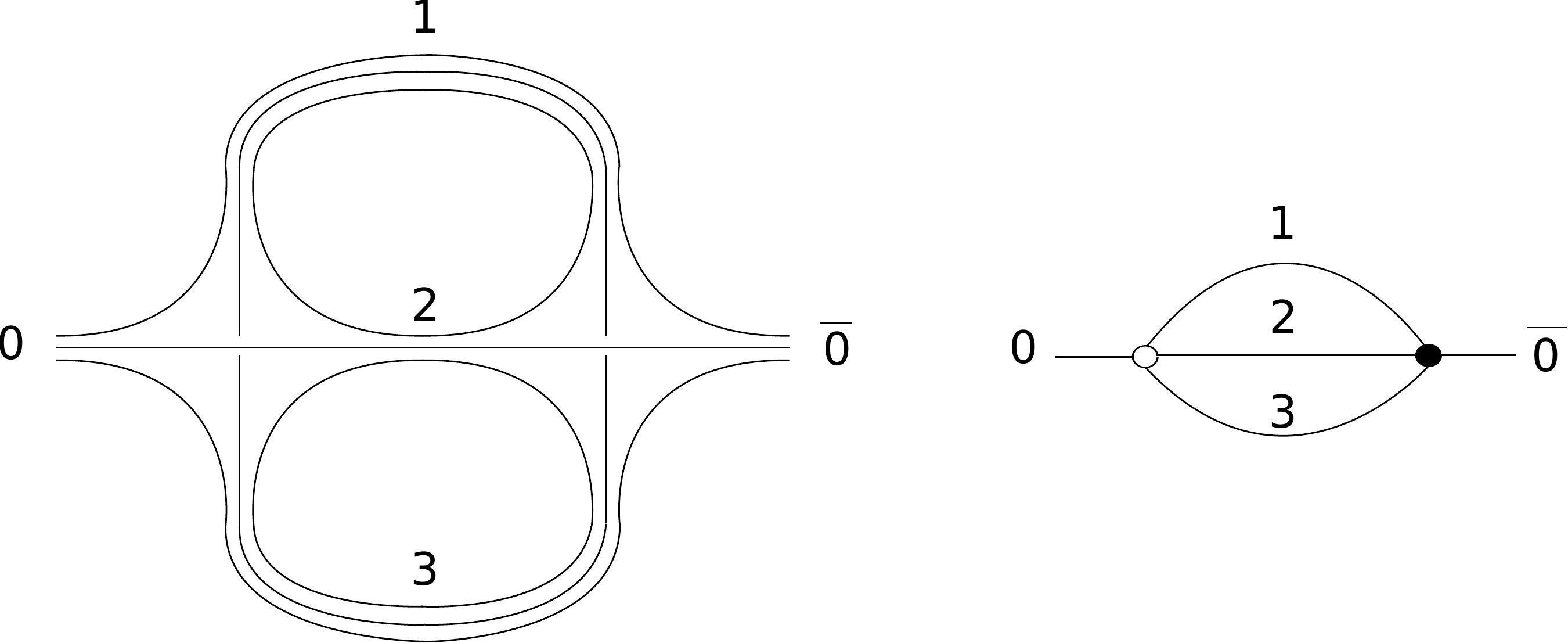} 
\hspace{0.5cm}
\includegraphics[angle=0, width=6cm, height=2.5cm]{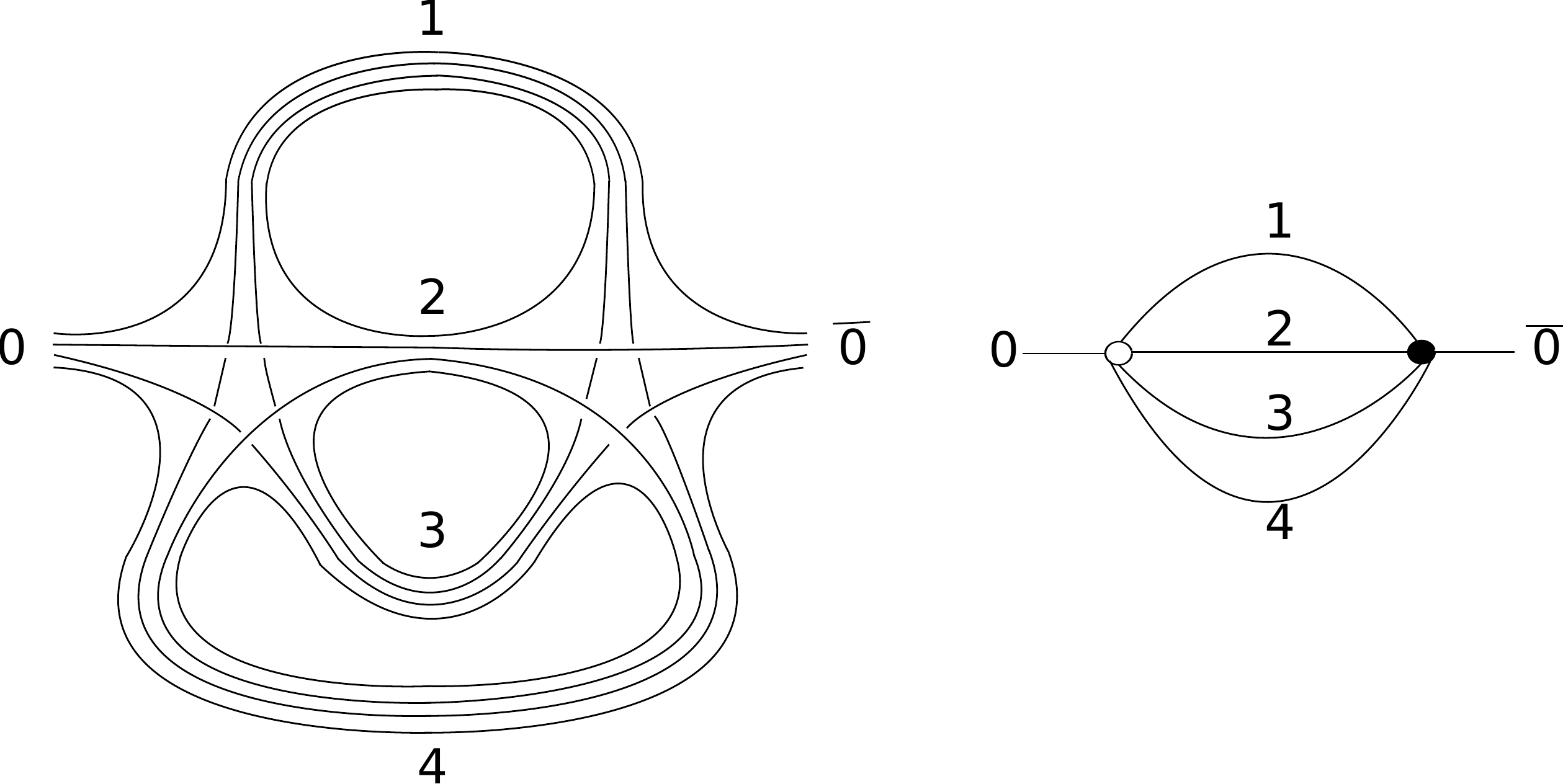} 
\vspace{0.5cm}
\caption{ {\small Rank 3 ($\cG_{3d}$) and 4 ($\cG_{4d}$) colored open tensor graphs and their compact  representation. }}\label{fig:open}
\end{minipage}
\put(-290,-10){$\cG_{3d}$}
\put(-100,-10){$\cG_{4d}$}
\end{figure}

\noindent{\bf Open and closed graphs.}
A graph is said to be open if it contains half-lines incident
to a unique vertex otherwise it is called closed. One refers such 
half-lines to as external legs representing, from the field theory  
point of view  external fields.
Examples of  open graphs  are given in Figure \ref{fig:open}.

\medskip 

\begin{figure}[h]
 \centering
     \begin{minipage}[h]{.7\textwidth}
     \centering
\includegraphics[angle=0, width=5cm, height=2cm]{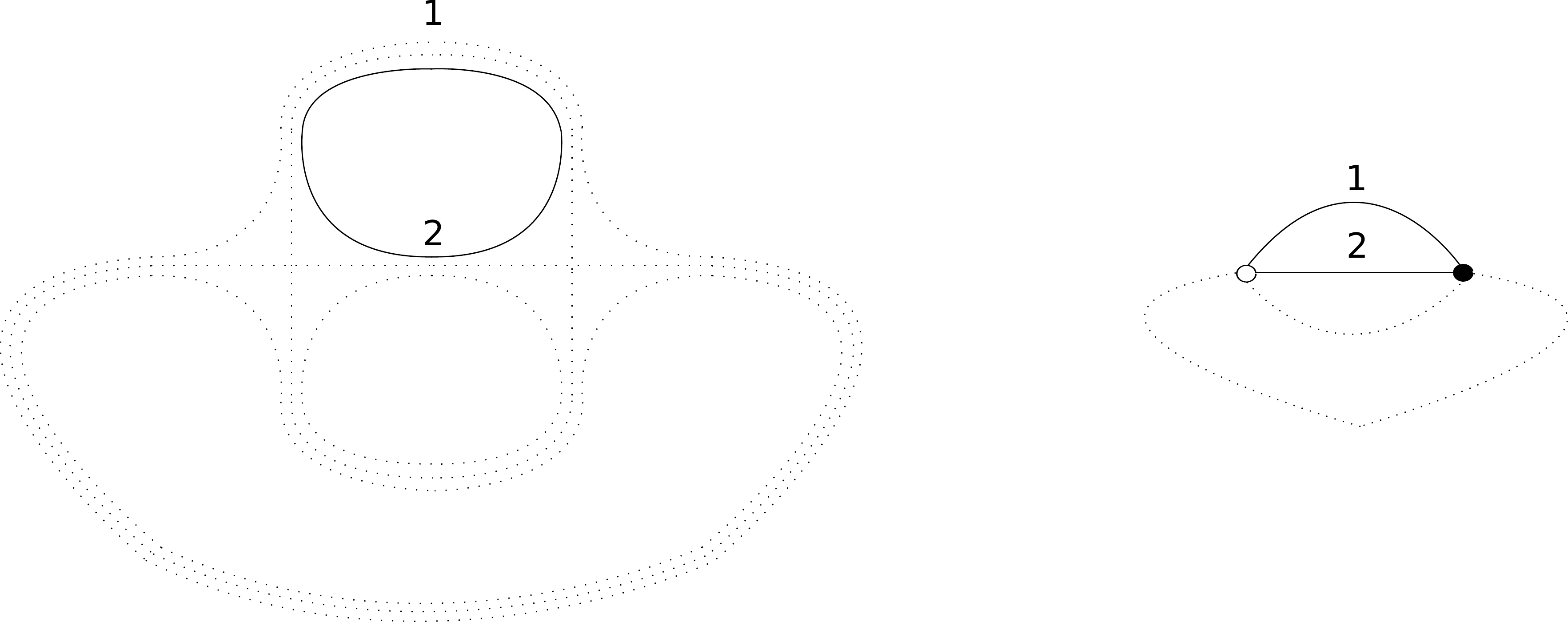} 
\hspace{1cm}
\includegraphics[angle=0, width=5cm, height=2cm]{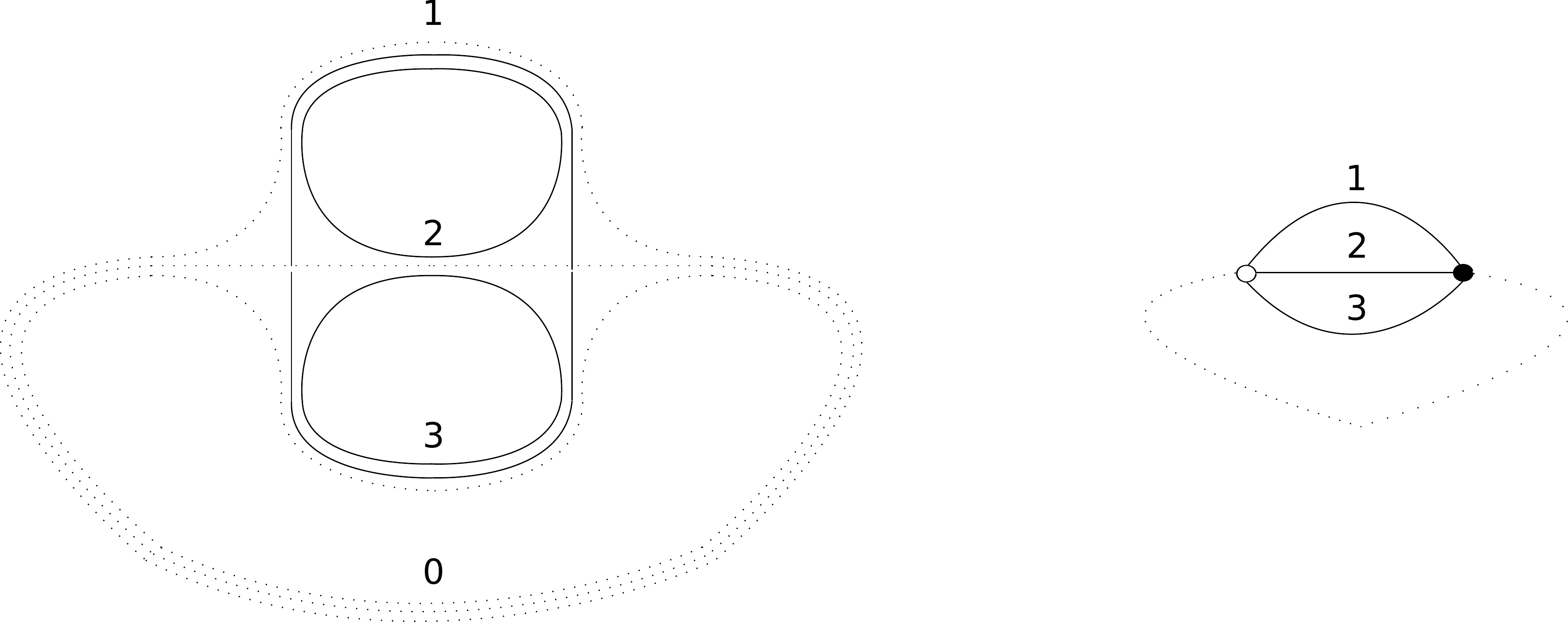}
\caption{ {\small Shading colors 0 and 3 in the  graph of Figure \ref{fig:com}, one gets a bi-colored face $f_{12}$ (left). Shading the color 
0, one obtains a 3-bubble $\bee_{123}$ (right).}} \label{fig:face}
\end{minipage}
\put(-235,35){\scriptsize{$f_{12}$}}
\put(-70,35){$\bee_{123}$}
\end{figure}

\noindent{\bf $p$-bubbles and faces.}
Colored tensor graphs in any rank $d$ have a  cellular structure \cite{color}. 
In rank $d\geq 3$, apart from vertices and edges, there exist several 
other components in the graphs. Call $p$-bubble a maximally connected
component subgraph\footnote{A subgraph of a graph $\cG$ is defined by 
a subset of edges of $\cG$ together with their incident vertices.} of the collapsed colored graph 
associated with a rank $d$ colored tensor graph, where $p$ is the number
of colors of edges used to define that subgraph. Maximally connected
because, given a colored graph, the set of $p$-bubbles
can be found by removing $(d+1-p)$ colors in the graphs
and simply observing the remaining connected components. 
Thus, $0$-bubbles are vertices,
$1$-bubbles are lines themselves. In a rank $d\geq 3$, there exist other
important components called faces which are $2$-bubbles. A {\bf face}
is a connected component in the graph made with 2 colors. Faces can be viewed in the simplified colored graph as cycles of edges with alternate colors. 
Next, {\bf $3$-bubbles} can be illustrated as connected subgraphs made
with 3 colors, etc. Examples are given in Figure \ref{fig:face}.

Some remarks and terminology can be introduced at this level: 

- Within this simplified colored picture, a line $l$ may be contained in a 
$p\geq 1$-bubble $\bee$ and we write $l \in \bee$. We say 
that ``$\bee$ passes through the line $l$.'' 

- Coming back to the full expansion of the colored graph using strands, a face 
is nothing but a connected component made with one strand. The color
of this strand alternates when passing through the edges which defines
the face.  

- A $p$-bubble is open if it contains an external half-line otherwise
it is closed. For instance, there exists three open 3-bubbles 
($\bee_{012}, \bee_{013}$ and $\bee_{023}$) and
one closed bubble ($\bee_{123}$) in the graph $\cG_{3d}$ of Figure \ref{fig:open}.

\medskip 

\begin{figure}[h]
 \centering
    \begin{minipage}[t]{.8\textwidth}
      \centering
\includegraphics[angle=0, width=10cm, height=2.5cm]{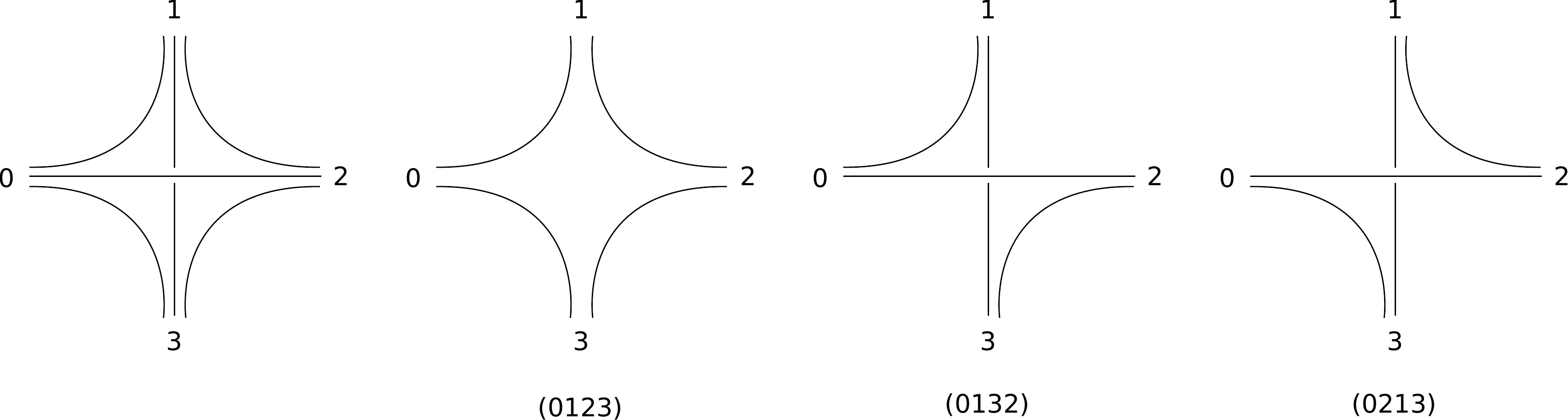} \\
\vspace{0.8cm}
\includegraphics[angle=0, width=8cm, height=2.5cm]{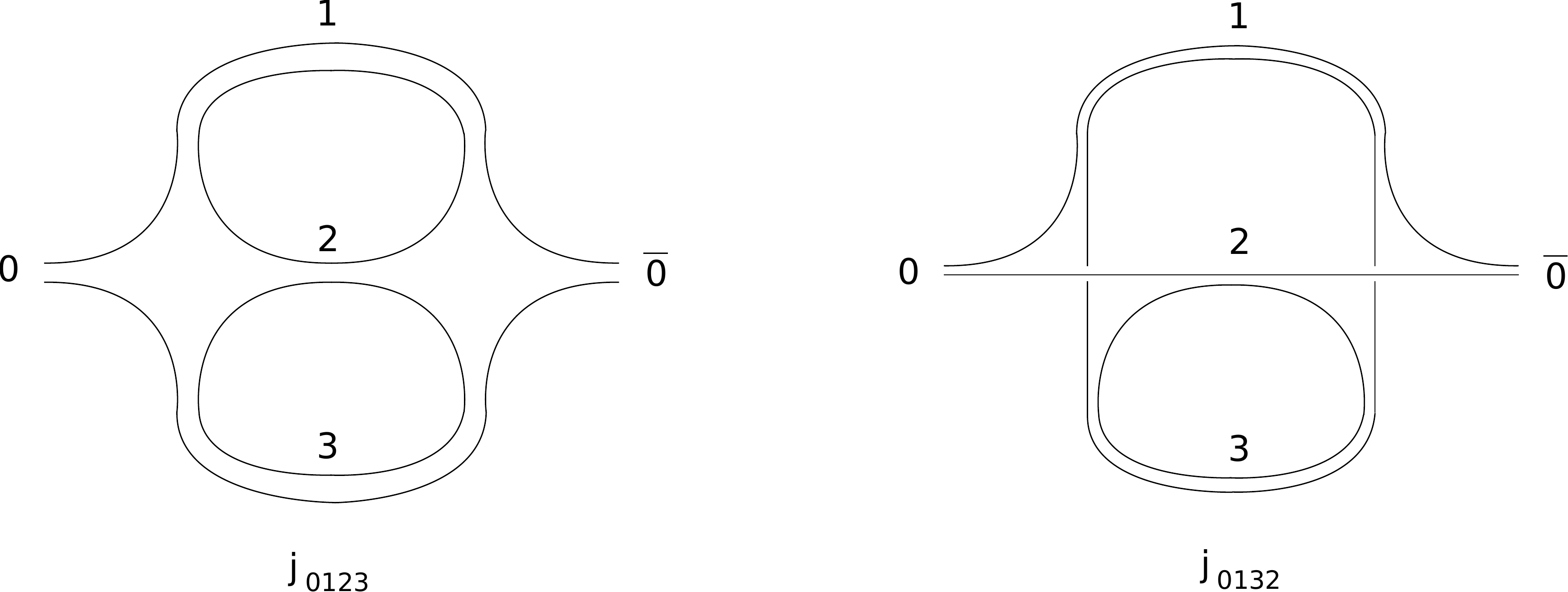}
\caption{ {\small In the rank 3 colored model, the vertex decomposes
in cycles $(0123)$, $(0132)$ and $(0213)$ (top). 
Two open jackets of the rank 3 colored graph $\cG_{3d}$ of Figure \ref{fig:open} and its jackets $J_{0123}$ and $J_{0132}$ associated with 
the color permutation $(0123)$ and $(0132)$, respectively.}}\label{fig:jack}
\end{minipage}
\end{figure}

\noindent{\bf Jackets.} Jackets are a ribbon graphs lying within 
a colored tensor graph. They are proved to be associated
with Heegaard splitting surfaces for the triangulation (simplicial 
complex) dual to the colored tensor graph \cite{Ryan:2011qm}. 
Combinatorially \cite{Gur4}, a jacket in a rank $d$ colored tensor graph
is defined by a permutation of $\{1,\dots, d\}$ namely $(0,a_1,\dots,a_d)$, $a_i\in [\![1,d ]\!]$, up to orientation. In practice, one splits the $(d+1)$-valent vertex into cycles of colors using only the strands with color pairs $(0a_1),(a_1a_2),\dots, (a_{d-1}a_{d})$.  See Figure \ref{fig:jack}, for an illustration. The same applies
for the edges. Hence, in a jacket, the above collection of pairs
defines the face bi-coloring.

A jacket of a colored tensor graph $\cG$ is nonlocal in the sense that, given a permutation, it only depends on the overall structure of $\cG$. 
For instance the number of jacket in a rank $d$ colored graph 
is given by $d!/2$ or simply the number of permutations of $[\![1,d ]\!]$
up to orientation, the number of vertices and the number of edges of a jacket 
equal the number of vertices and the number of  edges of its spanning graph, respectively.  
Meanwhile $p$-bubbles of $\cG$ are local in the sense that they depend on the local structure at each vertex of $\cG$. It is because of their 
nonlocal feature
that jackets turn out to be significant for the power-counting and divergence
analysis of graphs in tensor models. We will come back on this point later on. 

An open jacket keeps the above sense that it is a jacket touching
an external leg (see example in rank 3 in Figure \ref{fig:jack}). 

\medskip

\noindent{\bf Boundary graph.} We aim at studying tensor graphs with 
external legs. From the quantum field theory perspective, external legs
or fields are used as probes for events which might happen at much higher scale. In the present context, tensor graphs with external legs are viewed
as simplicial complexes with boundaries. The latter play the role of the above probes.
 There is a way to 
understand this boundary as  a simplicial complex itself in the colored
case  \cite{Gurau:2009tz}.  We can re-translate the boundary complex 
of a rank $d$ colored graph as a tensor graph with two peculiarity (1) its rank gets lowered to $d-1$ and (2) it possesses an vertex-edge coloring which will define in a moment \cite{avohou}.
The procedure for achieving this mapping (from boundary to rank $d-1$ colored graphs)  is known as ``pinching'' or closing open tensor graphs \cite{Gurau:2009tz}. This can be simply illustrated by the insertion of a $d$-valent vertex at each external leg of a rank $d$ tensor graph. As an effect of this $d$-valent vertex insertion,
we define the boundary $\bG$ of a rank $d$ colored tensor graph 
$\cG$ to be the graph 

- the  vertex set of which is one-to-one with the set of external 
legs of $\cG$;

- the edge set of which is one-to-one with open faces of $\cG$.

\begin{figure}[h]
 \centering
     \begin{minipage}[t]{.8\textwidth}
      \centering
\includegraphics[angle=0, width=6cm, height=2cm]{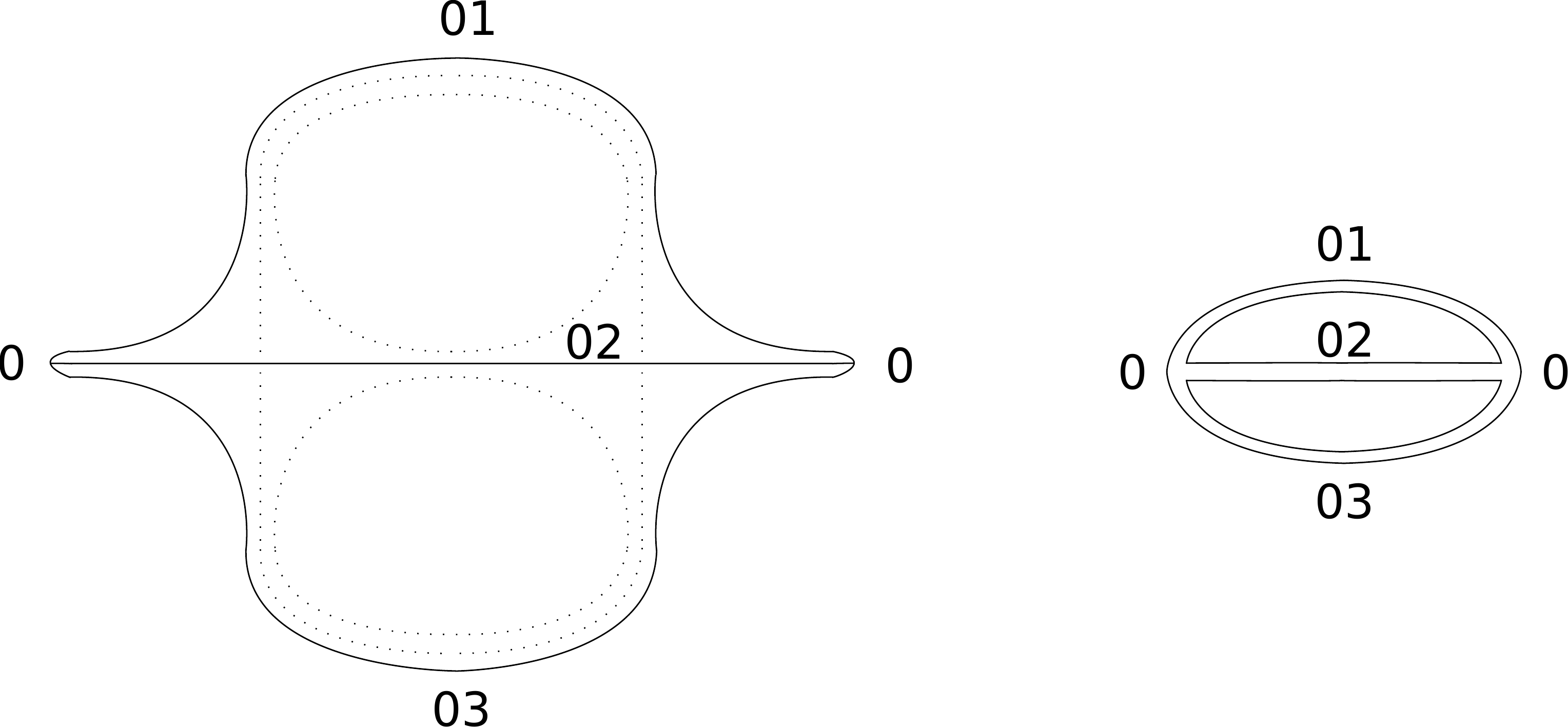} 
\hspace{0.5cm}
\includegraphics[angle=0, width=7cm, height=2cm]{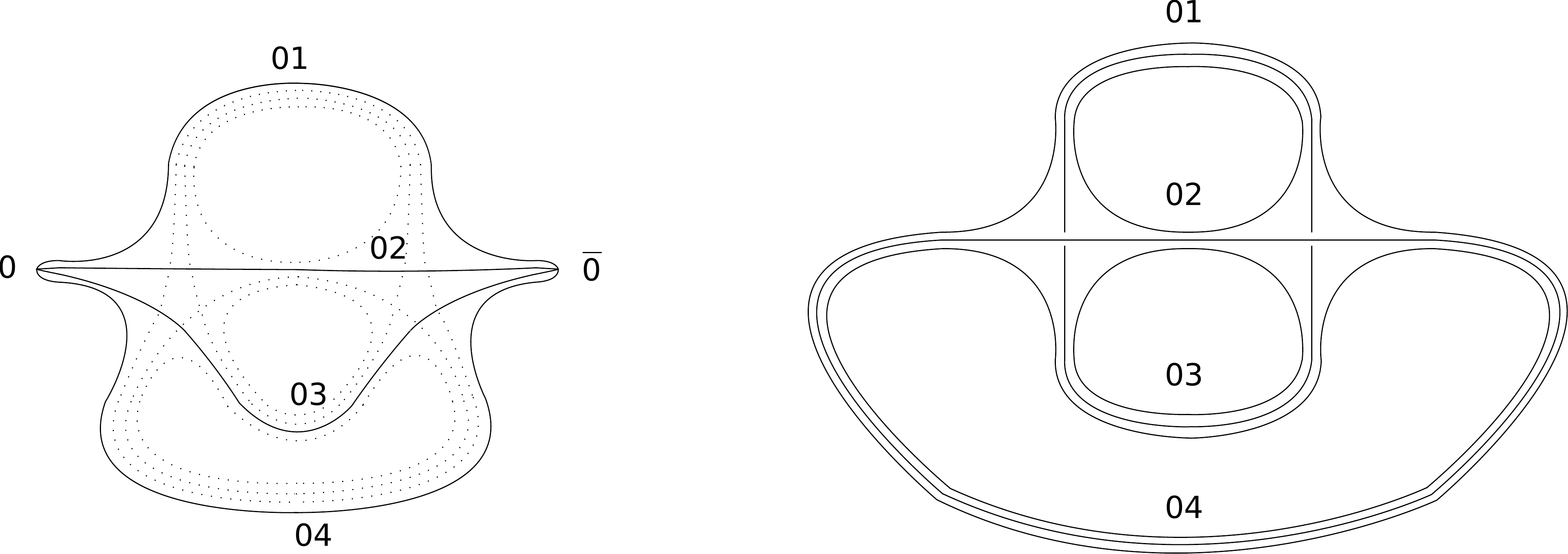}
\vspace{0.5cm}
\caption{ {\small The boundary graph $\bG_{3d}$ (and its ribbon structure) of $\cG_{3d}$ of Figure \ref{fig:open} is obtained by inserting a 3-valent vertex at each external leg and shading the closed internal faces. Similarly, $\bG_{4d}$ 
(and its internal rank 3 structure) is the boundary of $\cG_{4d}$
of Figure \ref{fig:open}.}} \label{fig:bound}
\end{minipage}
\put(-300,-10){$\bG_{3d}$}
\put(-130,-10){$\bG_{4d}$}
\end{figure}

The boundary graph has a vertex coloring inherited from the edge
coloring and has an edge bi-coloring coming from the bi-coloring
of the (external) faces of the initial graph. See Figure \ref{fig:bound} 
as illustrations of some boundary graphs. 
The boundary $\bG$ of a closed rank $d$ colored tensor graph $\cG$
is empty. The boundary graph is always closed.

Note that reducing to rank $d=3$, the boundary of a rank $3$ colored
tensor graph is a rank 2 tensor graph. Hence, it forms a ribbon graph coinciding
with its unique jacket. For rank $d\geq 4$, the boundary graph has a higher rank internal structure itself. For instance, it has $p$-bubbles and jackets that we will denote by $\bJ$. 

\medskip 

\noindent{\bf Degree of a colored tensor graph.} By organizing the
divergences occurring in the perturbation series of rank $d$ colored tensor graphs, the success of finding a 1/N expansion for amplitudes (here $N$ is some large 
size of the tensor labels) relies on the introduction of the quantity \cite{Gur4}
\bea
\omega(\cG) = \sum_{J} g_{J}\,,
\eea   
where $g_{J}$ is the genus of the jacket $J$ and the sum is performed over all jackets in the colored tensor graph $\cG$. 
The quantity $\omega(\cG)$ is called degree of $\cG$ and is useful 
to re-sum the perturbation series for the present class of models. 
Such a degree of a colored tensor graph replaces the genus of a ribbon graph in terms of which one organizes the partition function series in matrix models case. 
After rescaling of the coupling constant by a suitable power of $N$ the typical size of the 
tensor (cut-off), one finds that the amplitude $A(\cG)$ of some graph $\cG$ scales as $A(\cG) \sim N^{d-\frac{2}{(d-1)!}\omega(\cG)}$ (for a short survey see \cite{Gurau:2012vk,Gurau:2012vu}).  Gurau proves that the dominant amplitudes in the partition function of colored tensor models of any rank $d$ are of the sphere topology in dimension $d$. It is
direct to see that graphs associated with the 
most divergent amplitudes are such that $\omega(\cG)=0$. We will call these {\it melons} or {\it melonic graphs} \cite{Bonzom:2011zz}.

\subsection{Uncolored tensor graphs}
\label{sub:uncol}

Consider the partition function $Z$ of some rank $d$ colored tensor model defined by complex tensor fields denoted by $\varphi^{a}_{I}$,
where $a=0,\dots,d$ is called color of the tensor and the index $I$ collects the tensor indices. We have  
\bea
Z = \int d\nu_C(\{\varphi^{a}\}) e^{-S^{\col}[\{\varphi^a\}]}\,,
\eea
where $d\nu_C(\{\varphi^{a}\})$ is the iid Gaussian measure associated with the colored fields and related to a trivial kinetic term
of the form 
\bea
S^{\kin,\col} = \sum_{a=0}^d\sum_{I}\bar\varphi^a_{I} \varphi^a_{I}\,,
\eea 
and where 
\bea
S^{\col}=\sum_{I} \prod_{a=0}^d\varphi_{I_a}^a
 +\sum_{I} \prod_{a=0}^d\bar\varphi_{I_a}^a
\eea
is the colored interaction consisting only in identifications following the pattern of the colored vertex of rank $d$ as
discussed in Subsection \ref{subsect:colgra}. 

\begin{figure}[h]
 \centering
     \begin{minipage}[h]{.8\textwidth}
     \centering
\includegraphics[angle=0, width=10cm, height=4cm]{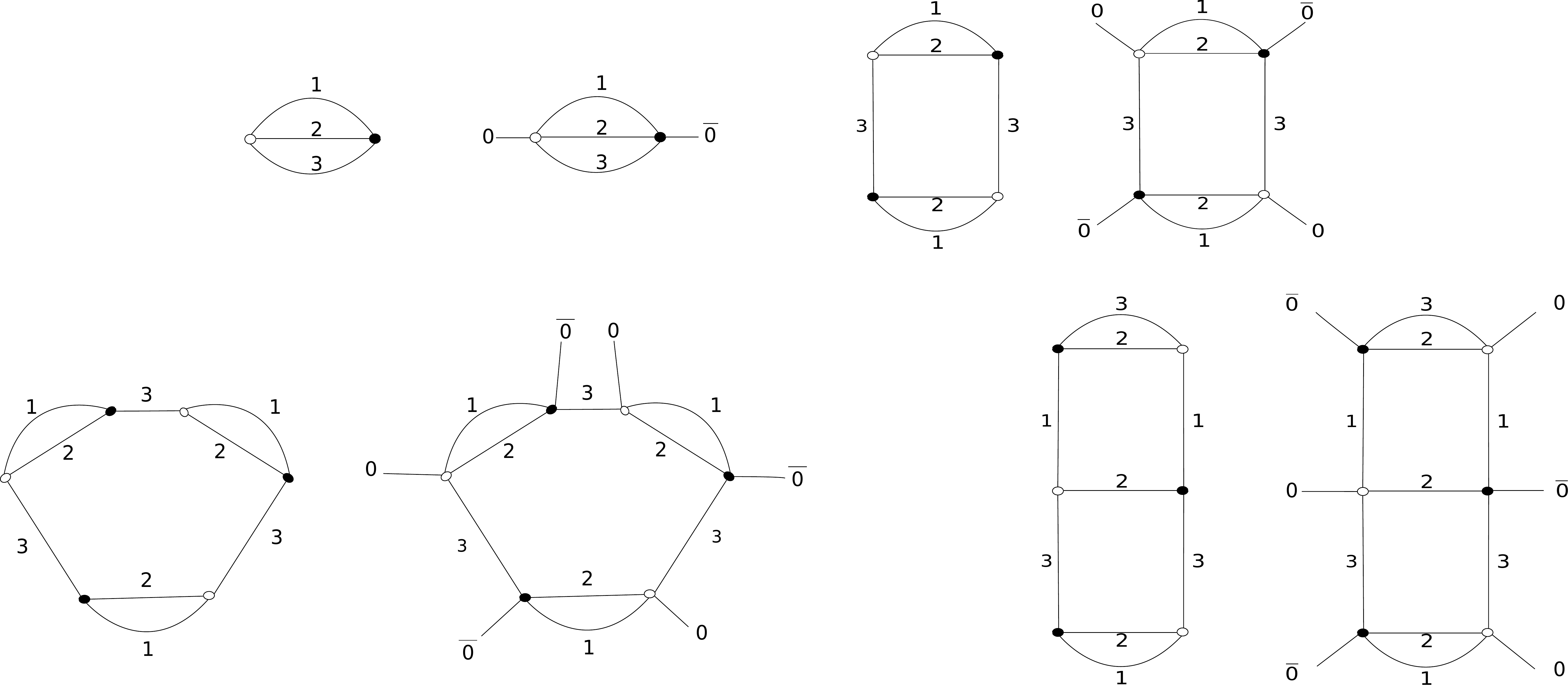} \\
\caption{ {\small Some rank 3 colored 3-bubbles and their corresponding tensor invariant. \label{fig:tensinv}}}
\end{minipage}
\end{figure}

One could partially integrate $Z$ on all but one field, say $\varphi^0$, and gets an effective
action in that remaining color: 
\bea
Z = \int d\nu_{\tilde C^0}(\{\varphi^0\}) \; e^{-S^{\uncol}[\{\varphi^0\}]}\,,
\eea
where $S^{\uncol}$ expresses in terms of the colored field $\varphi^0$ and $\bar\varphi^0$. 
The particular form of this action called ``uncolored'' \cite{Gurau:2012ix,Bonzom:2012hw} can be found elsewhere\footnote{The interested reader
is referred to one of the above references, see for instance Equation (49) in \cite{Gurau:2012ix}. }, but one can think about it as an action with infinite coupling interactions 
\bea
S^{\uncol}[\{\varphi^0\}] =\sum_{\bee \in \cB^{d}} \lambda_{\bee}\;  {\rm tr}_{\bee}[\bar\varphi^{0};\varphi^0]\,,
\label{infcoupl}
\eea
where the sum over $\bee$ is performed on the set of (vacuum) $d$-bubbles $\cB^{d}$ in the remaining colors but 0 with fixed number of vertices, 
$\lambda_{\bee}$ is some effective coupling constant associated with the tensor operator ${\rm tr}_{\bee}[\bar\varphi^{0};\varphi^0]$ which mainly implements the construction  of the $d$-bubble $\bee$ from contractions of a set of fields $\bar\varphi^0$ and $\varphi^0$. The quantity
${\rm tr}_{\bee}[\bar\varphi^{0};\varphi^0]$ is called  {\it connected tensor invariant}. In practice, the way that one understands this object
is in fact simple. 
Consider a bubble $\bee$, by increasing the valence of its vertices by one, there is a way to compose ${\rm tr}_{\bee}[\bar\varphi;\varphi]$ by adding a color to $\bee$. Thus, a connected tensor invariant is labeled by the bubble which defines it. Some connected tensor invariants in rank 3 have been illustrated in Figure \ref{fig:tensinv}.

\begin{figure}[h]
 \centering
    \begin{minipage}[t]{.8\textwidth}
\includegraphics[angle=0, width=10cm, height=2.5cm]{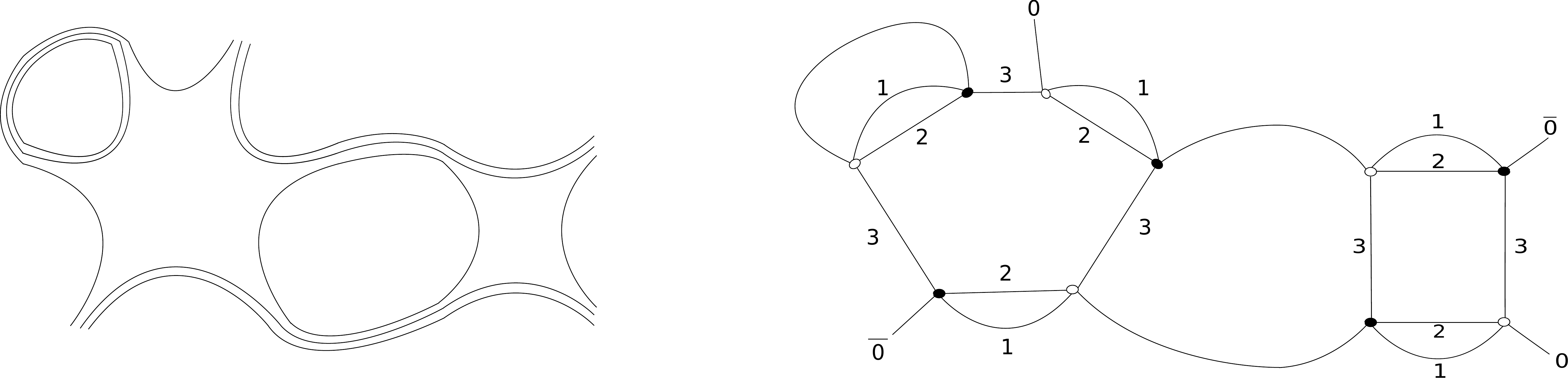} \\
\vspace{0.3cm}
\caption{ {\small A rank 3 uncolored graph $\cG$ and its associated colored extension $\cexG$. \label{fig:extcolo}}}
\end{minipage}
\put(-288,-12){$\cG$}
\put(-135,-12){$\cexG$}
\end{figure}

In the following, we shall  consider several tensor models of fixed
rank $d$ with action defined with a tensor field
associated with the last non integrated color $\varphi=\varphi^0$. 
This last color would be the one dynamical in the sense that 
there will be a nontrivial kinetic term associated with $\varphi$
so that the measure $d\nu(\{\varphi\})$ will be no longer
associated with an iid model. The interaction  will include all possible  tensor invariants in that rank $d$. An uncolored tensor graph 
in this setting is made with lines only of the last color 
and vertices consisting in tensor invariants. The other colored lines 
are integrated and should be regarded as fictitious. For instance,
see Figure \ref{fig:extcolo}. Thus, an uncolored tensor graph $\cG$ admits a rank $d+1$ color representation $\cexG$ obtained uniquely by restoration of colors. 
This procedure called ``color extension'' of a graph allows the passage
from the uncolored to the colored theories. 
 By the renormalization
prescription, we aim at truncating the infinite series \eqref{infcoupl} of interactions and keep only marginal and relevant coupling in the renormalization group (RG) sense and check that the model 
does not generate any other significant coupling. 
We mention also that a capital point in the proof of the renormalizability is the reintroduction of colors in order to get useful bounds and a clear understanding on the divergence degree of the graph.

\subsection{Rank 2, matrix or ribbon graphs}
\label{subsect:ribbon}

\begin{figure}[h]
 \centering
    \begin{minipage}[t]{.8\textwidth}
\includegraphics[angle=0, width=3cm, height=0.4cm]{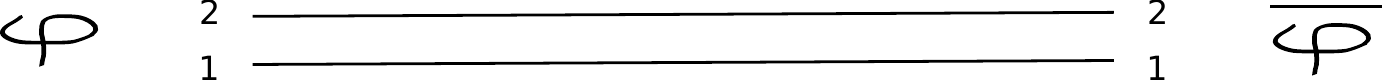} 
\hspace{0.5cm}
\includegraphics[angle=0, width=3.8cm, height=2cm]{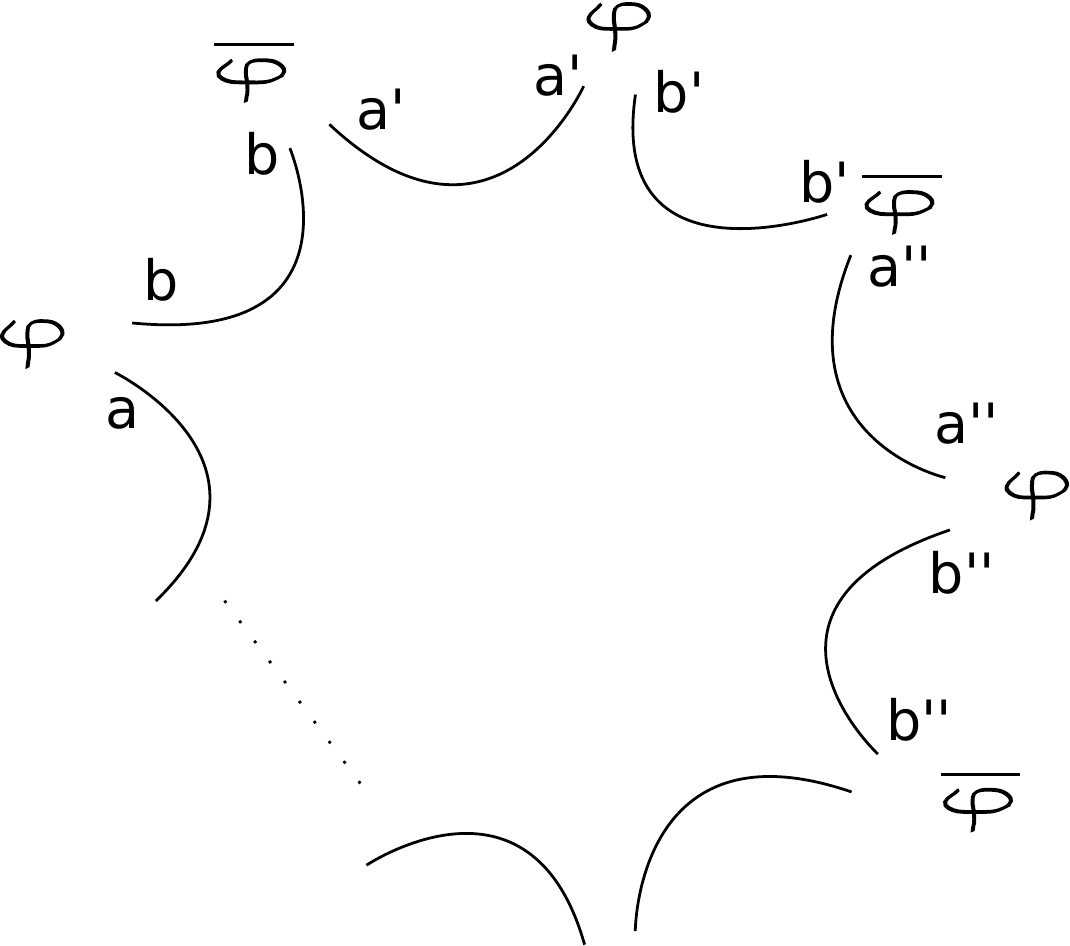} 
\vspace{0.3cm}
\caption{ {\small A ribbon edge and ribbon vertex with 
arbitrary valence. \label{fig:ribbon}}}
\end{minipage}
\put(-290,-12){ribbon line}
\put(-175,-12){ribbon vertex}
\end{figure}

In order to discuss the case of rank 2 graphs, 
we do not need  the above colored graph technology. 
A ribbon, matrix or rank 2 tensor graph is a graph made with 
lines which are ribbons and vertices which are cyclic objects
with arbitrary valence, see Figure \ref{fig:ribbon}. Note that, so formulated, one may not recognize the vertex as the same ingredient of the so-called ribbon (cyclic) graphs defined by standard combinatorics \cite{bollo}. 
In such a context, the vertex is a simple disc. We simply adopt here the quantum field theoretical perspective and put half-lines on this disc. 
Regarded as Feynman graphs of some matrix model, 
ribbon graphs are the gluing of lines corresponding to propagators
and vertices corresponding to the model interaction.  We will consider matrix models defined by complex matrix fields $\varphi_{AB}$.
The action of such models has the generic form 
\bea
S_{\text{mat}} =\sum_{AB,A'B'} \bar\varphi_{AB} K_{AB,A'B'} \varphi_{A'B'}
+ \sum_{p} \lambda_p {\text{tr}}[(\bar\varphi\varphi)^{p}]\,,
\eea
  where the kernel $K_{AB,A'B'}$ should have suitable properties so that
$S_{\text{mat}}$ makes sense, and $ {\text{tr}}$ is an ordinary matrix trace. Thus the interaction is defined as a sum of connected  unitary matrix invariants represented as in Figure \ref{fig:ribbon}.
We will not consider $K_{AB,A'B'}$ as the identity operator and, doing so
gives a non trivial dynamics for the fields and will lead us to non iid models. 

\begin{figure}[h]
 \centering
    \begin{minipage}[t]{.8\textwidth}
\includegraphics[angle=0, width=7cm, height=2.5cm]{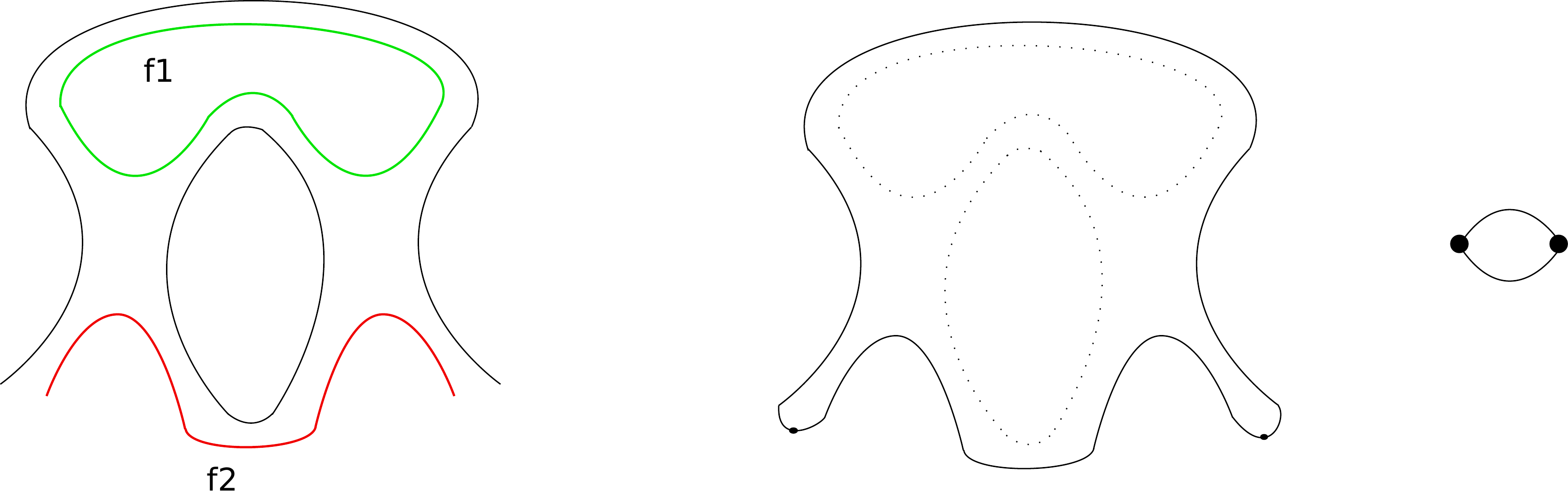} 
\vspace{0.3cm}
\caption{ {\small A open ribbon graph $\cG$ with $f_1$
a closed face and $f_2$ an open face. The boundary $\bG$ of $\cG$ after pinching and shading all closed faces of $\cG$. \label{fig:bounrib}}}
\end{minipage}
\put(-275,-12){$\cG$}
\put(-140,-12){$\bG$}
\put(-135,33){$=$}
\end{figure}

The notion of face of a ribbon face follows from the above 
description of face (forgetting the colors) as a connected component strand or, equivalently can be defined as the boundary of the ribbon graph when regarded as a geometric ribbon \cite{bollo}. Ribbon graphs can be closed  or open  if, in this last case, they have external legs (see Figure \ref{fig:bounrib}). A face can be also open or closed if it passes through an external leg. We can define a pinching procedure for ribbon graphs as well by inserting a 2-valent vertex at each external legs of the graph. The notion of boundary graph $\bG$ as the result of the pinching of some ribbon graph naturally restricts to the present situation as well (see Figure \ref{fig:bounrib}). 

\section{Seeking renormalizable models: Generic multi-scale analysis}
\label{sect:justren}

The goal of this section is to provide a list of potentially just- and super-renormalizable TGFTs models under some specific axioms. 
Our main tool for addressing this problem in full  generality is the multi-scale
analysis \cite{Rivasseau:1991ub}. We intend to give a general power-counting theorem 
and locality principle for a general class of models. 
The thorough renormalization analysis of some models detected as potentially renormalizable will be differed to next sections.

\medskip 

Constructing an action for the subsequent analysis, we do have some motivated restrictions: 

\begin{enumerate}

\item[(i)] Field are defined on a background which is a compact group manifold $G$. This is assumed for simplicity and any integral on the background position space generates an $O(1)$ factor. After Fourier transform, fields become tensors with labels in a discrete momentum space and amplitudes can still entail divergences at large momenta. Typically, we will restrict the study to $G=U(1)$ or $SU(2)$ or several copies of these groups. The case $U(1)^{\times p}\times SU(2)^{\times q}$, for $p,q$ strictly positive integers, could be deduced from this point. 

\item[(ii)] The propagator is stranded and should involve a sum momenta of the form $p^{2a}$ with $0 < a \leq 1$ associated to each field strand{\footnote{One may think about a ``duality'' between the models as will be investigated in this work and other types of
models by performing a symmetry $a \to a^{-1}$. This will deserve
a complete understanding.  }}. The upper bound $2a=2$ might be  essential in order to achieve Osterwalder-Schraeder positivity axiom \cite{Rivasseau:2012yp,Rivasseau:1991ub}. At $a=1$, one recovers an ordinary Laplacian dynamics acting on each field argument written in direct space. 

\item[(iii)] The interactions involved are unitary tensor invariant objects as discussed in Subsection \ref{sub:uncol} or unitary matrix invariants  (we will generally refer these to as ``trace invariants''). These objects belong to the sole class of interactions found so far as generating renormalizable rank $d\geq 2$ tensor theories. They provide a new and genuine notion of locality in TGFTs. Dually, the most dominant ones represent triangulations of simplicial complexes with spherical topology. This property could be of major importance in order to achieve the continuum limit of
these models as a spacetime with a large and regular topology and geometry \cite{Gurau:2013pca,Gurau:2013cbh}. 
 
\end{enumerate}

Apart from these restrictions, we should not 
exclude any possible model. We must emphasize that, 
because we are allowing quite an arbitrary power in momentum $p^{2a}$ 
in the kinetic term, a direct space formulation of several models that we shall discuss is still under investigation. For the interaction
however, the direct and momentum spaces are in perfect duality in all models discussed
below and both have a clear meaning as a basic simplexe. All models are nonlocal: the interactions occur in a region 
 rather than a definite point of the background space
and introducing an arbitrary power in momenta in the kinetic terms in these models leads to even more nonlocality. 
We finally emphasize that, the existence of such models and their significant renormalizability property are, to our modest opinion, undoubtedly useful in order to let open the door to  models which could lead to a better comprehension for a quantum theory of topology and may be even quantum gravity.

\subsection{Models}
\label{subsect:mods}

We consider a  rank $d\geq 2$ complex tensor field over a Lie compact group
$G$, $\varphi: G^{d} \to \mathbb{C}$. This field can be decomposed
in Fourier modes as
\bea
\varphi(h_1,h_2,\dots,h_d)
= \sum_{P_{I_l}} \tilde\varphi_{P_{I_1},P_{I_2},\dots,P_{I_d}} D^{P_{I_1}}(h_1) D^{P_{I_2}}(h_2) \dots D^{P_{I_d}}(h_d) \,,
\label{field}
\eea
where  the group elements $h_s \in G$ and the sum is performed on all momenta $P_{I_s}$ labeled by multi-indices $I_s$, $s=1,2,\dots,d$; $I_s$ defines the representation indices of the group element $h_s$ in the momentum space such that $D^{P_{I_s}}(h_s)$ plays the role of the plane wave in that representation. For the tensor $\tilde \varphi$, we will simply use the notation 
$\varphi_{[I]}:=\tilde\varphi_{P_{I_1},P_{I_2},\dots,P_{I_d}}$, where
the super index $[I]$ collects all momentum labels involved in 
the sum, i.e. $[I]=\{I_1,I_2,\dots,I_d\}$.  It is important to note that no symmetry under permutation of arguments 
is assumed for the tensor $\varphi_{[I]}$. We rewrite \eqref{field} 
in these shorthand notations as
\bea
\varphi(h_1,h_2,\dots,h_d) = \sum_{P_{[I]}} \varphi_{[I]}
D^{I_1}(h_1) D^{I_2}(h_2) \dots D^{I_d}(h_d) \,.
\eea
In the particular instance $d=2$, $\varphi_{I_1,I_2}$ will be referred to as 
a matrix.

For any $D\in \N\setminus \{0\}$, consider the group $G=G_D$, $h_k\in G_D$, we are interested in two cases: 

\begin{enumerate}
\item[(a)] $G_D=U(1)^D$: The representation and momentum indices 
are obtained as
\bea
&&
h_s=(h_{s,1},\dots, h_{s,D}) \in G_D\,,\;\,
h_{s,l} = e^{i \theta_{s,l}}\in U(1)\,,\;\,
D^{I_s}(h_s)= D^{P_{I_s}}(h_s) = \prod_{l=1}^D D^{s,l}(\theta_{s,l})\,,
\;\,
D^{s,l}(\theta)= e^{i p_{s,l} \theta}\,,\crcr
&&
P_{I_s} = \{p_{s,1},\dots,p_{s,D}\}\,,\;\,
I_s = \{(s,1),\dots,(s,D)\}\,,\;\,
\{I\}=\{(1,1),\dots,(1,D); \dots; (d,1),\dots, (d,D)\}\,.\crcr
&&
\label{repu1}
\eea
where $p_{s,l}\in \Z$. 

\item[(b)] $G_{D}=SU(2)^D$: In this case, the momentum space
is obtained by  the transform
\bea
&&
h_s=(h_{s,1},\dots, h_{s,D}) \in G_{D}\,,\;\,
h_{s,l} \in SU(2)\,,\;\,
D^{I_s}(h_s)= D^{P_{I_s}}(h_s) = \prod_{l=1}^D [D^{s,l}]^{j}_{mn}(h_{s,l})\,,
\crcr
&&
[D^{s,l}]^{j}_{mn}(h):=D^{j_{(s,l)}}_{m_{(s,l)}n_{(s,l)}}(h)\,,
\qquad 
D^j_{mn}(h):=\langle j,m| h |j,n\rangle\,,
\label{repsu2}
\eea
where, given $j \in \frac12\N$, $\{|j,m\rangle\}_{m,n}$ denotes the familiar basis of the spin $j$ representation space of $SU(2)$, $|m| \leq j$, $|n|\leq j$, and  $D^j_{mn}(h)$ the Wigner matrix element of $h$ in that space, so that 
\beq
P_{I_s} = \{(j_{s,1},m_{s,1},n_{s,1}),\dots,(j_{s,D},m_{s,D},n_{s,D})\}\,,
\label{repsu}
\eeq
whereas $I_s$ and $[I]$ keep their meaning as in \eqref{repu1}. 

\end{enumerate}

\begin{remark}\label{rem}
One notices that, although for $d=2$
we refer $\varphi_{I_1,I_2}$ to as a matrix, 
it can be equally regarded as a tensor itself due to 
the multi-indices carried by the representation of $G_D=U(1)^D$, 
$D >1$ or of $G_D=SU(2)^D$, $D\geq 1$. However, we shall 
not distinguish these cases to the matrix case because, mainly, 
the combinatorics and analysis as performed in the following 
of these coincide.  
\end{remark}

\medskip 

\noindent{\bf Kinetic term.} The initial task is to build a general action satisfying the above mentioned
restrictions (i)-(iii). Consider the kinetic term given in momentum space as
\beq
S^{\kin} =
 \sum_{P_{[I]}}
\bar\varphi_{P_{[I]}}
\Big(\sum_{s=1}^d |P_{I_s}|^{a} + \mu^2\Big)\varphi_{P_{[I]}}\,,
\label{skin}
\eeq
where $\mu$ is a mass term, $a \in (0,1]$, and 

- for case (a): 
$
 |P_{I_s}|^a  := \sum_{l=1}^D |p_{s,l}|^{2a}
$
and  the sum \eqref{skin} is performed over all momentum values $p_{s,l}\in \Z$;

- for case (b): 
$
|P_{I_s}|^a  := \sum_{l=1}^D [j_{s,l}(j_{s,l}+1)]^a
$
and the sum \eqref{skin} performed over all 
momentum triples $(j_{s,l},m_{s,l},n_{s,l})\in \frac12 N \times \{-j,\dots, j\}^2$. 

We will need also the following companion sums over momenta:
\beq
\text{case (a)}: \qquad 
|P_{*I_s}|^a  = |P_{I_s}|^a \,; \qquad \qquad 
\text{case (b)}: \qquad |P_{*I_s}|^a  =\sum_{l=1}^D (j_{s,l})^{2a}\,.
\eeq 

Clearly, at $a=1$, \eqref{skin} implies a Laplacian dynamics on $G_D$ and on each strand labeled by $s$. 
The corresponding Gaussian measure on tensors reads $d\nu_C(\varphi,\bar\varphi)$ and has a covariance given by
\beq
C[\{P_{I_s}\},\{\tilde P_{I_{s}}\}] = \Big[\prod_{s=1}^d \delta_{P_{I_s},\tilde P_{I_{s}}} \Big] \left(\sum_{s=1}^d  |P_{I_s}|^a + \mu^2\right)^{-1}\,,
\label{chew0}
\eeq
such that, for (a), $\delta_{P_{I_s},\tilde P_{I_{s'}}} := \prod_{l=1}^D \delta_{p_{s,l},\tilde p_{s,I} }$ and, when restricted to (b), 
$
\delta_{P_{I_s},\tilde P_{I_{s'}}} := \prod_{l=1}^D [\delta_{j_{s,l},\tilde j_{s,I} }
\delta_{m_{s,l},\tilde m_{s,I} }\delta_{n_{s,l},\tilde n_{s,I} } ]$. 
Using the Schwinger parametric integral, it is immediate to get
\beq
C[\{P_{I_s}\},\{\tilde P_{I_{s}}\}] =\Big[\prod_{s=1}^d \delta_{P_{I_s},\tilde P_{I_{s}}} \Big]\int_0^{\infty} d\alpha\,
 e^{-\alpha(\sum_{s=1}^d  |P_{I_s}|^a + \mu^2)}\,.
\label{chew}
\eeq

\noindent{\bf Interactions.}
The interactions of the model are effective interaction terms 
obtained after integrating $d$ colors in the rank $d+1$ colored model as detailed in Subsection \ref{sub:uncol}. The above kinetic term 
is defined over the remaining field $\varphi^0=\varphi$.  
The interaction is defined from unsymmetrized tensors as trace invariant objects as discussed in Subsection \ref{sub:uncol} and built from the particular contraction (or convolution) of arguments of some set of tensors $\varphi_{[I]}$ and $\bar\varphi_{[I']}$. 
This contraction is made in such a way that only the $s^{th}$ component
of some $\varphi_{[I]}$, i.e. some $P_{I_s}$, is allowed to be summed 
with the $s^{th}$ component of some $\bar\varphi_{[I']}$. Thus
the place of each index is capital in such a theory and, as a strong consequence,
the above trace invariants are connected $d$ colored graphs. A  general interaction reads: 
\beq
S^{\inter}(\varphi,\bar\varphi) = \sum_{\bee \in {\mathcal B}} \lambda_{\bee} I_{\bee}(\varphi,\bar\varphi)\,,
\label{intergen}
\eeq
where the sum is performed over a finite set ${\mathcal B}$ of rank 
$d$ colored tensor bubble graphs and  $\lambda_{\bee}$ is a coupling constant. 

One strong fact about TGFTs is that either written in 
momentum space or direct space, the vertex is the same. 
For any $I_{\bee}(\varphi,\bar\varphi)$, we associate a vertex
operator of the form of the product of delta functions
identifying entering and existing momenta.

In the case of rank $d=2$ or matrix  models, the type of interaction
considered is given by 
\beq
S^{\inter}(\varphi,\bar\varphi)=\sum_{p=2}^{p_{\max}}\lambda_p S^{\inter}_p(\varphi,\bar\varphi)\,,
\qquad 
 S^{\inter}(\varphi,\bar\varphi) = {\text{tr}} [(\bar\varphi\varphi)^{p}]\,,
\label{matrixinter}
\eeq
where $\lambda_p$ is a coupling constant. Graphically, they
are represented by cyclic graphs, see Figure \ref{fig:ribbon}. 

\medskip 
\noindent{\bf Amplitudes.}  The partition function of a generic model 
described by \eqref{skin} and \eqref{intergen} or \eqref{matrixinter} is of the form
\bea
Z = \int d\nu_C(\varphi,\bar\varphi) \,e^{-S^{\inter}(\varphi,\bar\varphi)}\,.
\label{part}
\eea
 To any connected graph 
$\cG$ made with set $\cL$ of lines and set $\cV$ of vertices, we
associate the amplitude
\beq
A_{\cG} = \kappa(\lambda)\sum_{P_{[I](v)}} \,\prod_{\ell \in \cL}
C_{\ell}[\{P_{I_s(\ell);\, v(\ell)}\},\{\tilde P_{I_{s}(\ell);\, v'(\ell)}\}] \prod_{v\in \cV;s}
\delta_{P_{I_s;\,v};P_{I_s;\, v}}\,,
\label{ampli}
\eeq
where the sum is performed on all momenta $P_{[I](v)}$ associated
with vertices $v$ where lines are hooked, and the propagator $C_\ell$ possesses line label $\ell \in \cL$. The function $\kappa(\lambda)$ includes
all coupling constants and symmetry factors. 
The specific form of the vertex operator and propagators
implies that the amplitude \eqref{ampli} factorizes in terms of connected
strand components which are the faces of the graph (in the sense
given in Subsection \ref{subsect:colgra}). There exist two types of faces: 
 open faces the set of which will be denoted $\cF_{\ext}$ and closed faces  the set of which will be denoted $\cF_{\inter}$. 
Using \eqref{chew}, we have from \eqref{ampli}:
\beq
A_{\cG} = \kappa(\lambda) \sum_{P_{I_f}} \int \Big[\prod_{\ell \in \cL} d\alpha_\ell\Big]\Big\{
\prod_{f \in \cF_{\ext}}\Big[
e^{-(\sum_{\ell \in f}\alpha_\ell) |P^{\ext}_{I_f}|^a} \Big]
\prod_{f \in \cF_{\inter}}
\Big[ 
e^{-(\sum_{\ell \in f}\alpha_\ell) |P_{I_f}|^a} \Big] 
\Big\} \,,
\label{amfa}
\eeq
where $P^{\ext}_{I_f}$ are external momenta and are not summed. One notices that the momenta $P_{I_f}$ depend
now only on closed faces. In the specific case of $G_{D}=SU(2)^D$,
since the summand is independent of momenta  $(m_{f,l},n_{f,l})$, for a face closed $f$, the sum over $P_{I_f}$ generates a factor $\dee^{2}_{P_{l_f}} $ where
\beq
\dee_{P_{l_f}} := \prod_{l=1}^D d_{j_{f,l}}\,,\qquad   d_j:=2j+1\,. 
\eeq
If the face is open, the sum over $(m_{f,l},n_{f,l})$ can be performed to the
very end and yields 1. Introducing, in the model $G_D=U(1)^D$, $\dee_{P_{I_f}}=1$ for all $I_f$, we get in full generality
\beq
A_{\cG} =  \kappa(\lambda)\sum_{P_{I_f}} \int \Big[\prod_{\ell \in \cL} d\alpha_\ell\Big]\Big\{
\prod_{f \in \cF_{\ext}}\Big[
e^{-(\sum_{\ell \in f}\alpha_\ell) |P_{I_f}^{\ext}|^a} \Big]
\prod_{f \in \cF_{\inter}}\Big[\dee_{P_{I_f}}^2
e^{-(\sum_{\ell \in f}\alpha_\ell) |P_{I_f}|^a}  \Big]
\Big\} \,,
\label{amfa2}
\eeq
where, though we still keep the notation $\sum_{P_{I_f}}$, this sum is now restricted only on spins $j_{f,l}$ in the case of a model over  $G_{D}=SU(2)^D$. 

The amplitude \eqref{amfa} is generally divergent because
of the first sum on arbitrary large momenta. Finding a well defined
regularization scheme is the purpose of the renormalization
program consisting in three steps \cite{Rivasseau:1991ub}: a multi-scale analysis
from which results a power counting theorem and the main locality principle of the model. This last point deals with the identification of the main features of the primitively diverging contributions and why they can be recast in term of initial terms in the Lagrangian. Then, one proceeds to
the proper subtractions of these divergences yielding a renormalized
theory.

\subsection{Multi-scale analysis and power counting theorem}
\label{subsect:multi}

We consider the model defined by the partition function \eqref{part}
introducing arbitrary trace invariant (rank $d\geq 3$) or planar cyclic
(rank $d=2$) polynomial interaction. 
The multi-scale analysis will be performed at this general level and, only
at the end, we will truncate the interaction series to relevant
and marginal terms supplemented, if necessary, by anomalous terms (not necessarily included in \eqref{intergen} or \eqref{matrixinter}), depending on the parameters of our theory, namely the rank $d$, the group dimension $\dim G_D$, the kinetic term parameter $a$.

The multi-scale analysis starts by a slice decomposition of the theory's 
propagator. The kernel \eqref{chew} expresses in the following way:
\bea
C=\sum_{i=0}^{\infty} \hat C_{i}\,, \qquad
\hat C_i[\{P_{I_s}\},\{\tilde P_{I_{s}}\}] &=& 
\Big[\prod_{s=1}^d \delta_{P_{I_s},\tilde P_{I_{s}}} \Big]
C_i[\{P_{I_s}\}] \,, \;\;
C_i[\{P_{I_s}\}]  =
\int_{M^{-2i}}^{M^{-2(i-1)}} d\alpha\,
 e^{-\alpha(\sum_{s=1}^d  |P_{I_s}|^a + \mu^2)}\,, \;\, \forall i \geq 0\,,\crcr
\hat C_0[\{P_{I_s}\},\{\tilde P_{I_{s}}\}] 
&=& \Big[\prod_{s=1}^d \delta_{P_{I_s},\tilde P_{I_{s}}} \Big]
C_0[\{P_{I_s}\}] \,, \;\;
C_0[\{P_{I_s}\}]  = \int_{1}^{\infty} d\alpha\,
 e^{-\alpha(\sum_{s=1}^d  |P_{I_s}|^a + \mu^2)}\,,
\label{chew2}
\eea
for some constant $M >1$. The regularization scheme requires to 
introduce a ultraviolet (UV) cut-off $\Lambda$ on the sum over $i$. 
The cut-offed propagator reads as $C^\Lambda = \sum_{i=0}^{\Lambda} C_{i}$. The following bounds hold 
\bea
\label{bslice}
\forall i \geq 1 ,&&\qquad 
C_i[\{P_{I_s}\}]\leq K_1 M^{-2i} 
 e^{-M^{-2i}(\sum_{s=1}^d  |P_{I_s}|^a + \mu^2)}
\leq K M^{-2i} 
 e^{-\delta M^{-2i}(\sum_{s=1}^d  |P_{*I_s}|^a + \mu^2)} \quad \,,\crcr
&&\qquad \qquad \qquad \;\, \leq   K M^{-2i} 
 e^{-\delta M^{-i}(\sum_{s=1}^d  |P_{*I_s}|^\frac{a}{2} + \mu^2)} \cr\cr
&&\qquad
C_0[\{P_{I_s}\}]\leq K 
 e^{-(\sum_{s=1}^d  |P_{*I_s}|^{\frac{a}{2}} + \mu^2)}
\leq K\,. 
\eea
 for some constant $K_1,K, \delta$.
Hence, for all $a\in (0,1]$, high $i$ probes high momenta $p_{s,l}$ or $j_{s,l}$
of order $M^{\frac{i}{a}}$ (or short distance on the group manifold)
and therefore,  the slice 0 refers to the infrared (IR) and the slice $\Lambda$ to the UV. 

The next stage is to find an optimal bound on the  amplitude 
$A_\cG$ for any graph $\cG$. From \eqref{ampli}, the following amplitude
is found
\bea
&&
A_{\cG} = \sum_{\bmu} \kappa_{\bmu} (\lambda) A_{\cG;\bmu} \crcr
&&
A_{\cG;\bmu} = \sum_{P_{[I](v)}} \,\prod_{\ell \in \cL}
C_{i_\ell}[\{P_{I_s(\ell);\, v(\ell)}\},\{\tilde P_{I_{s}(\ell);\, v'(\ell)}\}] \prod_{v\in \cV;s}
\delta_{P_{I_s;\,v};P_{I_s;\, v}}\,,
\label{ampli3}
\eea
where $\bmu=(i_\ell)_{\ell \in \cL}$ is a multi-index called momentum 
assignment which collects the momentum scales $i_\ell \in [0, \Lambda]$ from each propagator. 
From the point of view of the effective expansion, the constant $\kappa_{\bmu}$ collects
effective couplings corresponding to $\bmu$.  The important
quantity which needs to be analyzed is $A_{\cG;\bmu}$. 
The sum over assignments $\bmu$ will be only performed
after renormalization according to a standard procedure \cite{Rivasseau:1991ub}.

Optimal bounds on amplitudes very similar to \eqref{ampli3} have been analyzed recently, in several contexts and using different basis (direct or momentum basis) of TGFTs \cite{BenGeloun:2011rc,BenGeloun:2012pu,Carrozza:2012uv,Samary:2012bw,
Carrozza:2013wda}. The first optimal bounds have been sorted out in 
simple TGFT framework \cite{BenGeloun:2011rc} and \cite{BenGeloun:2012pu} in direct and momentum space 
and were restricted to $\dim G_D=1$ and $d=3,4$ and $a=1,\frac12$, respectively. Then, these amplitudes have been studied in   
Abelian \cite{Carrozza:2012uv,Samary:2012bw} and non Abelian \cite{Carrozza:2013wda} gi-TGFTs. Today, the state of the start is
given in \cite{Carrozza:2013wda}, the analysis has been carried out in direct space for any $\dim G_D$ and any rank $d \geq 3$. All of these works but \cite{BenGeloun:2012pu, Geloun:2012bz}, consider only Laplacian dynamics and so are defined at the point $a=1$. The purpose of the following is to gather all these results and to express 
in the momentum space and for general $\dim G_D$, $d$ and $a$, a power counting theorem for the present class of simple TGFTs. At the end, one should recover the result of \cite{Carrozza:2013wda}
but putting the contribution invoking the gauge invariance (which is the 
rank of the incident matrix between lines and faces) to 0 and one can still have a new effect induced by the parameter $a$ in the propagator. The parameter $a$ will allow us to explore the theory space without restricting to models with Laplacian dynamics.

Consider a graph $\cG$, with set of lines $\cL$ with cardinal $L=|\cL|$,
set of internal faces $\cF_{\inter}$ with cardinal $|\cF_{\inter}|=F_{\inter}$
and set of external faces $\cF_{\ext}$ with cardinal $|\cF_{\ext}|=F_{\ext}$.
Let $A_{\cG;\bmu}$ be  the associated amplitude as given in \eqref{ampli3}. 
This divergence degree of this amplitude will be expressed in terms of specific
subgraphs of $\cG$ which make transparent the notion of locality. These
subgraphs are called quasi-local (or dangerous) and are defined by a subset of lines of $\cG$ with internal scale much higher than any external scale. In symbol, a subgraph $\cG^i$ of $\cG$ is defined by
a set of lines such that $\forall \ell \in \cL(\cG^i) \cap \cL(\cG)$, $i_\ell \geq i$.  
Noting that $\cG^i$ may have several connected components, we denote  each component by $G^i_{k}$. The set $\{G^i_k\}_{i,k}$ defines quasi-local subgraphs
and will be used to rewrite the amplitude of any graph. Seeking
if a subgraph $g$ is quasi-local there is the following specific criterion: define $\cL(g)$
and $\cL_e(g)$ the set of internal and set external lines of $g$, respectively,
given a momentum assignment $\bmu$ in $\cG$, 
$i_g(\bmu)= \inf_{\ell \in \cL(g)}i_\ell$ and $e_{g}(\bmu)=\sup_{\ell \in \cL_e(g)}$,
then $g$ is quasi-local if and only if $i_g(\bmu) > e_g(\bmu)$. 

\begin{figure}[h]
 \centering
    \begin{minipage}[t]{.8\textwidth}
\includegraphics[angle=0, width=14cm, height=4cm]{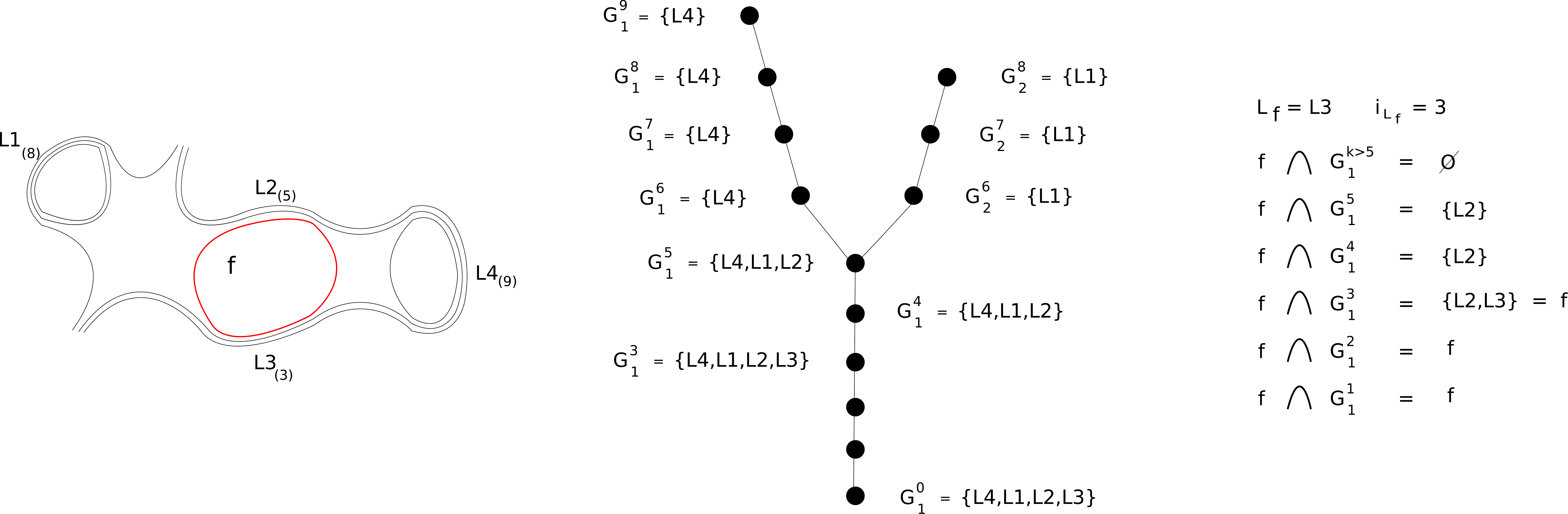} \\
\vspace{0.3cm}
\caption{ {\small A rank 3 graph $\cG$ with lines $L1,L2,L3$ and $L4$
at a given momentum attribution $(8,5,3,9)$, respectively, 
and a face composed by lines $L2$ and $L3$. 
The corresponding Gallavotti-Nicol\`o (GN) tree 
and the face optimization procedure.}} \label{fig:GN}
\end{minipage}
\put(-350,20){$\cG$}
\put(-195,-12){GN tree}
\put(-85,-12){Face optimization}
\end{figure}

The set $\{G^i_k\}_{i,k}$ is partially ordered by inclusion and forms a tree if
$\cG$ is  connected. 
In such situation, the tree has a root which is the entire graph itself $\cG=\cG^0$. This tree is called the Gallavotti-Nicol\`o tree \cite{Galla}. 
Figure \ref{fig:GN} gives an example of such a tree for a graph. The optimal bound on the amplitude of a connected graph will be found by integrating internal momenta along a tree $T$ (in  the graph) in a specific way to be compatible with the abstract Gallavotti-Nicol\`o tree.

Using the bounds \eqref{bslice}, 
 \eqref{ampli3} can be evaluated in a similar way of \eqref{amfa2} as
\beq
|A_{\cG;\bmu}| \leq K^{L} K_1^{F_{\ext}}\Big[\prod_{\ell \in \cL} M^{-2i_\ell}\Big]
\sum_{P_{I_f}} 
\Big\{ \prod_{f \in \cF_{\inter}}\Big[\dee_{P_{I_f}}^2
e^{-\delta(\sum_{\ell \in f}M^{-i_\ell}) |P_{*I_f}|^\frac{a}{2}}  \Big]
\Big\}.
\label{ampli4}
\eeq
Each sum over an internal $P_{I_f}$ used in an exponential yields 
(see Appendix \ref{app:euler} for details pertaining to the following
identities): 

- for case (a) $U(1)^D$:
\bea
\sum_{P_{I_f}}  e^{-\delta' M^{-i} |P_{*I_f}|^\frac{a}{2}} = 
\sum_{p_{f,1},\dots, p_{f,D}}  
e^{-\delta' M^{-i} (\sum_{l=1}^D |p_{f,l}|^a)} = 
\Bigg[ \sum_{p \in \Z}  
e^{-\delta' M^{-i} |p|^a} \Bigg]^D = 
 c M^{\frac{D}{a}i} (1+ O(M^{-\frac{i}{a} }))\,,
\label{facea1}
\eea
for some constant $c,\delta'$ and some scale $i$;

- for case (b) $SU(2)^D$:
\bea
\sum_{P_{I_f}}  \dee_{P_{I_f}}^2e^{-\delta' M^{-i} |P_{*I_f}|^\frac{a}{2}} &=& 
\sum_{j_{f,1},\dots, j_{f,D}}  \Big[ \prod_{l=1}^D (2j_{f,l}+1)^2
e^{-\delta' M^{-i}  |j_{f,l}|^a}\Big] 
 = \Bigg[ \sum_{p \in \frac12 \N}  (2p+1)^2 
e^{-\delta' M^{-i} |p|^a} \Bigg]^D \crcr
&=& 
 c M^{\frac{3D}{a}i} (1+ O(M^{-\frac{i}{a} })) \,,
\label{facea2}
\eea
for some constants $c,\delta'$ and scale $i$.

Finding an optimal bound on $A_{\cG;\bmu}$  requires to 
sum over $P_{I_f}$ such that each integral of an exponential  will bring
a minimal divergence. In order to satisfy this, we must choose in
the  face evaluation \eqref{facea1} and \eqref{facea2} the scale $i$ such that it corresponds to  $i_f =\min_{\ell \in f} i_\ell$. Call $\ell_f$ the line such that $i_{\ell_f}=i_f$. 
We must show that the end result after all optimal integrations of this kind is compatible with the Gallavotti-Nicol\`o tree in the sense that  result formulates in terms of the set $\{G^i_k\}$. 

Let us remark that a face $f$ becomes closed in some $G^i_k$ only if all its lines belong to that quasi-local subgraph. This means that $f$ is closed in some $G^i_k$ if $i\leq i_{l_f}$ which further implies that the set of lines contributing to $f$ close exactly in the $G^{i_{\ell_f}}_k$ and for all $i_\ell$,  
$0\leq i \leq i_{\ell_f}$, $f \in \cF_{\inter}(G^i_k)$ (this is illustrated in Figure \ref{fig:GN}).  
 Using this remark, introducing $\rho_{D,a}=\dim G_D/a$ 
and integrating \eqref{ampli4} in a optimal way using $i_f$, we expand the result using the set of quasi-local graphs (in the way of \cite{Rivasseau:1991ub}) as
\bea
|A_{\cG;\bmu}| &\leq& K^{L}  K_1^{F_{\ext}}K_2^{F_{\inter}}\Big[\prod_{\ell \in \cL} M^{-2i_\ell}\Big]
\prod_{f \in \cF_{\inter}}\Big[ M^{\rho_{D,a} i_f}  \Big] \crcr
&\leq&
K^L  K_1^{F_{\ext}}K_2^{F_{\inter}}\Big[\prod_{\ell \in \cL }\prod_{(i,k)/ \ell \in \cL(G^i_k)} M^{-2}\Big]
\prod_{f \in \cF_{\inter}}\prod_{(i,k)/ \ell_f \in \cL(G^i_k)}\Big[ M^{\rho_{D,a}}  \Big] \crcr
&\leq& K^L   K_1^{F_{\ext}}K_2^{F_{\inter}}\Big[\prod_{(i,k)} \prod_{\ell \in \cL(G^i_k) }M^{-2}\Big]
\prod_{(i,k)}\prod_{f \in \cF_{\inter}(G^i_k)}\Big[ M^{\rho_{D,a}}  \Big] \crcr
&\leq&  K^L   K_1^{F_{\ext}}K_2^{F_{\inter}}\Big[\prod_{(i,k)}  M^{-2L(G^i_k)+\rho_{D,a}F_{\inter}(G^i_k) }  \Big] .
\label{ampli5}
\eea
Changing $M$ for $M^{a}$ in the \eqref{ampli5}
leads to following statement. 
 
\begin{theorem}[Power counting theorem]\label{powcont}
Let $\cG$ be a connected graph of the model \eqref{part}, there exist
some large constants $K$ and $n$, such that
\beq\label{degdiv}
|A_{\cG, \bmu}| \leq K^n \prod_{(i,k)\in \N^2} M^{\omega_d(G^i_k)}\,,
\qquad 
\omega_d(G^i_k) = -2aL(G^i_k)  +\dim G_D\,F_{\inter}(G^i_k) \,.
\eeq
\end{theorem}

The quantity $\omega_d(\cG)$ is called the divergence degree of the graph $\cG$.
Setting $a=1$, we get from \eqref{degdiv}, as expected, the power counting theorem of 
\cite{Carrozza:2013wda} after putting to zero the rank of the incidence matrix 
associated line-face with the gauge invariance. Setting $\dim G_D=1$, $a=1$,
we obtain the power counting of \cite{BenGeloun:2011rc}, and $a=1/2$
yields the power counting of \cite{BenGeloun:2012pu}.
We also understand that, introducing a dynamics depending on $a$ has the same effect as dilating the group dimension $\dim G_D$ by the factor $a^{-1}$. If we introduced $a>1$, then the effect would be naturally to decrease the group dimension  or enhancing the damping effect that has the propagator on the amplitude. 
Note also that the above power counting is valid for matrix models.
We must then emphasize that the introduction of a
non integer power $a$ in propagator momenta might lead
to a non integer divergence degree. We will see however that, seeking renormalizable models,  the possible values of $a$ are limited to rational numbers. 

\medskip

\noindent{\bf A remark on $G_D= U(1)^{p} \times SU(2)^q$, $\dim G_D=p+3q$.} 
Considering $G_D$ as a product of $U(1)^p$ and $SU(2)^q$
is straightforward: the kinetic term builds
as sums of kinetic terms of the form \eqref{skin} in each sector. 
The slice decomposition can be performed as in \eqref{bslice}
and, after the multi-scale analysis, the final degree of divergence \eqref{degdiv} splits as
\beq
\omega_d(G^i_k) = -2aL(G^i_k)  +p\,F^{1}_{\inter}(G^i_k)
+3q\,F^{2}_{\inter}(G^i_k)\,,
\eeq
where two types of faces have to be introduced according 
to the fact that these can be generated in the $U(1)$ sector
or in the $SU(2)$ sector. This study will be  postponed
to a subsequent work.

\subsection{Divergence degree and list of potential renormalizable models}
\label{subsect:deg}

\noindent{\bf Divergence degree.}
In form \eqref{degdiv}, the divergence degree $\omega_d$ is not very 
insightful for the determination of all primitively divergent graphs. 
To fully understand this quantity, we must introduce further details on the graph which are related to the underlying color structure corresponding to the trace invariants. 

Consider a connected graph $\cG$
and its colored extension $\cexG$ as introduced in
Subsection \ref{sub:uncol}. The following result has been established for a reduced case $d=4$ \cite{BenGeloun:2011rc} and then  extended to any rank in \cite{Samary:2012bw}. 

\begin{proposition}[Number of internal faces in a rank $d\geq 3$ model]
\label{prop:face}
Let $\cG$ be a rank $d$ connected graph, $\cexG$ its colored extension,
$\bG$ its boundary with number $C_\bG$  of connected components, $V_k$ its number of vertices of coordination $k$, $V= \sum_{k} V_k$ its total number of vertices,  $n \cdot V = \sum_{k} k V_k$ its number of half lines exiting from vertices, $N_{\ext}$ its number of external legs. 
 
The number of internal faces of $\cG$ is given by 
\beq
\label{faceinter}
F_{\inter}(\cG) = -\frac{2}{(d-1)!}(\omega(\cexG) - \omega(\bG)) - (C_\bG -1)
- \frac{d-1}{2} N_{\ext} + d-1 - \frac{d-1}{4}(4-2n)\cdot V\,,
\eeq
\end{proposition}\noindent
where $\omega(\cexG)=\sum_{J}g_{\tJ}$ is the degree of $\cexG$, 
$\tJ$ is the pinched jacket associated with $J$ a jacket of $\cexG$, 
$\omega(\bG)=\sum_{\bJ}g_{\bJ}$ is the degree of $\bG$.

\proof See Proposition 3.7 in \cite{Samary:2012bw}.  \qed

Note that the number of internal faces does not depend
on the dimension of the group or the gauge invariance of the theory 
but only on the combinatorics and topology of the graph itself. 
From this proposition, we are in position to reformulate the divergence degree
of a graph. 

\begin{proposition}[Divergence degree]\label{prop:tens}
The divergence degree of a graph $\cG$ is given by 
\bea
\omega_d(\cG) &=& - \frac{2\dim G_D}{(d-1)!}(\omega(\cexG) - \omega(\bG)) - \dim G_D(C_\bG -1) \crcr
&&
 - \frac12\Big[(\dim G_D(d-1) -2a) N_{\ext} - 2\dim G_D(d-1)\Big] -\frac12\Big[ 2\dim G_D(d-1) + (2a-\dim G_D(d-1))n \Big] \cdot V\,.
\crcr
&&
\label{degree}
\eea
\end{proposition}
\proof This formula $\omega_d(\cG) $ can be easily 
obtained after substituting the combinatorial relation (we omit the dependence in the graph $\cG$)
\beq
-2L =-(n\cdot V  - N_{\ext})
\label{linvert}
\eeq
and \eqref{faceinter}  in the divergence degree \eqref{degdiv}. 
  
\qed

Since we will be also interested in matrix models, it is relevant
to understand the above power counting in the rank 2 case. 
In that situation, the following proposition holds (in the same notations). 
\begin{proposition}[Divergence degree of matrix models]
\label{prop:mat}
The divergence degree of a graph $\cG$ is given by 
\beq
\omega_d(\cG) = - 2\dim G_D g_{\tilde\cG} - \dim G_D(C_\bG -1) \
 - \frac12\Big[(\dim G_D -2a) N_{\ext} - 2\dim G_D\Big] -\frac12\Big[ 2\dim G_D+ (2a-\dim G_D)n \Big] \cdot V  \,, 
\label{dmat}
\eeq
where $\tilde\cG$ is the closed (pinched) graph associated with $\cG$.
\end{proposition}
\proof  Introducing the closed graph $\tilde \cG$
(closing all external faces by inserting a two-leg vertex at each external half-line, see Figure \ref{fig:bounrib}) and get from the
Euler characteristic of $\tilde \cG$ 
\beq
F_{\inter} = 2-2 g_{\tilde \cG} -(V-L +C_{\bG})\,.
\eeq
Substituting the last result and \eqref{linvert} in \eqref{degdiv}
yields the desired relation.

\qed 

\medskip 

\noindent{\bf Criteria for potential renormalizable models.}
There is a proof that, for any graph in this category of models (\cite{BenGeloun:2011rc}
and its addendum \cite{Geloun:2012fq})
\beq
\omega(\cexG) - \omega(\bG) \geq 0\,.
\eeq 
Also for any graph with external legs (as those of interest in the renormalization procedure) $C_\bG \geq 1$. Therefore, for any connected graph with $N_{\ext}\geq 2$, the following is valid  (introducing $d^-=d-1$)
\beq\label{ome}
\omega_d(\cG) \leq - \frac12\Big[(\dim G_Dd^- -2a) N_{\ext} - 2\dim G_Dd^-\Big] -\frac12\Big[ 2\dim G_Dd^- + (2a-\dim G_Dd^-)n \Big] \cdot V\,.
\eeq
Furthermore, in any theory rank $d\geq 3$, melonic graphs (recalling that these are defined such that $\omega(\cexG)=0$) with melonic boundary (such that $\omega(\bG)=0$) with a unique connected component on the boundary saturate this bound. Thus \eqref{ome} is  optimal. 
The particular rank $d=2$ situation is similar. 
The class of dominant graphs in power counting are planar graphs
$g_{\tilde \cG}=0$ with $C_{\bG}=1$ for which \eqref{ome} saturates
as well. Given the above bound, we are in position to investigate
$(\dim G_D, a, d)$ for super- and just-renormalizable models
having as dominant contributions the graphs saturating the bound \eqref{ome}.

Consider $D\geq 1$, $d \geq 2$, $a\in (0,1]$, and 
$k_{\max}\geq 4$ the maximal coordination among all interactions
of a graph $\cG$ ($k_{\max}=2$ leads to a quadratic trivial interactions; $k_{\max}$ odd is possible for matrix models but not when using complex matrices; $k_{\max}=3$ is also impossible in the tensor case because there is no trace invariant built from contractions of an odd number of tensors \cite{Gurau:2012ix}). 

Let us first inspect the coefficient of $N_{\ext}$. If that coefficient
turns out to vanish then
\bea
&&
\dim G_D d^- =2a \qquad \Leftrightarrow \qquad 
0 <\dim G_D (d-1) \leq 2\,,
\crcr
\dim G_D (d-1) =1 \,, &&\qquad  \dim G_D=1\,, \quad G_D =U(1)\,, \quad  d=2 \,, \qquad a=\frac12 \,;\cr\cr
\dim G_D (d-1) =2 \,, &&\qquad  \dim G_D=2\,, \quad G_D =U(1)^2\,, \quad  d=2 \,, \qquad a=1 \,,\cr\cr
&&\qquad  \dim G_D=1\,, \quad G_D =U(1)\,, \quad  d=3 \,, \qquad a=1\,.
\label{fort}
\eea
Consider graphs $\cG$ in the models \eqref{fort} such that $\omega(\cexG)=0=\omega(\bG)$ and $C_{\bG}=1$, the degree of divergence of these graphs is given by 
\beq
\omega_d(\cG) = 2(1 - V)\,.
\label{sup0}
\eeq
Then, in these theory only graphs $\cG$ 
such that $\omega(\cexG)=0=\omega(\bG)$ (or $g_{\tilde \cG}=0$ for $d=2$, resp.) and $C_{\bG}=1$ made 
with 1 vertex with arbitrary number of external legs are diverging.
In other words, only melonic (planar, resp.) tadpoles  with arbitrary number of legs are possibly logarithmically divergent (log--divergent) in this theory. If one performs a truncation in the interaction \eqref{intergen}, 
choosing all trace invariants from  valence 2 up to order $k_{\max}$, then 
contracting any of these interactions of coordination larger than 4 in order to form melonic (planar, resp.) tadpoles, might give another graph the boundary of which is again a trace invariant of lower order or 
a disjoint union of such trace invariants. The latter case has been 
called anomalous terms in anterior studies \cite{BenGeloun:2011rc, Samary:2012bw,Carrozza:2013wda}. Thus,  at a maximal order 
$k_{\max}$, including all lower order trace invariants (the set of which should be finite) and their possible anomalous terms by successive contractions (the set of which should be finite too), yields a possible class of stable (does not generate any other vertex than the one included) and super-renormalizable theory.  
At the end, the number of diverging tadpole graphs in such a theory is always determined by the number of vertices which is finite. 

\medskip

Henceforth, let us assume that the coefficient of $N_{\ext}$ is never
vanishing. A set of necessary conditions for having a just-renormalizable 
model is given by

- $N_{\ext}= k_{\max}$, then $\omega_d(\cG)= 0$,

- $N_{\ext} >k_{\max}$, then $\omega_d(\cG) < 0$.

\medskip 

We are also interested in specifying which models are super-renormalizable. Seeking a set of conditions determining super-renormalizability is not a simple task since it should cover the quite broad definition that ``a super-renormalizable model only has a finite number of divergent contributions.'' We propose to call
super-renormalizable tensor model a model generating only graphs $\cG$
such that

- $N_{\ext} \geq k_{\max}$, then $\omega_d(\cG) < 0$.

The new point here is to  require that, for $N_{\ext} =k_{\max}$,
the graph amplitude is always converging.

\medskip

\noindent{\bf Potential just-renormalizable models.}
Just renormalizable models requires $N_{\ext}= k_{\max}$ and  $\omega_d(\cG)=0$,  such that, given graphs such that $\omega(\cexG)=0=\omega(\bG)$ (or $g_{\tilde \cG}=0$ for $d=2$) and $C_{\bG}=1$, we have:
\bea
&& 0= 
- \frac12\Big[(\dim G_Dd^- -2a)  k_{\max} - 2\dim G_Dd^-\Big] -\frac12\Big[ 2\dim G_Dd^- (\sum_{k<k_{\max}}V_k + V_{k_{\max}}) \crcr
&&+
 (2a-\dim G_Dd^-)\Big[\sum_{k<k_{\max}}kV_k + k_{\max}V_{k_{\max}}\Big] \Big]\crcr
&&= - \frac12\Big[(\dim G_Dd^- -2a)  k_{\max} - 2\dim G_Dd^-\Big](1-V_{k_{\max}} )
 -\frac12\Big[ 2\dim G_Dd^- +
 (2a-\dim G_Dd^-)\tilde n\Big]\cdot \tilde V \,, 
\label{intem}
\eea
where are introduced $\tilde V = \sum_{k< k_{\max}}V_{k}$
and $\tilde n \cdot \tilde V= \sum_{k< k_{\max}}k V_{k}$.

\medskip

Let us  focus on graphs which only contain maximal
coordination vertices $V_{k_{\max}}>0$ and 
$V_{k < k_{\max}}=0$. In this category of graphs, considering those
for which $V_{k_{\max}}=1$ and $N_{\ext}=k_{\max}$,  necessarily yields an open convergent 1-vertex graph. Then, $V_{k_{\max}}>1$ is the interesting case for which a necessary condition for just-renormalizability
can be achieved as
\bea
(\dim G_Dd^- -2a)  k_{\max} - 2\dim G_Dd^- = 0
\quad& \Leftrightarrow &\quad k_{\max} = \frac{2\dim G_D (d-1)}{\dim G_D (d-1) -2a}\,,\crcr
&&  \quad  k_{\max} = 2+\gamma_{a,D}\,, \quad  \gamma_{a,D}  = \frac{4a}{\dim G_D (d-1) -2a}\,.
\eea
We immediately see that this class of simple TGFT models radically differs
from the class of gauge invariant ones \cite{Carrozza:2013wda} for which 
a similar condition is obtained using $d^-=d-2$ and $a=1$. 
Focusing on the second term in \eqref{intem}, one has
\beq
\Big[ 2\dim G_Dd^- +
 (2a-\dim G_Dd^-)\tilde n\Big]\cdot \tilde V
 =-\sum_{k<k_{\max}}\Big[ (\dim G_Dd^--2a) k -2\dim G_Dd^-
\Big] V_{k}\,.
\eeq
From this point, two cases occur. First, one may have
\bea
&& 
\dim G_D d^- -2a> 0 \,, \qquad  \dim G_Dd^- >2a >0\crcr
&&
0 = (\dim G_Dd^- -2a)  k_{\max} - 2\dim G_Dd^- >
(\dim G_Dd^--2a) k -2\dim G_Dd^- \,,
\eea
such that, keeping in mind \eqref{intem}, any graph having
$N_{\ext}=k_{\max}$, $V_{k_{\max}}>0$ and $V_{k<k_{\max}}>0$ is converging. This gives us the interesting property that 
{\it a graph with $N_{\ext}=k_{\max}$ is log-divergent if and only if 
it is built with vertices of maximal valence $k_{\max}$.}

Second, one may also have
\bea
&&
  1 \leq \dim G_Dd^- <2a \leq 2 \,;
\quad   \dim G_Dd^- =1  \;\, \Leftrightarrow \;\,
[G_D = U(1)\,, \; d=2]\,,\cr\cr
&&
0 = (\dim G_Dd^- -2a)  k_{\max} - 2\dim G_Dd^- <
(\dim G_Dd^--2a) k -2\dim G_Dd^- \,,
\eea
such that, from \eqref{intem}, any graph having
$N_{\ext}=k_{\max}$, $V_{k_{\max}}>0$ and $V_{k<k_{\max}}>0$ yields $\omega_d(\cG) >0$ which violates our definition of 
just-renormalizability. 

For this study, we have a first condition for obtaining 
just-renormalizable models and it reads
\bea
&
\left\{ 
  \dim G_D(d-1) >2a \right\};
\quad \left\{k_{\max} = 2 + \gamma_{a,D} > 2 \quad  \Leftrightarrow  \quad
 \dim G_D (d-1) \leq 6a  \right\};
\cr\cr
&\text{so that} \qquad 
2a< \dim G_D (d-1) \leq 6a\,.
\label{just}
\eea
Assuming $d=2$, the above condition translates as
\bea
&& \dim G_D >2a\, ,\cr\cr
&&
a =  1 \,, \qquad \dim G_D >2a \geq 2 \,,\quad \dim G_D > 2 \,, \crcr
&&
1> a\geq 1/2 \,, \qquad \dim G_D >2a \geq 1 \,,\quad \dim G_D \geq 2 \,,\crcr
&&
a < 1/2 \,, \qquad \dim G_D \geq 1 >2a  \,, \quad\dim G_D \geq  1 \,.
\label{rd2}
\eea
Assume now that $d\geq 3$, then we have
\bea
&&
 \dim G_D >a\, ,\cr\cr
&&
a= 1 \,, \qquad \dim G_D > 1=a \,,\quad \dim G_D \geq 2 \,,\crcr
&&
a < 1 \,, \qquad \dim G_D \geq 1 >a  \,, \quad\dim G_D \geq  1 \,.
\label{rd3}
\eea
For an arbitrary $N_{\ext}$,  \eqref{ome} translates now as 
\bea
\omega_d(\cG)\leq -\frac12(\dim G_Dd^- -2a)(N_{\ext} - k_{\max}) 
 -\frac12\Big[ 2\dim G_Dd^- +
 (2a-\dim G_Dd^-)\tilde n\Big]\cdot \tilde V \,.
\eea
Such that for $N_{\ext} > k_{\max}$ gives a convergent amplitude
provided the fact that $\dim G_D d^- >2a$. For the situation, such that
$N_{\ext} < k_{\max}$, the amplitude may or may not diverge. 

We are now in position to determine models which are potentially just-renormalizable.  
If $d=2$ \eqref{rd2}, from \eqref{just}, one realizes that
\bea
&&
a=1 \,, \qquad  2  < \dim G_D  \leq 6  \cr\cr
&&
\dim G_D=3 \,,\quad k_{\max} = 6 \,, \crcr
&&
\dim G_D=4 \,,\quad k_{\max} = 4 \,, \crcr
&&
\dim G_D=5 \,,\quad k_{\max} = \frac{10}{3} \notin \N \,, \crcr
&&
\dim G_D=6 \,,\quad k_{\max} = 3   \,.
\eea
The last case $\dim G_D=6$ would not be retained because $k_{\max}\geq 4$ for complex matrix models. However, this case could
still induce a potential renormalizable  real matrix model. 
 
Recalling that 
\bea
\quad k_{\max} = 2+ \gamma_{a,D} \,, 
\qquad  \gamma_{a,D}=\gamma=\frac{4a}{\dim G_D-2a}\in \N\setminus\{0\}\,,\qquad
a(4+2\gamma) = \dim G_D \gamma \,,
\eea
in the second sector, the following is satisfied:
\bea
&&
1>a\geq 1/2 \,,  \qquad 2 \leq \dim G_D < 6 \,,
 \qquad  \frac{4}{\dim G_D -2} > \gamma \geq \frac{2}{\dim G_D -1}\,,
\cr\cr
&&
\dim G_D= 2 \,, \qquad \forall \gamma \geq 2 \text{ and } \gamma \in \N\,,\quad a = \frac{\gamma}{2+\gamma}\,,\quad
\quad k_{\max} = 2+\gamma >2  \,;\crcr
&&
\dim G_D= 3 \,, \qquad  4>\gamma \geq 1 \text{ and } \gamma\in \N\,,\quad a = \frac{3\gamma}{4+2\gamma}\,,\quad
\quad k_{\max} = 2+\gamma >2 \,; \crcr
&&
\dim G_D= 4,5 \,, \qquad   \gamma =1,\quad a = \frac{\dim G_D}{6}\,,\quad
\quad k_{\max} = 3 >2  \,.
\label{rdd2}
  \eea
In the last sector, we have
\bea
&&
0<a < 1/2 \,,  \qquad    1 \leq \dim G_D <3 \,,
\qquad  0 < \gamma < \frac{2}{\dim G_D-1}\,,
\cr\cr
&&
\dim G_D= 1 \,, \qquad \forall \gamma \geq 1 \text{ and } \gamma \in \N\,,\quad a = \frac{\gamma}{4+2\gamma}\,,\quad
\quad k_{\max} = 2+\gamma >2  \,;\crcr
&&
\dim G_D= 2 \,, \qquad  \gamma = 1\,,\quad a = \frac{1}{3}\,,\quad
\quad k_{\max} = 3>2  \,.
\label{rdd3}
\eea
This exhausts potential just-renormalizable models in the rank $d=2$ case. We  point out that several of the above models in \eqref{rdd2} and \eqref{rdd3} defined with $k_{\max}$ an odd positive integer are potentially interesting only in the case
of real matrix models. 
Table \ref{table:modrk2} gives a summary of the previous analysis. 

\begin{table}[h]
\begin{center}
\begin{tabular}{  l||llllll }
$\dim G_D$  
&$a=1$&  & $a\in (0,1/2)$ & &$a\in [1/2,1)$&\\
 \hline \hline 
1 & $\times$  &  &$\Phi^{2k>2}$& &$\times$& 
\\
2&$\times$& & $\times$ &&$\Phi^{2k\geq 4}$&  \\
3&$\Phi^6$&& $\times$ &&$\Phi^{4}$, $a=\frac{3}{4}$&   \\
4&$\Phi^4$&&$\times$&&$\times$&\\
\hline\hline
1 & $\times$  &  &$\Phi^{2k+1>2}$& &$\times$& 
\\
2&$\times$& & $\Phi^{3}, a=\frac13$ &&$\Phi^{2k+1\geq 4}$&  \\
3&$\times$&& $\times$ &&$\Phi^{ k=3,5}$, $a=\frac{1}{2},\frac{9}{10}$&   (resp.)\\
4&$\times$&&$\times$&&$\Phi^{3}$, $a=\frac{2}{3}$&\\
5&$\times$&&$\times$&&$\Phi^{3}$, $a=\frac{5}{6}$&\\
6&$\Phi^3$&&$\times$&&$\times$&\\
 \hline
\end{tabular}
\caption{List of all potential just-renormalizable matrix models $_{\dim G_D}\Phi^{k}_2$. All interactions of the $\Phi^{2k+1}$ type may only 
be considered in a real model. }
\label{table:modrk2}
\end{center}
\end{table}

Next, we study the rank $d \geq 3$ situation.  It is important to keep
in mind the following feature: for any $d \geq 3$, $k_{\max}$ cannot be an odd integer as it turns out to be impossible to built a trace invariant out off a odd number of tensors in any theory rank $d\geq 3$.
One has:
\bea
&&
 \dim G_D >a\, ,\cr\cr
&&
a= 1 \,, \qquad \dim G_D > 1=a \,,\quad \dim G_D \geq 2 \,,\crcr
&&
a < 1 \,, \qquad \dim G_D \geq 1 >a  \,, \quad\dim G_D \geq  1 \,.
\label{rd33}
\eea
We define 
\bea
\gamma = \frac{4a}{\dim G_D d^- - 2a} \in \N \setminus\{0\}\,, \qquad
a(4+2\gamma) = \dim G_D d^- \gamma\,,
\eea
such that $a=1$, yields
\bea
&&
2 < \dim G_D d^- \leq 6 \,, \qquad 
(4+2\gamma) = \dim G_D d^- \gamma\,, \cr\cr
&&
 \dim G_D d^- =3 \,, \qquad \gamma =4\,, \quad k_{\max} = 6\,,\crcr
&&
 \dim G_D d^- =4 \,, \qquad \gamma =2\,, \quad k_{\max} = 4\,,\crcr
&&
 \dim G_D d^- =5 \,, \qquad 3\gamma =4\,, \quad k_{\max} \notin \N\,,\crcr
&&
 \dim G_D d^- =6 \,, \qquad \gamma =1\,, \quad k_{\max} = 3\,.
\eea
The last case should be excluded because of the same above reason. 
Next, one focuses on $0<a<1$, for which the sole relevant situations
are obtained as
\bea
&& \dim G_D d^- \geq 2 \,,\qquad 
0<(\dim G_D d^- -2)\gamma <4 \,\cr\cr
&&
\dim G_D d^- = 2\,, \qquad \forall \gamma \geq 2 \text{ and } \gamma\in 2\N\,, \quad a = \frac{\gamma}{2+\gamma}\,,
\qquad 
k_{\max} = 2 + \gamma >2\,, \crcr
&&
\dim G_D d^- = 3\,, \qquad  \gamma =2 \,, \quad a = \frac{3}{4}\,,
\qquad 
k_{\max} = 4\,, 
\eea
where in the above cases, the cases of 
odd $\gamma$ giving an odd $k_{\max}$ have been precluded (same above remark). 
Let us summarize at this point the data in Table \ref{table:modrk3}. 
 
\begin{table}[h]
\begin{center}
\begin{tabular}{  l||lll}
$\dim G_Dd^-$  &$a=1$&  & $a\in (0,1)$ \\
 \hline \hline  
2&$\times$& & $\Phi^{2k>2}$   \\
3&$\Phi^6$&& $\Phi^{4}$, $a=\frac34$  \\
4&$\Phi^4$&&$\times$\\
 \hline\hline
\end{tabular}
\caption{List of  rank $d\geq 3$ tensor models potentially just-renormalizable.}
\label{table:modrk3}
\end{center}
\end{table}

Compiling Table \ref{table:modrk2} and Table \ref{table:modrk3} and considering the group dimension $\dim G_D$ and theory rank $d$, yields Table \ref{table:list23} (each model depends also on $a$)
giving a summary of all  potential renormalizable models 
(including real in the matrix case). 

\begin{table}[h]
\begin{center}
\begin{tabular}{  l||llllll }
$\dim G_D \downarrow$ $\rotatebox[origin=c]{45}{\rule{0.1mm}{6mm}} ^{d-1 \rightarrow}$ 
&1&2&3&4&5&6\\
 \hline  \hline  \\
1 & $_1\Phi^{k>2}_2$  & $_1\Phi^{2k>2}_3$ 
&$_1\Phi^{6,4}_4$ &$_1\Phi^{4}_5$&$\times$& $\times$
\\
2&$_2\Phi^{k>2}_2$ & $_2\Phi^{4}_3$& $\times$ &$\times$ &$\times$& $\times$\\
3&$_3\Phi^{3,4,5,6}_2$ &$\times$&$\times$&$\times$&$\times$& $\times$ \\
4&$_4\Phi^{3,4}_2$&$\times$&$\times$&$\times$&$\times$& $\times$ \\
5&$_5\Phi^{3}_2$&$\times$&$\times$&$\times$&$\times$& $\times$ 
\\
6&$_6\Phi^{3}_2$&$\times$&$\times$&$\times$&$\times$& $\times$ \\
 \hline\hline
\end{tabular}
\caption{List of  rank $d\geq 2$ tensor models which are potentially just-renormalizable. }
\label{table:list23}
\end{center}
\end{table}

\vspace{0.2cm}

Some comments are in order: 

a) First, one notices that there is no model over $SU(2)$
which could be just-renormalizable when $d\geq 3$. 
 $G_1 = SU(2)$ can only be the background group manifold for the
$_3\Phi^{3,4,5,6}_2$ models  and $G_2= SU(2)^2$ for $_6\Phi^3_2$.

b) The tensor models $_1\Phi^6_4$ and $_1\Phi^4_3$ (in first row) have been already proved to be renormalizable in \cite{BenGeloun:2011rc}
and \cite{BenGeloun:2012pu}, respectively.

c) The matrix models $_1 \Phi^{k_1>2}_2$ and $_2\Phi^{k_2\geq 4}_2$
may be in fact very similar when both $k_1$ and $k_2$ coincide
after mapping $a\to 2a$, if $2a \in (1/2,1)$, that is if $a \in (1/4,1/2)$. 
Otherwise these models are actually different and, in the following, we will treat them  as such unless otherwise explicitly stated. 
 
d) Another strong fact is that  there are, a priori, three  
towers of potentially interesting models: 
any $\Phi^{2k \geq 4}$ over $G=U(1)$
in rank $d=3$, and $\Phi^{k \geq 3,4}$ over $G\in \{U(1),U(1)^2\}$ in rank $d=2$. 

e) Interestingly, one notices that the $\Phi^4$ interaction appears
several times in that list up to the group
dimension $\dim G_D = 4$. It is a kind of ``privileged'' interaction
for the tensor and matrix field models. 

f) The complex Grosse-Wulkenhaar (GW) model in 4 dimensions  written in the matrix basis and at its self-dual point \cite{Grosse:2004yu} is a matrix model
which can be written in terms of rank two tensors $\bar\varphi_{\vec m, \vec n}$
and $\varphi_{\vec m, \vec n}$, $\vec n=(n_1,n_2) \in \N^2$, as
\bea
&&
S_{GW} = \frac12\sum_{\vec p, \vec q \in \N^2}
\bar\varphi_{\vec p, \vec q} \Big[ |p| +|q|  + \mu \Big] \varphi_{\vec q, \vec p } +
\frac{\lambda}{4}
\sum_{\vec m, \vec n, \vec p, \vec q \in \N^2}
\bar\varphi_{\vec m, \vec n}\,\varphi_{\vec n, \vec p}\,
\bar\varphi_{\vec p, \vec q}\,\varphi_{\vec q, \vec m}\,, 
\label{gw4d}
\eea
where we introduce the notation, for any $\vec n \in \N^2$, $|n| =n_1 + n_2$. 

\begin{figure}[h]
 \centering
    \begin{minipage}{.8\textwidth}
\includegraphics[angle=0, width=7cm, height=1.5cm]{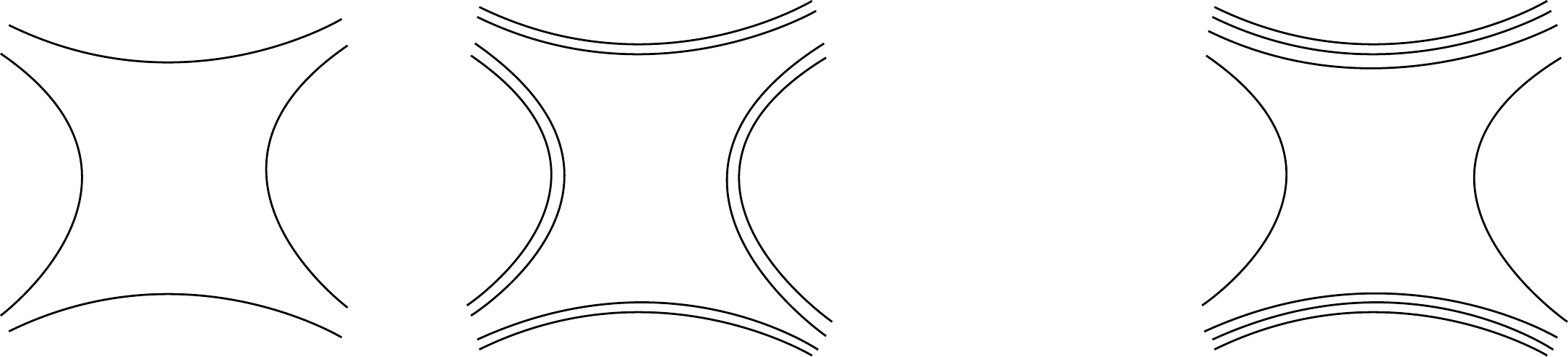}
\vspace{0.3cm}
\caption{ {\small The interaction of GW model in 4D viewed as a matrix model is $I_{GW}$ and, viewed as a rank 4 model, the interaction reads $I'_{GW}$. Both have to be distinguished with the melonic rank 4 interaction $I''$.}} \label{fig:GW}
\end{minipage}
\put(-295,-12){$I_{GW}$}
\put(-230,-12){$I'_{GW}$}
\put(-132,-12){$I''$}
\end{figure}

One should pay attention to the fact that, although 
the GW model can be naively considered as defined with rank 4 tensors $\varphi_{\vec p, \vec q}=\varphi_{p_1,p_2,q_1,q_2}$ then, according to the previous
developments, it is not the $_1\Phi^4_4$ tensor 
 model with $a=\frac{1}{2}$ (according to Table \ref{table:list23}, 
only the $_1\Phi^4_4$ tensor model with $a=\frac{3}{4}$ is potentially just-renormalizable). The reason why this is not the case comes from the particular form of the GW interaction. In the above analysis, we strongly use the fact that the divergence degree \eqref{ome}  saturates for melonic graphs. 
However, in the GW model viewed as a rank 4 tensor model, the vertex
is of the form $I'_{GW}$ of Fig. \ref{fig:GW}, and cannot generate such category of melonic graphs
(see $I''$of  Figure \ref{fig:GW} representing a rank 4 melonic $\Phi^4$ interaction).  
Thus, the GW model should be strictly considered as a matrix model
and its power counting theorem should follow from Proposition \ref{prop:mat} and not
from  Proposition \ref{prop:tens}.
The way to embed the 4D GW in the above formalism comes from the fact that the tensor indices in that model are one to one with representation indices of $U(1)^2$. Hence, the GW model as a matrix model might read $_{2}\Phi^4_{2}$ for $a=1/2$. This model is included in Table \ref{table:list23}. 
Using the projection mapping introduced in \cite{Geloun:2012bz}, 
which allows to reduce the rank of the tensor and still to preserve the
power counting, we can map $\varphi_{\vec n,\vec m}\to \varphi_{n, m}$, 
$n,m \in \N$, so that we might fully represent the GW model over $U(1)$ as $_1\Phi^4_{2}$ with $a=1/4$ again a matrix model included in the table.

\medskip 

Concentrating on rank $d\geq 2$ complex tensors, 
Table \ref{table:listcomplex} provides the list of models that we will 
discuss in the following. 

\begin{table}[h]
\begin{center}
\begin{tabular}{  l||llllll }
$\dim G_D \downarrow$ $\rotatebox[origin=c]{45}{\rule{0.1mm}{6mm}} ^{d-1 \rightarrow}$ 
&1&2&3&4&5&6\\
 \hline  \hline  \\
1 & $_1\Phi^{2k>2}_2$  & $_1\Phi^{2k>2}_3$ 
&$_1\Phi^{6,4}_4$ &$_1\Phi^{4}_5$&$\times$& $\times$
\\
2&$_2\Phi^{2k>2}_2$ & $_2\Phi^{4}_3$& $\times$ &$\times$ &$\times$& $\times$\\
3&$_3\Phi^{4,6}_2$ &$\times$&$\times$&$\times$&$\times$& $\times$ \\
4&$_4\Phi^{4}_2$&$\times$&$\times$&$\times$&$\times$& $\times$ \\
 \hline\hline
\end{tabular}
\caption{ List of potential just-renormalizable rank $d\geq 2$ complex
tensor models.}
\label{table:listcomplex}
\end{center}
\end{table}

\medskip

\noindent{\bf On other potential super-renormalizable models.}
We now investigate conditions on super-renormalizability. 
The condition $\dim G_D d^- -2a=0$ has already led
to some likely super-renormalizable models, we now focus 
on other types of conditions. 

The divergence degree \eqref{ome} obeys the bound
\bea
&&
\omega_d(\cG)\leq -\frac12(\dim G_Dd^- -2a)(N_{\ext} - k_{\max})  \crcr
&&
 - \frac12 \Big[(\dim G_Dd^- -2a)  k_{\max} - 2\dim G_Dd^-\Big](1-V_{k_{\max}} )
 -\frac12\Big[ 2\dim G_Dd^- +
 (2a-\dim G_Dd^-)\tilde n\Big]\cdot \tilde V \,.
\eea
Let us first assume that $\dim G_Dd^- > 2$, then 
\bea
&&
(\dim G_Dd^- -2a)  k_{\max} - 2\dim G_Dd^- \geq  0
\quad \Leftrightarrow \quad k_{\max} \geq \frac{2\dim G_Dd^- }{(\dim G_Dd^- -2a)}\,; \crcr
&&(\dim G_Dd^- -2a)  k_{\max} - 2\dim G_Dd^- \leq  0
\quad \Leftrightarrow \quad k_{\max} \leq \frac{2\dim G_Dd^- }{(\dim G_Dd^- -2a)}\,. 
\eea
Then, for $k<k_{\max}$, 
\bea
0 \geq (\dim G_Dd^- -2a)  k_{\max} - 2\dim G_Dd^- 
> (\dim G_Dd^- -2)  k - 2a\dim G_Dd^- \,.
\eea
Meanwhile, if $\dim G_Dd^- < 2$, then 
\bea
(\dim G_Dd^- -2a)  k_{\max} - 2\dim G_Dd^- \leq  0
\quad \Leftrightarrow \quad k_{\max} \geq 0 >\frac{2\dim G_Dd^- }{(\dim G_Dd^- -2a)}\,. 
\eea
Thus assuming 
\beq
(S1)\qquad N_{\ext} > k_{\max} 
 \,, \qquad 
\dim G_Dd^- > 2\,, \qquad 
\{ 0 \geq (\dim G_Dd^- -2a)  k_{\max} - 2\dim G_Dd^- 
 \,; \quad k_{\max} \leq 2+ \gamma_{a,D} \}
\eeq
leads clearly to a strictly negative divergence degree. 
However,  assuming 
\bea
(S2)&&\qquad 
N_{\ext} = k_{\max}\,,  \qquad 
\dim G_Dd^- > 2a\,, \qquad 
\{ 0 \geq (\dim G_Dd^- -2a)  k_{\max} - 2\dim G_Dd^- 
 \,; \quad k_{\max} \leq 2+ \gamma_{a,D} \}
\cr\cr
&& \qquad 
\omega_d(\cG)\leq 
 - \frac12\Big\{ \Big[(\dim G_Dd^- -2a)  k_{\max} - 2\dim G_Dd^-\Big](1-V_{k_{\max}} )\Big\} \crcr
&& \qquad\qquad 
 -\frac12\Big[ 2\dim G_Dd^- +
 (2a-\dim G_Dd^-)\tilde n\Big]\cdot \tilde V \leq 0
\eea
hence does not obey the necessary condition $\omega_d(\cG)<0$. Hence
nothing more can be said if $N_{\ext}=k_{\max}$
because of the number of vertices which 
can be totally arbitrary and can depend on the types of graphs. 
In conclusion, the only condition known at this stage which could 
ensure super-renormalizability as defined above is given by $\dim G_D d^-=2a$.

\section{Just renormalizable rank $d \geq 3$ tensor models}
\label{sect:rank3}

The previous section determines
the maximal valence $k_{\max}$ of the vertices which may actuate renormalizability. 
In this section, we intend to build models which are indeed
renormalizable given the above data. The models are built
in a standard way: we include all vertices of lower valence up 
to $k_{\max}$ and, due to some specific higher rank structure,
we also should pay attention to the appearance 
of peculiar anomalous terms which should be included as well. Afterwards, given a model $_{\dim G_D}\Phi^{k}_{d}$, we provide a list of all  divergent amplitudes and their associated graphs which must be renormalized in the subsequent section.

One proceeds according to the general recipe: Consider any model 
susceptible to be renormalizable as given by Table \ref{table:listcomplex}.
This will determine $a$, the kinetic term and covariance, 
$k_{\max}$ and from this, use all trace invariants of lower order
up to 4. In the rank $d \geq 3$ case, among the trace invariants, use only melonic ones as the interaction terms. 

\subsection{Tensor models and their renormalizability}
\label{subsect:just3}

\noindent{\bf Truncating the rank $d=3$ tower.}
The tower $(_1\Phi^{k_{\max}}_{3}, a=1-\frac{2}{k_{\max}})$,
$k_{\max}\in 2\N\setminus \{0,2\}$, is worth discussing in more details. At a given order $k_{\max}$,
the problem of finding specific criteria for listing $k_{\max}$ 
connected unitary tensor invariants is not known to the best of our knowledge. In fact, the problem is subtle because we do
not need to list all 3-bubbles, but only specific 3-bubbles of the melonic kind. Melonic bubbles, in general, will govern the locality principle 
of tensor models. Returning to the above tower dealing with models of rank $d=3$, we did not successfully identify an algorithm for generating all possible connected melonic $3$-bubbles made with $k_{\max}$ vertices up to a line-coloring.  It is known 
\cite{Bonzom:2011zz} that melonic two-point functions made with $q$ vertices maps to rooted colored trees with $q$ unlabeled vertices. 
The number of such combinatorial species is given by a generalized $(d+1)-$Catalan number. The problem addressed now is to identify all possible melonic $k_{\max}$-point
function made with $k_{\max}$ vertices. This is more
intricate because of the equivalence of several configurations if there is
not a unique root. Certainly the number of these objects is bounded by the generalized Catalan number because a $k_{\max}$-point function can be obtained by cutting more lines (of color 0 for instance) in the 2-point function. The problem then consists to quotient and/or subtract from this Catalan an unknown number of equivalent configurations coming from the fact we can expand a graph from the point of view of one root or another. Also, even knowing the number of different configurations does
not necessarily gives a way to list them according to some criteria. This delicate study will require
more combinatorial tools. For this reason, we will address only the 
renormalizability of models such that $k_{\max}=4,6$
in this tower.

At a given $k_{\max}=4,6$, we aim at studying all rank $d \geq 3$ tensor models
 $_{\dim G_D} \Phi^{k_{\max}}_d$. 
Note that the $_1\Phi^6_4$ and $_1\Phi^4_3$ have been already proved
renormalizable \cite{BenGeloun:2011rc} \cite{BenGeloun:2012pu}.
We provide in the following here a unifying perspective towards 
the study of renormalizability of TGFTs. Hence, in addition to
the aforementioned models,  we will address the renormalizability 
proof of the following models
\beq
_1 \Phi^6_3 \,, \qquad _2\Phi^4_3 \,, \qquad _1 \Phi^4_4 \,, 
\qquad _1\Phi^4_5\,.
\eeq

\noindent{\bf Block index notations.}
Dealing with an arbitrary rank $d=3,4,5$ and 
a group $G_D=U(1)^D$ with $D =1,2$, 
we must find adequate notations for representing
the different types of interactions. We have already introduced
$\varphi_{[I]}$ the rank $d$ tensor field where
$[I]=\{ I_1,\dots, I_d \}$ collects all momentum labels. The interactions are built from particular contractions of the indices $I_s$ 
between tensor fields. This is done in such a way that we can further decompose $[I]$ in sub-blocks contracted with other 
sub-blocks in other tensor fields. We write  
\beq
\varphi_{[I]}= \varphi_{[1][2]\dots[q]}\,, \quad [s]=(I_{s,1},\dots,I_{s,l})\,,
\quad 
s=1,\dots, q\,,
\eeq
 where $q$ may vary according the contraction we are interested in. 
For any rank $d$ tensor field $\varphi_{[I]}=\varphi_{12\dots d}$,  
there are two particular decompositions of the tensor labels of interest
in the following. The first can be called the identical decomposition
and is given by 
\beq
\varphi_{12\dots d} = \varphi_{[1][2]\dots [d]} \,, \quad 
[s]=s\,, 
\label{iden}
\eeq
and the second is the so-called matrix type decomposition 
of the tensor entries which can be written as
\bea
\varphi_{123\dots d} = \varphi_{[a]\{\check{a}\}}\,, 
\quad 
[a]=a \,,\quad \{\check{a}\} = (1,2,\dots, \check{a},\dots d)\,,
\label{matr}
\eea
where we distinguish this decomposition using brace brackets
for at least one block. Note that one block contains a single element and, 
though not explicit, the brace bracket depends on the rank. 
For matrix models, both decompositions \eqref{iden}
and \eqref{matr} of the matrix field coincide. Furthermore, given some index $a$, 
the matrix decomposition is ``canonical'' in the sense that,
even though its seems that we have changed the place of the index $a$ in the
tensor as $\varphi_{...a....}=\varphi_{[a]....}$, the previous position
of that index is still encoded in $\{\check{a}\}$. Thus this allows us
to preserve the (colored-like) gluing rule for graphs depending on the position of labels
in the tensor.

For all models $_{(\cdot)}\Phi^{4}_{d\geq 3}$, a matrix type decomposition will be used. For the model $_{1}\Phi^{6}_{3}$, 
we will use a matrix and an identical decomposition to describe the model. Last,  for $_{1}\Phi^{6}_{4}$, we introduce
the non trivial decomposition for the model characterized by $k_{\max}=6$ and $d=4$:
\beq
\varphi_{1234}= \varphi_{[1][2][3]}\,, \quad [1]=1\,, \quad  [2]=(2,3)\,,
\quad  [3]=4\,.
\eeq
One should pay attention to the fact that, in  the above notations, discussing either $_{1}\Phi^{6}_{3}$ or $_{1}\Phi^{6}_{4}$, the tensor field will be denoted as $\varphi_{[1][2][3]}$, but according to the model context, these block notations of tensor do not refer to the same quantities. The point for introducing such notations
comes from the fact that the model properties and its renormalization
analysis only depend on these decompositions of the tensors. 

\medskip 

\noindent{\bf Propagators.}
Propagators has been already discussed. They are given by 
\eqref{chew0} and are represented by stranded lines as
in Figure \ref{fig:prop}. 

\medskip

\noindent{\bf Melonic interactions.} The interactions of the models must be chosen, as in the ordinary situation, according to the truncation of the series of all possible interactions from relevant to marginal terms. Here the interactions are generated by the series $S^{\inter}(\varphi,\bar\varphi)$ \eqref{intergen} of all connected melonic contractions for a certain rank $d$ theory. From the power-counting
theorem, Proposition \ref{prop:tens}, and subsequent analysis 
of the divergence degree, for a given $_{(\cdot)}\Phi^{k_{\max}}_{d}$ model, 
we have a specific criterion to truncate
the series $S^{\inter}(\varphi,\bar\varphi)$. We will only consider melonic interactions with at most $k_{\max}= N_{\ext}$ external legs. 
It may happen that terms called anomalies \cite{BenGeloun:2011rc} appear because generated by the RG flow without being initially present in \eqref{intergen}. In such instance, one must  add these terms in 
the initial action and check that the resulting action does not generate
any further term. 

\medskip 

\begin{figure}[h]
 \centering
    \begin{minipage}[t]{.8\textwidth}
\includegraphics[angle=0, width=4cm, height=2.5cm]{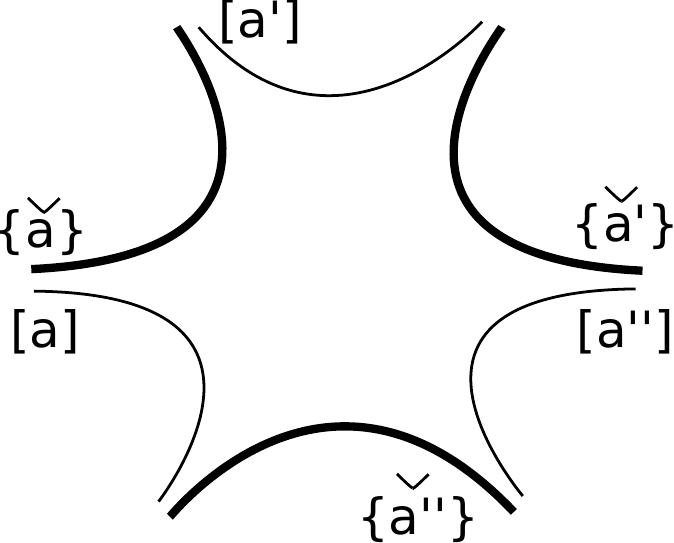}
\hspace{0.5cm}
\includegraphics[angle=0, width=4cm, height=2.5cm]{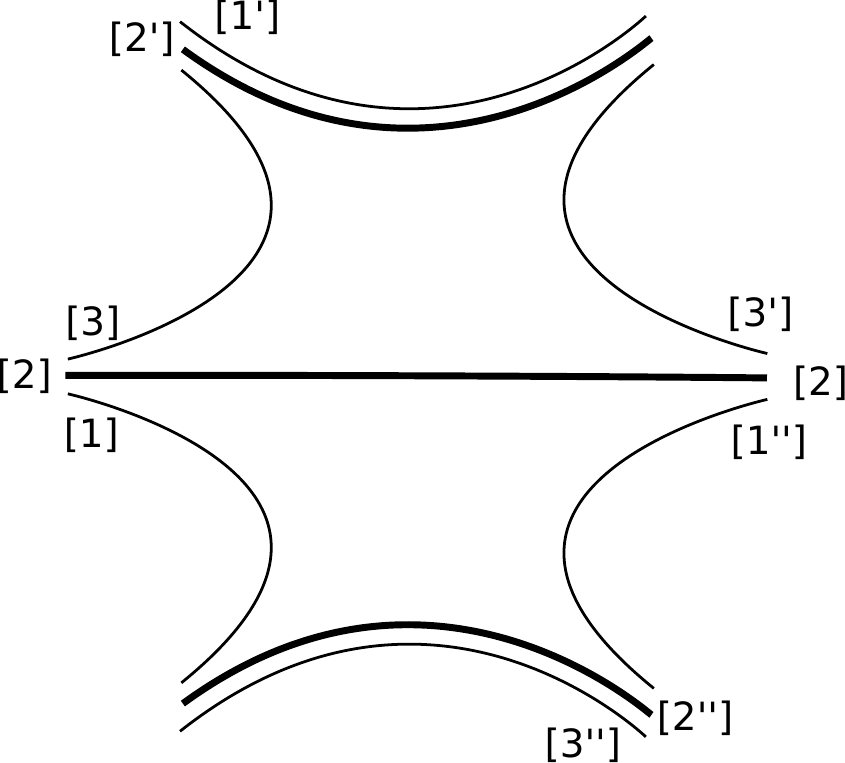}
\vspace{0.3cm}
\caption{ {\small The vertices of the types $V_{6;1}$ (A) and $V_{6;2}$ (B).}} \label{fig:Inter6}
\end{minipage}
\put(-275,-12){A }
\put(-140,-12){B}
\end{figure}

- For $k_{\max}=6$ ($d=3,4$ and $\dim G_D =1$),
 two types of interactions $S^{\inter}_{6;1}$  and $S^{\inter}_{6;2}$
of the $\Phi^6$-form can be constructed. These
are given by
\bea
&&
S^{\inter}_{6;1} = 
\sum_{P_{[I]}} 
\varphi_{[1]\{\check{1}\}} \bar \varphi_{[1']\{\check{1}\}}
 \varphi_{[1']\{\check{1}'\}} \bar \varphi_{[1'']\{\check{1}'\}}
 \varphi_{[1'']\{\check{1}''\}} \bar \varphi_{[1]\{\check{1}''\}} \label{vphi61}\quad + \quad \text{permutations} \,,\\
&&
S^{\inter}_{6;2} = 
\sum_{P_{[I]}} 
\varphi_{[1][2][3]} \bar \varphi_{[1]'[2]'[3]}
 \varphi_{[1]'[2]'[3]'} \bar \varphi_{[1]''[2][3]'}
 \varphi_{[1]''[2]''[3]''} \bar \varphi_{[1][2]''[3]''} \quad + \quad \text{permutations} \,,
\label{vphi62} 
\eea
where the sum of ``permutations'' is performed on 
colors. For the $_1\Phi^6_{d=3,4}$ model, \eqref{vphi61} contains 
exactly $d$ terms, meanwhile \eqref{vphi62} contains 
$d(d-1)/2$ terms. Using the colored extension of these
vertex (as discussed in Subsection \ref{sub:uncol}),
one may check that the resulting colored graphs satisfy 
the condition that their degree are vanishing.
A drawing of these interactions is given in Figure \ref{fig:Inter6} where bold lines may encapsulate several strands
depending on the model. Within a model, from 
$V_{6;1}$ to $V_{6;2}$, these bold lines may not
contain the same number of strands.
As one realize, the block index notation is simply
handy to write simple vertices (and not always Feynman graphs,
because in the gluing the propagator may not follows the block index) and to perform the subsequent calculations.

\medskip

\begin{figure}[h]
 \centering
    \begin{minipage}[t]{.8\textwidth}
\includegraphics[angle=0, width=7cm, height=2cm]{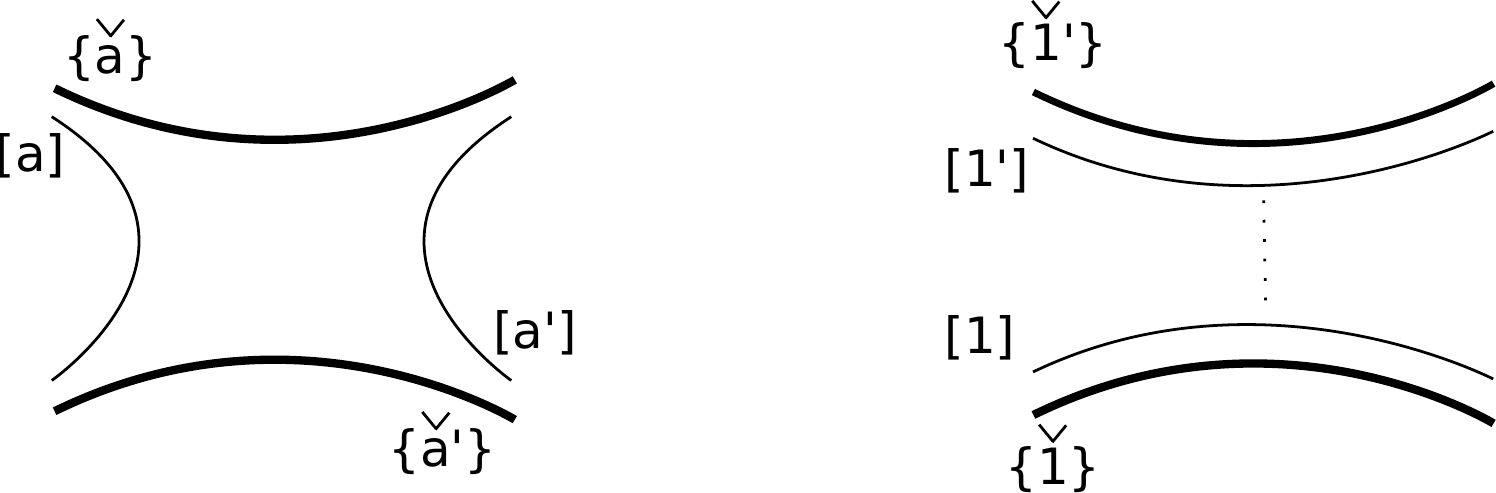}
\vspace{0.3cm}
\caption{ {\small Vertices $\Phi^4$  of the $V_{4}$ (A) and $V_{4;{\rm a}}$ (B) types.}} \label{fig:Int4}
\end{minipage}
\put(-270,-12){A}
\put(-145,-12){B}
\end{figure}

In addition to these interactions, we must introduce a $\Phi^4$ interaction as 
\bea
S^{\inter}_{4} = 
\sum_{P_{[I]}} 
\varphi_{[1]\{\check{1}\}} \bar \varphi_{[1']\{\check{1}\}}
 \varphi_{[1']\{\check{1}'\}} \bar \varphi_{[1]\{\check{1}'\}}
\quad + \quad \text{permutations} 
\label{vphi64}
\eea
which contains $d$ terms (Figure \ref{fig:Int4} A shows a general term in 
this interaction). It turns out that the 
$_1\Phi^6_4$ model generate an anomalous term of the
$(\Phi^2)^2$ type given by 
\beq
S^{\inter}_{4;{\rm a}} = 
\sum_{P_{[I]}} 
(\varphi_{[1]\{\check{1}\}} \bar \varphi_{[1]\{\check{1}\}})
(\varphi_{[1']\{\check{1}'\}} \bar \varphi_{[1']\{\check{1}'\}})\,.
\label{vphi622}
\eeq
The graphical representation of the vertex associated with 
this interaction is given by Figure \ref{fig:Int4} B. 

\medskip

- Turning our attention to the  $_{\dim G_D}\Phi^{4}_d$ models ($(d, \dim G_D)\in \{(3,1), (3,2), (4,1), (5,1)\}$), there is a unique
type of interaction of the same form given by $S^{\inter}_4$ \eqref{vphi64}
hence possesses the same graphic as given by Figure \ref{fig:Int4} A. We will use the same notation $S^{\inter}_4$ for this interaction because no confusion will occur discussing
one model or the other. No anomalous can be generated here.

Introducing a UV cut-off in the propagator $C$ becomes $C^\Lambda$, 
we must consider bare and renormalized couplings and their 
difference known as coupling constant counter-terms $CT$. 
In particular,  we must introduce mass and wave function counter-terms
for each model as
\beq
S_{2;1} = \sum_{P_{[I]}} \bar \varphi_{[I]}\varphi_{[I]} \;, 
\qquad 
S_{2;2} = \sum_{P_{[I]}}  \bar \varphi_{[I]}\Big(\sum_{s=1}^d |P_{I_s}|^{a}\Big) \varphi_{[I]}\;.
\label{cterms}
\eeq
The action that we will consider are given by

- For $k_{\max}=6$, $d=3,4$, 
\beq
S^{\Lambda} = \frac{\lambda_{6;1}^{\Lambda}}{3} S^{\inter}_{6;1} + 
 \lambda_{6;2}^{\Lambda}S^{\inter}_{6;2} + 
\frac{\lambda_{4;1}^{\Lambda}}{2} S^{\inter}_{4} 
 + \frac{\lambda_{4;2}^{\Lambda}}{2} \delta_{d,4}\,S^{\inter}_{4;{\rm a }}
+ CT^{\Lambda}_{2;2}  S_{2;2} + 
CT^{\Lambda}_{2;1}  S_{2;1} \,,
\label{actphi6}
\eeq
where $\delta_{d,a}$ is a Kronecker symbol. 

- For $k_{\max}=4$, $(d, \dim G_D)\in \{(3,1), (3,2), (4,1), (5,1)\}$,
\beq
S^{\Lambda} = \frac{\lambda_{4;1}^{\Lambda}}{2} S^{\inter}_{4} 
 + CT^{\Lambda}_{2;2}  S_{2;2} + CT^{\Lambda}_{2;1}  S_{2;1} \,.
\label{actphi4}
\eeq

The following statement holds
\begin{theorem}[Renormalizable tensor models]\label{theo:rentensor}
The models $(_{\dim G_D}\Phi^{k_{\max}}_d,a)$ such that 
\bea
&&
(_1\Phi^{6}_{4}, a=1) \text{ over } U(1)\,, \quad 
(_1\Phi^{6}_{3}, a=\frac23) \text{ over } U(1)\,, \cr\cr
&&
(_1\Phi^{4}_{4}, a=\frac34) \text{ over } U(1)\,, \quad 
(_1\Phi^{4}_{3}, a=\frac12) \text{ over } U(1)\,, \quad 
(_1\Phi^{4}_{5}, a=1) \text{ over } U(1)\,, \quad 
(_2\Phi^{4}_{4}, a=1) \text{ over } U(1)^2\,, \quad 
\eea
with action defined by \eqref{actphi6} or \eqref{actphi4} are all just-renormalizable at all orders of perturbation.
\end{theorem}

The proof of this theorem has been nearly achieved through the multi-scale analysis and analysis of the divergence degree in the generic
situation of Subsection \ref{subsect:multi} and Subsection \ref{subsect:deg}.   Our remaining task is to introduce wave function
counter-terms in the divergence degree, list all possible divergent
amplitudes for each case and perform the renormalization of these
divergences.  

Introducing in a graph, a number of $V_{2;2}$ wave function counter-term vertices \eqref{cterms} each of which bringing a factor of $|P_{I_s}|^a \sim M^{2i}$
in a slice $i$, it is simple to find from the previous multi-scale analysis
the degree of divergence a connected graph $\cG$ as 
\beq
\omega_d(\cG)  = -2aL(\cG)   +\dim G_D \,F_{\inter}(\cG) 
 +2a V_{2;2}\,.
\eeq
If one includes the anomalous term $S^{\inter}_{2;{\rm a}}$, one pays
attention to the fact that, from the point of view of the external 
legs, the anomalous vertex is disconnected. We will consider only
half of the anomalous vertex when discussing connected graphs.  

The formula of number of internal faces $F_{\inter}(\cG)$ given by 
Proposition \ref{prop:face} remains barely of the same form but 
in the definition of 
$V=\sum_{k} V_k$ and
$n \cdot V= \sum_{k} k V_k$, one should
incorporate the number $V_{2;2}$ of wave function vertices, the number $V_{2;1}$ of mass vertices and number $\delta_{d,4} V_{2;{\rm a}}$ of half-anomalous vertices. It is direct to realize from \eqref{faceinter} 
that $F_{\inter}(\cG)$ does not depend on 2-valent vertices, 
in particular on $V_{2;2}$. We can finally express the degree of divergence  $\omega_d(\cG)$ of a connected graph again as \eqref{degree} with 
a number of vertices $V$ which does not include the wave function counter-term vertices but still includes mass and anomalous vertices. 
Thus, as expected, the degree of divergence does not
depends on wave function counter-terms. 

\medskip

\noindent{\bf List of primitively divergent graphs.}
We consider graphs with an even number of external legs such that 
$\bG \neq \emptyset$ and $C_{\bG} \geq 1$.
Moreover, there are some important features satisfied the difference 
$\omega(\cexG) - \omega(\bG)$. In particular in \cite{BenGeloun:2011rc},
it has been proved that either this quantity is vanishing  or it 
obeys
\beq
\frac{2}{(d-1)!}\big(\omega(\cexG) - \omega(\bG) \big)\geq d-2 \geq 0\,.
\label{mag}
\eeq
We can now fully address the list of divergent amplitudes for the different models by introducing
\bea
\,\omega_d(\cG) &=& - \frac{2\dim G_D}{(d-1)!}(\omega(\cexG) - \omega(\bG)) - P_a(\cG)\,, 
\label{omegaa}\\
P_a(\cG) &=&\dim G_D (C_\bG -1)  + \frac12\Big[(\dim G_D(d-1) -2a) N_{\ext} - 2\dim G_D(d-1)\Big]\crcr
&& +\frac12\Big[ 2\dim G_D(d-1) + (2a-\dim G_D(d-1))n \Big] \cdot V
\label{degree3}
\eea
and then by seeking conditions under which $\omega_d(\cG) \geq 0$.

\medskip 

$\bullet$ For $k_{\max}=6$, $\dim G_D =1$, 
$(d,a)\in \{(3, \frac{2}{3}),(4,1)\}$, respectively, 
then $(d-1-2a)\geq 0$ and, given a connected
graph $\cG$ with $V_4$ number of vertices of
the $\Phi^4$ type, $\delta_{d,4}V_{2;{\rm a}}$ number of half anomalous vertices 
and $V_{2;1}$ number of mass vertices, 

\medskip

- if $N_{\ext} > 6$,
\bea
&&
((d-1) -2a) N_{\ext} - 2(d-1) >  ((d-1) -2a)6 - 2(d-1)=0 \,, \crcr
&& \forall  2 \leq k \leq 4\,,\qquad 
2(d-1) + (2a-(d-1))k = (2-k)(d-1) + 2ak \geq 0
\eea
so that 
\beq
P_a(\cG) =  (C_\bG -1)  + \frac12\Big[((d-1) -2a) N_{\ext} - 2(d-1)\Big] +\Big[ (d-1) +2 (2a-(d-1)) \Big] V_{4}
+ 2a (V_{2;1}+ \delta_{d,4}V_{2;{\rm a}})  >0
\eeq
which proves that $\omega_d(\cG) <0$ and hence the graph amplitude converges; 

\medskip 

- if $N_{\ext} = 6$, the graph amplitude is at most log--divergent, i.e.
$\omega_d(\cG) \leq 0$. Focusing on $\omega_d(\cG) =0$, this can be satisfied if and only if 
$\omega(\cexG)=0=\omega(\bG)$ and $P_a(\cG)=0$ that is  $C_\bG =1$, $V_4=0=V_{2;1}=V_{2;{\rm a}}$; 

\medskip

- if $N_{\ext} = 4$, we have
\bea
P_a(\cG) &=& (C_\bG -1)  + \Big[ (d-1)-4a\Big] +\Big[ (d-1) +2 (2a-(d-1)) \Big] V_{4}+ 2a (V_{2;1}+ \delta_{d,4}V_{2;{\rm a}})  
\crcr
&=& 
 (C_\bG -1)  +a\Big(-1 +V_{4} + 2 (V_{2;1}+ \delta_{d,4}V_{2;{\rm a}})\Big)  \,.
\eea
Then the divergence degree is at most $a$. 

\noindent (a) We can have $P_a(\cG)=0$, and in this case the graph amplitude is at most log--divergent. Then, $\omega_d(\cG)=0$,
if 

(a1) $V_{4}=1$, $C_\bG=1$, $V_{2;1}=0 =V_{2;{\rm a}}$, 
and  $\omega(\cexG)=0=\omega(\bG)$;

(a2) $V_{4}=0$, $C_\bG=1+a$, $V_{2;1}=0 =V_{2;{\rm a}}$, 
and $\omega(\cexG)=0=\omega(\bG)$. However $C_\bG=1+a$
must be an integer, then only the situation for which $(d,a)=(4,1)$ 
is consistent. This case incorporates indeed the anomaly discovered in \cite{BenGeloun:2011rc}.

\medskip 

\noindent (b) One may also have $P_a(\cG)=-a$ entailed by 
$C_\bG=1$, $V_4=0$ and $V_{2;1}=0 =V_{2;{\rm a}}$.
In this case, the graph amplitude is at most $a$. We seek for cases such that $0 \leq \omega_d(\cG)\leq a$,

(b1) $\omega_d(\cG)=a$, if $\omega(\cexG)=0=\omega(\bG)$. 

(b2) Let us assume now that $\omega(\bG) >0$, then 
Proposition 1 in \cite{Geloun:2012fq} ensures that 
$\frac{2}{(d-1)!}(\omega(\cexG) - d\omega(\bG)) \geq d-2$
then, from \eqref{mag}, we have $\omega_d(\cG)\leq a-(d-2) <0$.

(b3) Let us assume now that $\omega(\bG) =0$ and $\omega(\cexG)>0$, then we use \eqref{mag} in order to have 
$\frac{2}{(d-1)!}\omega(\cexG) \geq d-2$,
then $\omega_d(\cG)\leq a-(d-2)) <0$.

Thus, both (b2) and (b3) gives convergent amplitudes;

- if $N_{\ext} = 2$ necessarily leads to $C_{\bG}=1$ and $\omega(\bG)=0$ (there is a unique configuration of the boundary of a graph with two external legs such that this equality holds: for $d=4$,
all boundary jackets are planar; in $d=3$ the boundary graph 
is itself a planar graph), and
\beq
P_a(\cG) =  a\Big(-2 +V_{4} + 2 (V_{2;1}+ \delta_{d,4}V_{2;{\rm a}})\Big)  \,.
\eeq
The divergence degree is at most $2a$. 

\noindent
(c) We can set $P_a(\cG)=0$, once again the graph amplitude is at most
log--divergent. Finding configurations for which $\omega_d(\cG)=0$
leads to

(c1)  $V_{4}=2$, $V_{2;1}=0=V_{2;{\rm a}}$, 
and $\omega(\cexG)=0=\omega(\bG)$;

(c2)  $V_{4}=0$, $V_{2;1}+\delta_{d,4}V_{2;{\rm a}}=1$, 
and $\omega(\cexG)=0=\omega(\bG)$;

\medskip 

\noindent
(d) Setting $P_a(\cG)=-a$, the graph divergence degree  is $\omega_d(\cG) \leq a$, and 

(d1) $V_{4}=1$, $V_{2;1}=0=V_{2;{\rm a}}$, and $\omega(\cexG)=0=\omega(\bG)$, for which $\omega_d(\cG)=a$.

(d2) $V_{4}=1$, $V_{2;1}=0=V_{2;{\rm a}}$, and $\omega(\bG)\geq 0$, 
then using similar arguments as in (b2) and (b3), one proves that  
 $\omega_d(\cG)<0$.

\medskip 

\noindent
(e) Choosing $P_a(\cG)=-2a$, the amplitude is such that $\omega_d(\cG)\leq 2a$. We have

(e1) $V_{4} =0$, $V_{2;1}=0=V_{2;{\rm a}}$, 
let us assume that $\omega(\cexG)>0$, according to \eqref{mag} $\frac{2}{(d-1)!}\omega(\cexG)\geq 2$ so that 
$\omega_d(\cG) \leq -2(1-a)$. Then $\omega_d(\cG)<0$
if $(d,a)=(3,\frac23)$. We can only have $\omega_d(\cG)=0$
for $(d,a)=(4,1)$ if $\omega(\cexG)=(d-1)!=6$.

(e2) $V_{4} =0$, $V_{2;1}=0=V_{2;{\rm a}}$,  considering
$\omega(\cexG)=0$, then we obtain $\omega_d(\cG)=2a$.

\medskip 

A summary of the list of divergent graphs of the model
$_1\Phi^6_3$ is given in the Table \ref{tab:listprim} (the list of divergent
graph for the model $_1\Phi^6_4$ can be found in \cite{BenGeloun:2011rc}).

\begin{table}[h]
\begin{center}
\begin{tabular}{lccccccccccccc|}
\hline
$N_{\ext}$ && $V_2 + V_2''$  && $V_4$ && $\omega(\bG)$ && $C_{\bG}-1$ && $\omega(\cexG)$ && $\omega_d(\cG)$  \\
\hline
6 && 0 && 0 && 0 && 0&& 0&&0\\
4 && 0 && 0 && 0 && 0 && 0 && $a$ \\
4 && 0  && 1 && 0 && 0 && 0 && 0\\
2 && 0 && 0 && 0 && 0 && 0 && $2a$\\
2 && 0 && 1 && 0 && 0 && 0 && $a$\\
2 && 0 && 2 && 0 && 0 && 0 && 0\\
2 && 1 && 0 && 0 && 0 && 0 && 0\\
\hline
\end{tabular}
{\caption{List of primitively divergent graphs of the 
$_1\Phi^6_{d=3}$.\label{tab:listprim}}}
\end{center}
\end{table}

 Note that the list of  graphs with
divergent amplitudes of the $_1\Phi^6_4$ contains those of $_1\Phi^6_3$
plus two more lines defined by (a2) and (e1).
From (e1), one notices the fact that only the model 
$_1\Phi^6_{4}$ generates  sub-leading divergent non-melonic contributions. 

\medskip

$\bullet$ For $k_{\max}=4$, $(d, \dim G_D,a)\in \{(3,1,\frac12), (3,2,1),  (4,1,\frac34), (4,1,1), (5,1,1)\}$, consider a connected graph $\cG$ 
with $V_{2;1}$ number of mass terms. Then \eqref{omegaa} 
still holds with  
\beq
P_a(\cG) = \dim G_D (C_\bG -1)  + \frac12\Big[(\dim G_D(d-1) -2a) N_{\ext} - 2\dim G_D(d-1)\Big]+ 2a   V_{2;1} \,.
\label{degree4}
\eeq

- If $N_{\ext} > 4$, noting that $(\dim G_D(d-1) -2a) \geq 0$, one has
\beq
(\dim G_D(d-1) -2a) N_{\ext} - 2\dim G_D(d-1)
 >(\dim G_D(d-1) -2a) 4- 2\dim G_D(d-1)=0\,,
\eeq
so that $P_a(\cG) >0$ and $\omega_d(\cG) <0$. Thus all 
graphs of this kind have a convergent amplitude; 

- if $N_{\ext} =4$, the amplitude of $\cG$ is at most log--divergent
and so $\omega_d(\cG)=0$, if $V_{2;1}=0$, $C_\bG=1$,
and $\omega(\cexG) =0= \omega(\bG)$; 

- $N_{\ext} =2$ necessarily gives $C_\bG =1$ and $\omega(\bG)=0$
as the boundary graph here becomes the standard one. We get 
\beq
P_a(\cG) = 2a(-1+  V_{2;1})\,,
\eeq
and the divergence degree is at most $2a$. 
The following relevant cases can be read off:

\noindent (f)  $P_a(\cG)=0$, if this case the amplitude is at best
$\omega_d(\cG)=0$ occurring if $V_{2;1}=1$ and $\omega(\cexG) =0$. 

\noindent (g) $P_a(\cG)=-2a$, that means $V_{2;1}=0$, and in this
situation 

(g1) $\omega(\cexG) =0$ yields $\omega_d(\cG)=2a$.

(g2) $\omega(\cexG) >0$, by \eqref{mag}, we have
 $\frac{2\dim G_D}{(d-1)!}\omega(\cexG) \geq \dim G_D(d-2)$, so that 
\bea
\omega_d(\cG) \leq  -  \dim G_D(d-2) + 2a
\eea
and this leads to only log--divergent amplitude, namely  $\omega_d(\cG)=0$,  if  $\omega(\cexG) =\frac12 (d-1)!(d-2)$, which only occurs for 
$(d,\dim G_D, a)\in \{(3,1,\frac12), (3,2,1), (4,1,1)\}$. 

In summary, Table \ref{tab:listprim2} gives the list of divergent
graphs for each model

\medskip 

\begin{table}[h]
\begin{center}
\begin{tabular}{lccccccccccc|}
\hline
$N_{\ext}$ && $V_{2;1}$   && $\omega(\bG)$ && $C_{\bG}-1$ && $\omega(\cexG)$ && $\omega_d(\cG)$  \\
\hline
4 && 0 && 0 && 0&& 0&&0\\
2 && 0  && 0 && 0 && 0 && $2a$ \\
2 && 1   && 0 && 0 && 0 && 0\\
\hline 
2 && 0  && 0 && 0 && $\frac12 (d-1)!(d-2)$ && 0\\
\hline
\end{tabular}
{\caption{List of primitively divergent graphs of the 
$_{\dim G_D}\Phi^4_{d}$. 
The last line with $N_{\ext}=2$ only concerns the cases
$(d,\dim G_D, a)\in \{(3,1,\frac12), (3,2,1), (4,1,1)\}$. 
\label{tab:listprim2}}}
\end{center}
\end{table}

\subsection{Renormalization in tensor models}
\label{subsect:rentensr}

The subsequent part of the renormalization program consists 
in the proof that the divergent and local part of all amplitudes can be
recast in terms present in the Lagrangian of the models studied so far. This is the purpose
of this section where we perform the Taylor expansion of the amplitudes
of graphs listed in  Table \ref{tab:listprim} and Table \ref{tab:listprim2}. We will not study separately the renormalization of the $N$-point functions for each model but rather perform the renormalization 
in more general notations valid for any model.  

\medskip 

\noindent {\bf Renormalization of marginal 4- and 6-point functions.}
 6-point functions are at most marginal and encountered only in the $_1\Phi^6_{d=3,4}$ models. Marginal 4-point functions occur in both
$\Phi^6$ and $\Phi^4$ models. 

The significant 6-point functions must be characterized by the first line of Table \ref{tab:listprim} (which is identical in both models $_1\Phi^6_{d=3,4}$).  The external momenta data of such graphs follows
necessarily the pattern of vertices $V_{6;1}$ or $V_{6;2}$ (see Figure \ref{fig:Inter6}). This is
the locality principle in such models. 
On the other hand, marginal 4-point functions are given by the third
line of Table \ref{tab:listprim} for $_1\Phi^6_{d=3,4}$ models, and the first
line of Table \ref{tab:listprim2} for all $\Phi^4$ models. The pattern of their
external momenta should follow from $V_4$ vertices (see 
Figure \ref{fig:Int4}). 

In the following, since the analysis can be carried out for any other external momentum configurations of the form given by vertices given by $V_{6;1}$, $V_{6;2}$, $V_4$ and $V_{4;{\rm a}}$ and will yield a similar result, we will treat in a row

- a 6-point graph with external data of the same form of one vertex of the $V_{6;1}$ type, namely the one given in Figure \ref{fig:Inter6} A.

- a 4-point graph the external momenta of which 
is given by the particular vertex $V_4$ given in Figure \ref{fig:Int4} A.

To be precise, consider a 6-point graph (respectively, a 4-point graph) with 6 external propagator lines (respectively, 4 propagators)
attached to it with momenta dictated by the pattern of the $V_{6;1}$
(respectively, $V_4$) vertex. For any rank $d$ model, each external field $\varphi_{[1]\{\check{1}\}}$ is written in the block matrix index notation as introduced in the beginning of Subsection \ref{subsect:just3}. 
The notation $f=f_{\{\check{1}\}}$ refers to $d-1$ external faces  whereas $f=f_{[1]}$ always refers to a unique external face in all models. 
We denote $\{P^{\ext}_{f}\}$ the set of external face momenta associated with $f\in\cF_{\ext}=\{f_{[1]},f_{\{\check{1}\}},f_{[1']},f_{\{\check{1}'\}},f_{[1'']},f_{\{\check{1}''\}}\}$ (respectively, 
$f\in \cF_{\ext}=\{f_{[1]},f_{\{\check{1}\}},f_{[1']},f_{\{\check{1}'\}}\}$).

In a compact notation, consider $A_{6/4}(\{P^{\ext}_f\})$ the amplitude of a $G^i_k$ graph of the form described above. There are
two types of scale indices to be considered in this amplitude: external scales $j_l$ associated to each external field corresponding to an external propagator line denoted $l$ and the internal scale $i$ related to all internal propagator lines. $G^i_k$ being quasi-local, this 
means that $j_l \ll i$.

The amplitude of $G^i_k$ is given by (in simplified notations)
\bea
A_{6/4}(\{P^{\ext}_f\})=\sum_{P_f} \int [\prod_{\ell}d\alpha_\ell e^{-\alpha_\ell \mu^2}]
\prod_{f \in \cF_{\ext}}\Big[
e^{-(\sum_{\ell \in f}\alpha_\ell) |P_{f}^{\ext}|^a} \Big]
\prod_{f \in \cF_{\inter}}\Big[\dee_{P_{f}}^2
e^{-(\sum_{\ell \in f}\alpha_\ell) |P_{f}|^a}  \Big],
\eea
where $\alpha_\ell \in [M^{-2ai_\ell}, M^{-2a(i_\ell-1)}]$ if $\ell$ 
is internal, and if $\ell$ is external, we denote it by $l$, such that $\alpha_l \in [M^{-2aj_l}, M^{-2a(j_l-1)}]$. We have $j_l \ll i \leq i_\ell$. 
 
The next stage is to perform a Taylor expansion 
of an external face amplitude using the fact that  $\sum_{\ell \in f; \ell \neq l}\alpha_\ell$ is  small such that 
\bea
e^{-(\sum_{\ell \in f}\alpha_\ell) |P_{f}^{\ext}|^a} 
&=& e^{-(\alpha_l+\alpha_{l'}) |P_{f}^{\ext}|^a} [1- R_f] 
\crcr
R_f &=&  \big(\sum_{\ell \in f; \ell \neq l}\alpha_\ell\big) |P_{f}^{\ext}|^a
\int_0^1 e^{-t(\sum_{\ell \in f; \ell \neq l}\alpha_\ell) |P_{f}^{\ext}|^a} dt \,.
\label{tayface}
\eea
We substitute this external face expansion in the amplitude 
and get
\bea
A_{6/4}(\{P^{\ext}_f\})&=&\sum_{P_f} \int [\prod_{\ell}d\alpha_\ell e^{-\alpha_\ell \mu^2}]\Big[
\prod_{f\in \cF_{\ext}}
e^{-(\alpha_l+\alpha_{l'}) |P_{f}^{\ext}|^a} \Big]\Big[1- \sum_{f \in \cF_{\ext}} R_f  + \sum_{f,f' \in \cF_{\ext}} R_f R_{f'} + \dots \Big] \crcr
&& \times 
\prod_{f \in \cF_{\inter}}\Big[\dee_{P_{f}}^2
e^{-(\sum_{\ell \in f}\alpha_\ell) |P_{f}|^a}  \Big]
\label{a6full}
\eea
where the dots invoke terms involving higher order products of the
Taylor remainders $R_f$. 

The divergence of $A_{6}(\{P^{\ext}_f\})$ come
from the 0th order term of this expansion which is given by
\bea
A_{6/4}(\{P^{\ext}_f\};0)= \sum_{P_f} \int [\prod_{\ell}d\alpha_\ell e^{-\alpha_\ell \mu^2}]\prod_{f\in \cF_{\ext}}\Big[
e^{-(\alpha_l+\alpha_{l'}) |P_{f}^{\ext}|^a} \Big]
\prod_{f \in \cF_{\inter}}\Big[\dee_{P_{f}}^2
e^{-(\sum_{\ell \in f}\alpha_\ell) |P_{f}|^a}  \Big]
\eea
and which factors as
\beq
A_{6/4}(\{P^{\ext}_f\};0)= \Big[\int [\prod_{l}d\alpha_l e^{-\alpha_l \mu^2}]
\prod_{f\in \cF_{\ext}}
e^{-(\alpha_l+\alpha_{l'}) |P_{f}^{\ext}|^a} \Big]\sum_{P_f} \int [\prod_{\ell\neq l}d\alpha_\ell e^{-\alpha_\ell \mu^2}]
\prod_{f \in \cF_{\inter}}\Big[\dee_{P_{f}}^2
e^{-(\sum_{\ell \in f}\alpha_\ell) |P_{f}|^a}  \Big].
\label{a60}
\eeq
In this expression, the first factor of $A_{6}(\{P^{\ext}_f\};0)$ can be fully expanded as
\bea
&&
\int [\prod_{l}d\alpha_l e^{-\alpha_l \mu^2}]
\prod_{f\in \cF_{\ext}}
e^{-(\alpha_l+\alpha_{l'}) |P_{f}^{\ext}|^a}= \crcr
&&\int [\prod_{l}d\alpha_l e^{-\alpha_l \mu^2}]
e^{-(\alpha_{l_{[1]}}+\alpha_{l'_{[1]}}) |P_{f_{[1]}}^{\ext}|^a}
e^{-(\alpha_{l_{\{\check{1}\}}}+\alpha_{l'_{\{\check{1}\}}}) |P_{f_{\{\check{1}\}}}^{\ext}|^a} \crcr
&& \times 
e^{-(\alpha_{l_{[1']}}+\alpha_{l'_{[1']}}) |P_{f_{[1']}}^{\ext}|^a}
e^{-(\alpha_{l_{\{\check{1}'\}}}+\alpha_{l'_{\{\check{1}'\}}}) |P_{f_{\{\check{1}'\}}}^{\ext}|^a} 
e^{-(\alpha_{l_{[1'']}}+\alpha_{l'_{[1'']}}) |P_{f_{[1'']}}^{\ext}|^a}
e^{-(\alpha_{l_{\{\check{1}''\}}}+\alpha_{l'_{\{\check{1}''\}}}) |P_{f_{\{\check{1}''\}}}^{\ext}|^a} 
\eea
Now using the pattern of the $V_{6;1}$ for each external momenta,
we have 
\bea
&&
\alpha_{l_{[1]}}=\alpha_{l_{\{\check{1}\}}}= \alpha_{l_1}\,,\quad  
\alpha_{l_{[1']}}= \alpha_{l'_{\{\check{1}\}}} = \alpha_{l_2}\,, \quad
\alpha_{l'_{[1']}}=\alpha_{l_{\{\check{1}'\}}}= \alpha_{l_3}\,,\quad  \crcr
&& 
\alpha_{l_{[1'']}}= \alpha_{l'_{\{\check{1}'\}}} = \alpha_{l_4}\,, \quad
\alpha_{l'_{[1'']}}= \alpha_{l_{\{\check{1}''\}}} = \alpha_{l_5}\,, \quad
\alpha_{l'_{[1]}}= \alpha_{l'_{\{\check{1}''\}}} = \alpha_{l_6}
\eea
from which we identify that the first factor of \eqref{a60} 
determines nothing but a product of 6 propagators glued together
in the pattern of a $V_{6;1}$ interaction.  
Using a similar kind of expansion, such that 
\bea
&&
\alpha_{l_{[1]}}=\alpha_{l_{\{\check{1}\}}}= \alpha_{l_1}\,,\quad  
\alpha_{l_{[1']}}= \alpha_{l'_{\{\check{1}\}}} = \alpha_{l_2}\,, \quad
\alpha_{l'_{[1']}}=\alpha_{l_{\{\check{1}'\}}}= \alpha_{l_3}\,,\quad
\alpha_{l'_{[1]}}= \alpha_{l'_{\{\check{1}'\}}} = \alpha_{l_4}\,,
\label{4ptdata}
\eea
we can identify in the amplitude
$A_{4}(\{P^{\ext}_f\};0)$ that the first factor is a product of 
4 propagators glued together as a $V_{4}$ vertex. 
The second factor in \eqref{a60} is a log--divergent term. In all cases, this term should contribute to the renormalization of the coupling constant 
associated with either a $V_{6;1}$ or a $V_4$ interaction for the corresponding situation. 

Next we must prove that the remainders appearing in $A_{6/4}$ \eqref{a6full}
improve in a significant way the power counting. We have the first
order remainder:  
\bea
R_{6/4} &=&\sum_{P_f} \int [\prod_{\ell}d\alpha_\ell e^{-\alpha_\ell \mu^2}]\Big[
\prod_{f\in \cF_{\ext}}
e^{-(\alpha_l+\alpha_{l'}) |P_{f}^{\ext}|^a} \Big]\Big[- \sum_{f \in \cF_{\ext}} R_f  \Big] 
\prod_{f \in \cF_{\inter}}\Big[\dee_{P_{f}}^2
e^{-(\sum_{\ell \in f}\alpha_\ell) |P_{f}|^a}  \Big]\crcr
&=& 
\sum_{P_f} \int [\prod_{\ell}d\alpha_\ell e^{-\alpha_\ell \mu^2}]\Big[
\prod_{f\in \cF_{\ext}}
e^{-(\alpha_l+\alpha_{l'}) |P_{f}^{\ext}|^a} \Big]\Big[- \sum_{f \in \cF_{\ext}}\big(\sum_{\ell \in f; \ell \neq l}\alpha_\ell\big) |P_{f}^{\ext}|^a
\int_0^1 e^{-t(\sum_{\ell \in f; \ell \neq l}\alpha_\ell) |P_{f}^{\ext}|^a} dt  \Big] \crcr
&& \times
\prod_{f \in \cF_{\inter}}\Big[\dee_{P_{f}}^2
e^{-(\sum_{\ell \in f}\alpha_\ell) |P_{f}|^a}  \Big]
\label{rem64}
\eea
Using $i(G^i_k)=\inf_{\ell\in G^i_k} i_\ell $ and $e(G^i_k)=\sup_{l \in G^i_k}j_l$, the last quantity can be optimally bounded as
\bea
|R_{6/4}| \leq K M^{-2a(i(G^i_k)- e(G^i_k))}
\sum_{P_f} \int [\prod_{\ell \neq l}d\alpha_\ell e^{-\alpha_\ell \mu^2}]
\prod_{f \in \cF_{\inter}}\Big[\dee_{P_{f}}^2
e^{-(\sum_{\ell \in f}\alpha_\ell) |P_{f}|^a}  \Big]
\label{a6remain}
\eea
for some constant $K$ and where the integral in $t$ brings simply a $O(1)$ factor. The factor $M^{-2a(i(G^i_k)- e(G^i_k))}$ improves
the power counting and will bring enough decay in such way that the last sum on momentum scale attributions can be performed  \cite{Rivasseau:1991ub}.  
One can prove in a similar way that higher order remainders in \eqref{a6full} will be even more convergent. 

\medskip

\noindent {\bf Renormalization of $a$--divergent 2- and 4-point functions.}
This type of 4-point and 2-point functions appears in the $_1\Phi^6_{3,4}$ models. We call them $a$--divergent 
for their property  $\omega_d(\cG)=a$. 
Such 4-point graphs should be characterized by 
the second row of Table \ref{tab:listprim} and of the external form given again by one of the $V_{4}$ vertex. Concerning 2-point functions, these are determined
by the fifth row of the same table. These should appear in both models $_1\Phi^6_{3,4}$. 

In the following, dealing with the 4-point function, 
we will concentrate on the situation of Figure \ref{fig:Int4} A again. The following
developments can be easily reported accordingly for other types
of configurations. Meanwhile for the 2-point function, 
there is a unique set of data encoding the external face configuration for
a graph with 2-external legs: $\cF_{\ext}=\{f_{[1]},f_{[2]},f_{[3]}\}$, 
where $[1]$ and $[3]$ are 1-index, and where $[2]$ is either a block containing two indices for $_1\Phi^6_{4}$, or a 1-index for $_1\Phi^6_{3}$.  

In a similar way as in the previous case, we perform a Taylor expansion of the face amplitude as given 
in \eqref{tayface} and write the amplitude expansion $A_{4/2}(\{P^{\ext}_f\})$ for a $G^i_k$ graph with 4 and 2 external legs the external data of which follow the pattern of a $V_4$ and $V_2$ vertex configuration, respectively. One should obtain the expression \eqref{a6full}.

The 0th order term $A_{4}(\{P^{\ext}_f\};0)$ factorizes in the same way given in \eqref{a60} and, using still \eqref{4ptdata},
provides a divergent term of degree $a$ contributing to the renormalization of the coupling constant associated with the $V_4$ interaction in all 
models. The 0th order term $A_{2}(\{P^{\ext}_f\};0)$ factorizes as well
as \eqref{a60} with its first factor recast as 
\beq\label{massterm}
 \int [\prod_{l=1,2}d\alpha_l e^{-\alpha_l \mu^2}]\Big[\prod_{f\in \cF_{\ext}}
e^{-(\alpha_l+\alpha_{l'}) |P_{f}^{\ext}|^a} \Big]
 =  \int [\prod_{l=1,2}d\alpha_l]
e^{-\alpha_{1} (\sum_{s} |P_{s}^{\ext}|^a+\mu^2) }
e^{-\alpha_{2} (\sum_{s} |P_{s}^{\ext}|^a+\mu^2) } 
 \Big]
\eeq
corresponding to the gluing of two propagators. Thus, $A_{2}(\{P^{\ext}_f\};0)$ is associated with a mass renormalization term. 

Concerning the remainders that we denote $R_{4/2}$, noting their similarity with \eqref{rem64},  they are bounded as 
\bea
| R_{4/2}| \leq  K M^{-2a(i(G^i_k)- e(G^i_k))}
\sum_{P_f} \int [\prod_{\ell \neq l}d\alpha_\ell e^{-\alpha_\ell \mu^2}]
\prod_{f \in \cF_{\inter}}\Big[\dee_{P_{f}}^2
e^{-(\sum_{\ell \in f}\alpha_\ell) |P_{f}|^a}  \Big]\,.
\eea
Now since the last integral  provides a divergence degree
of $a$, it is direct to get
\bea
| R_{4/2}| \leq  K M^{-2a(i(G^i_k)- e(G^i_k))} M^{\omega_d(G^i_k)=a}
\eea
ensuring already the convergence of all of the remainders and the summability in the attributions.

\medskip

\noindent {\bf Renormalization of $2a$--divergent 2-point  functions.}
Such 2-point functions occur in all model and they satisfy $\omega_d(\cG)=2a$. 

A second order Taylor expansion of the face amplitude is performed
as
\bea
e^{-(\sum_{\ell \in f}\alpha_\ell) |P_{f}^{\ext}|^a} 
&=& e^{-(\alpha_l+\alpha_{l'}) |P_{f}^{\ext}|^a} [1+ R_f+ Q_f] 
\crcr
R_f &=& -\big(\sum_{\ell \in f; \ell \neq l}\alpha_\ell\big) |P_{f}^{\ext}|^a\,,  \qquad 
Q_f =  [\big(\sum_{\ell \in f; \ell \neq l}\alpha_\ell\big) |P_{f}^{\ext}|^a]^2
\int_0^1 (1-t)e^{-t(\sum_{\ell \in f; \ell \neq l}\alpha_\ell) |P_{f}^{\ext}|^a} dt\,, 
\label{tayface2}
\eea
and the amplitude expansion for a $G^i_k$ graph satisfies
 \bea
A_{2}(\{P^{\ext}_f\})& =& \sum_{P_f} \int [\prod_{\ell}d\alpha_\ell e^{-\alpha_\ell \mu^2}]\Big[
\prod_{f\in \cF_{\ext}}
e^{-(\alpha_l+\alpha_{l'}) |P_{f}^{\ext}|^a} \Big] \crcr
&& \times \Big[1+ \sum_{f \in \cF_{\ext}} (R_f + Q_f)+ \sum_{f,f' \in \cF_{\ext}} (R_f + Q_f)(R_{f'} + Q_{f'}) + \dots \Big]
\prod_{f \in \cF_{\inter}}\Big[\dee_{P_{f}}^2
e^{-(\sum_{\ell \in f}\alpha_\ell) |P_{f}|^a}  \Big]
\eea
The 0th order term $A_{2}(\{P^{\ext}_f\};0)$ factorizes in the way of \eqref{a60} and, using \eqref{massterm}, provides a $2a$ divergent term contributing to the renormalization of the mass coupling.  

The remainders are now treated. The first order remainder involving  the sum $\sum_{f}R_f$ factorizes as  
\bea
A'_{2}(\{P^{\ext}_f\};0) &=& -\Big[ 
\int [\prod_{l}d\alpha_l e^{-\alpha_l \mu^2}]
\prod_{f\in \cF_{\ext}}
e^{-(\alpha_l+\alpha_{l'}) |P_{f}^{\ext}|^a} \Big]
 \sum_{f \in \cF_{\ext}}|P_{f}^{\ext}|^a 
\int [\prod_{\ell\neq l }d\alpha_\ell e^{-\alpha_\ell \mu^2}] 
\Big[ 
\big(\sum_{\ell \in f; \ell \neq l}\alpha_\ell\big) 
\crcr 
&& \times 
\sum_{P_f} \prod_{f \in \cF_{\inter}}\Big[\dee_{P_{f}}^2
e^{-(\sum_{\ell \in f}\alpha_\ell) |P_{f}|^a}  \Big]\,.
\eea
One notices that the first factor is again the product 
of two propagators using \eqref{massterm}. Hence, this  term should correspond to a wave function renormalization
if and only if the last integral over $\alpha_\ell$, $\ell \neq l$,  should give the same result for all  $|P^{\ext}_{f_{[1]}}|^a,|P^{\ext}_{f_{[2]}}|^a,$ and $|P^{\ext}_{f_{[3]}}|^a$. 
It may happen that for a given graph $G^i_k$, 
the integral is not the same at fixed $f \in \cF_{\ext}$. 
Because the model is symmetric in all colors,
it is simple to define in this case another graph $\tilde{G}^i_k$ so that the sum of contributions of $G^i_k$ and 
$\tilde{G}^i_k$ appears now to be symmetric for all $P^{\ext}_{f}$. 
A unique factor or wave function renormalization $A'$ can be defined 
from that colored symmetric quantity and yields $\Big(\sum_{s}|P_{s}^{\ext}|^a\Big)A'$. The sum $\sum_{\ell \in f; \ell \neq l}\alpha_\ell
\leq c M^{-2a i(G^i_k)}$ and the last integrals give $M^{\omega(G^i_k)=2a}$ so that the overall contribution to 
the wave function renormalization is $A' \sim \log M$. 

Focusing on the sum $\sum_fQ_f$, we can work out a bound as
\bea
|R_{2}| &\leq & K \Big[ 
\int [\prod_{l}d\alpha_l e^{-\alpha_l \mu^2}]
\prod_{f\in \cF_{\ext}}
e^{-(\alpha_l+\alpha_{l'}) |P_{f}^{\ext}|^{a}} \Big]
 \sum_{f,f' \in \cF_{\ext}}|P_{f}^{\ext}|^a |P_{f'}^{\ext}|^a \crcr 
&& \times 
\int [\prod_{\ell\neq l }d\alpha_\ell e^{-\alpha_\ell \mu^2}] 
\Big[ 
\big(\sum_{\ell \in f; \ell \neq l}\alpha_\ell\big)
\big(\sum_{\ell \in f'; \ell \neq l}\alpha_\ell\big)
\sum_{P_f} \prod_{f \in \cF_{\inter}}\Big[\dee_{P_{f}}^2
e^{-(\sum_{\ell \in f}\alpha_\ell) |P_{f}|^a}  \Big]\crcr
&\leq & K' M^{-4a(i(G^i_k)- e(G^i_k))}
M^{\omega_d(G^i_k)=2a}\,,
\eea
for some constants $K$ and $K'$. This last expression manifests the fact that the remainder  is convergent and will bring enough decay for the summability over the momentum assignment. One can show that, in the same vein,  higher order remainders are convergent as well. 

In conclusion, 

- the expansion of marginal 6- and 4-point functions around their local part gives a log--divergent term which renormalize the coupling constant associated with the 6- and 4-valent vertices, respectively.

- the expansion of $a$--divergent 4- and 2-point amplitudes around
their local part gives a $a$--divergent term which renormalize the coupling of 4-valent and mass vertices, respectively; 

- the expansion of a $2a$--divergent 2-point graph around its local
part yields a $2a$--divergent term renormalizing the mass
and a log--divergent term which contributes to the wave function
renormalization; 

- all remainders are convergent and will bring enough decay  for ensuring the final summability over scale attribution. From this point, the procedure for performing this last sum over attributions is standard  and will secure the renormalization at all orders of perturbation theory according to techniques developed in \cite{Rivasseau:1991ub}. 
Thus, Theorem \ref{theo:rentensor} holds. 

\section{Just renormalizable matrix models}
\label{sect:matrix}

\subsection{Matrix models and their renormalizability}
\label{subsect:just2}

Table \ref{table:modrk2} provides a list of matrix model interactions and kinetic terms susceptible to generate renormalizable actions. These are defined by 
\beq
(\,  _1 \Phi^{2+\gamma}_2\,, a=\frac{\gamma}{2(2+\gamma)}\leq \frac12) \,, \quad (\,   _2 \Phi^{2+\gamma}_2 , a=\frac{\gamma}{2+\gamma}\geq \frac12)\,, \quad
(\,  _3 \Phi^{4,6}_2\,, a=\frac34,1) \,, \quad (\,   _4 \Phi^{4}_2 , a=1)\,,
\label{matrixmod}
\eeq
where $\gamma$ is an even integer.

\medskip 

\noindent{\bf Propagators.} The propagator keeps its form 
\eqref{chew0} and can be pictured like a ribbon line as found in Figure \ref{fig:ribbon}.

\medskip 

\noindent{\bf Planar (and cyclic) interactions.} The interactions that we will introduce in the following will govern the locality principle of the matrix  models designed simply by a planarity condition. These are matrix trace invariants represented by planar graphs with $p$ legs. 

For any $_{\dim G_D}\Phi^{k_{\max}}_{2}$ model, consider the interactions giving by, for all $k=4,6, \dots, k_{\max}$,  
\beq
S^{\inter}_k= \sum_{P_{[I]}} {\rm tr} \Big[ (\bar\varphi_{[I]}\varphi_{[I]})^{k}\Big]= \sum_{P_{[I]}} 
\bar\varphi_{12}\,\varphi_{1'2}\,\bar\varphi_{1'2'}\,\varphi_{1''2'}\dots
\bar\varphi_{1'''2'''}\,\varphi_{12'''} \,.
\label{inmat}
\eeq
Figure \ref{fig:ribbon} illustrates $S^{\inter}_k$ as a cyclic and planar ribbon diagram with $k$ external fields. 

We introduce a cut-off $\Lambda$ on large momenta, so that the propagator in the UV reads $C^\Lambda$. Counter-term couplings $CT$ define as usually as the difference between bare and renormalized couplings. Mass and wave function counter-terms keeps their form
$S_{2;1}$ and $S_{2;2}$ \eqref{cterms}, where $\varphi_{[I]}$ may be simply written 
as a matrix $\varphi_{12}$. 

Given a matrix model $_{\dim G_D}\Phi^{k_{\max}}_{2}$, we introduce
the action defined by 
\beq
S^\Lambda = \sum_{k=2}^{k_{\max}/2} \frac{\lambda_k^{\Lambda}}{k}
S^{\inter}_k + CT_{2;1}^{\Lambda} S_{2;1} +  CT_{2;2}^{\Lambda} S_{2;2}\,, 
\label{inmat2}
\eeq
and the following statement is valid.

\begin{theorem}[Renormalizable matrix models]\label{theo:renormat}
The models $(_{\dim G_D}\Phi^{k_{\max}}_{2},a)$ 
defined by
\bea
\forall k \geq 2,\quad &&
(_1\Phi^{2k}_{2}, a=\frac12(1-\frac{1}{k})) \text{ over } G=U(1)\,,\qquad  
(_2\Phi^{2k}_{2}, a=1-\frac{1}{k})  \text{ over }  G=U(1)^2\,, \cr\cr
&&
(_3\Phi^6_{2}, a=1) \text{ over }  G=U(1)^3 \text{ or } G=SU(2)\,,
\qquad 
(_3\Phi^4_{2}, a=\frac34) \text{ over }  G=U(1)^3 \text{ or } G=SU(2)\,,
\cr\cr
&&
(_4\Phi^4_{2}, a=1) \text{ over }  G=U(1)^4, 
\eea
with actions defined by \eqref{inmat2}
are renormalizable at all orders of perturbation. 
\end{theorem}

The multi-scale analysis has been already performed and gives
a power counting theorem stated in Theorem \ref{powcont}.
This provides in return a divergence degree as described by Proposition \ref{prop:mat}.  In the same way proved earlier, adding mass and 
wave function counter-terms in the action and brings $V_{2;1}$ and $V_{2;2}$ vertices, respectively, but does not affect the divergence degree \eqref{dmat}. One pays attention that this divergence degree  includes now $V_{2;1}$ mass counter-term vertices. 

\medskip

\noindent{\bf List of divergent graphs.} We consider a connected
graph $\cG$ with external legs such that we always have 
$C_{\bG}\geq 1$. We consider also $V_k$ the number of vertices
of coordination $k$ and, in particular, $V_2=V_{2;1}$ the number of mass
vertices.  Using the same strategy developed in the previous section,  the divergence degree is recast in the form 
\bea
 \omega_d(\cG) &=& - 2\dim G_D g_{\tilde\cG} - P_a(\cG) \,,
\crcr
P_a(\cG)  &=&
\dim G_D(C_\bG -1) 
+  \frac12\Big[(\dim G_D -2a) N_{\ext} - 2\dim G_D\Big] +\frac12\sum_{k=2}^{k_{\max}-2}\Big[ 2\dim G_D+ (2a-\dim G_D)k\Big]  V_k \,,  \crcr
&& 
\label{pamat}
\eea
where the sum $\sum_{k=2}^{k_{\max}-2}$ is performed over even 
integers.

Given the fact that $\dim G_D -2a> 0$, one keeps in mind that,
for $2\leq k < k' \leq k_{\max}$,  
\beq
 2\dim G_D+ (2a-\dim G_D)k > 2\dim G_D+(2a-\dim G_D) k' \geq 0 \,.
\label{kkprim}
\eeq
In the last inequality, the upper bound is only saturated  at $k'=k_{\max}$. Another useful relation is provided by the following:  if $C_{\bG}>1$, 
then, for all $2\leq k< k_{\max}$, 
\beq
\dim G_D(C_\bG -1) 
-  \frac12\Big[ 2\dim G_D+(2a-\dim G_D) k \Big] 
\geq 
\frac12(\dim G_D-2a) k    >0\,.
\label{cgk}
\eeq

We are now in position to analyze the divergent contributions. 

\medskip

$\bullet$ If $N_{\ext}> k_{\max}$, then 
\beq
(\dim G_D -2a) N_{\ext} - 2\dim G_D > (\dim G_D -2a) k_{\max} - 2\dim G_D =0
\label{ngk}
\eeq
so that $P_a(\cG) >0$ and the amplitude converges;

\medskip 

$\bullet$  if $N_{\ext}=k_{\max}$, then the amplitude is at most
log--divergent and  $\omega_d(\cG)=0$ holds if 
$ g_{\tilde\cG}=0$ and $P_a(\cG)=0$. This latter condition
occurs if 
$C_{\bG}=1$, $V_k=0$, for $2\leq k<k_{\max}$;

\medskip 

$\bullet$ if $N_{\ext}<k_{\max}$, 
 we are interested in the divergent amplitudes with 
$N_{\ext} = k_{\max}-q$, $2\leq q \leq k_{\max}-2$, $q$ even.
The quantity \eqref{pamat} can be recast in the following way
\bea
P_a(\cG) & =&
\dim G_D(C_\bG -1) 
+  \frac12\Big[ 2\dim G_D+(2a-\dim G_D) (k_{\max}-q) \Big](V_{k_{\max}-q}-1)\crcr
&& +\frac12\sum_{k\in S_{k_{\max},q}}\Big[ 2\dim G_D+ (2a-\dim G_D)k\Big]  V_k \,,
\label{pa1}
\eea
where $S_{k_{\max},q}= \{2,\dots, k_{\max}-2\} \setminus\{ k_{\max}-q\}$ is a set of even integers. Note that $S_{4,2}=\emptyset$ for
all $\Phi^4$ models. 

(h) Let us assume that $P_a(\cG)=0$, then the amplitude is at most
divergent. We have $\omega_d(\cG)=0$ in the following cases:

(h$q$) For $2\leq q \leq k_{\max}-2$, $V_{k_{\max}-q}=1$,  $V_k=0$, $k \in S_{k_{\max},q}$, and $C_\bG =1$;

($\bar{\rm h}q$) For $2\leq q \leq k_{\max}-2$, the cases $V_{k_{\max}-q}>0$ or $C_\bG >1$, from \eqref{pa1} and \eqref{cgk}, yield a convergent amplitude.
 Hence, we must have $V_{k_{\max}-q}=0$ and
$C_\bG =1$. The second equality entails that there is no anomalous term in the above matrix models. Then, necessarily, one has
\bea
P_a(\cG)  =
  - \frac12\Big[ 2\dim G_D+(2a-\dim G_D) (k_{\max}-q) \Big]
 +\frac12\sum_{k\in S_{k_{\max},q}} \Big[ 2\dim G_D+ (2a-\dim G_D)k\Big]  V_k   \,.
\label{pamat2}
\eea
For $2\leq k \leq k_{\max}-q<k_{\max}$, by \eqref{kkprim}, we know that
\beq
 2\dim G_D+ (2a-\dim G_D)k \geq 2\dim G_D+(2a-\dim G_D) (k_{\max}-q)>0 \,,
\eeq
so that if there is some $V_k>0$ with $2\leq k \leq k_{\max}-q$
then $P_a(\cG)>0$ leading to a convergent amplitude.  We  therefore focus on $V_{k}=0$ for $2\leq k \leq k_{\max}-q$ giving  
\beq
P_a(\cG) =
  - \frac12\Big[ 2\dim G_D+(2a-\dim G_D) (k_{\max}-q) \Big]
 +\frac12\sum_{k\in S'_{k_{\max},q}}\Big[ 2\dim G_D+ (2a-\dim G_D)k\Big]  V_k \,,
\label{pa3}
\eeq
where $S'_{k_{\max},q}=\{k_{\max}-q+2, \dots, k_{\max}-2\}$
including only even integer elements and is non empty only for $q\geq 4$. 
Note that for $k_{\max}=4$,
$S'_{4,2}=\emptyset$ and given $k_{\max}=6$,
$S'_{6,2}=\emptyset$ and $S'_{6,4}=\{4\}$. 

Whenever $S'_{k_{\max},q}=\emptyset$, then $P_{a}(\cG)<0$ should 
be treated in the sequel point. Thus, looking for conditions
such that $P_a(\cG)=0$
for graphs with $N_{\ext}=k_{\max}-2=2$ models reduces to find solutions for (h2) for $_{\dim G_D}\Phi^4_2$ models. 
Note also that the solutions of $P_a(\cG)=0$ in both models 
$_1\Phi^{2+\gamma}_2$ or $_2\Phi^{2+\gamma}_2$ should coincide since the quantities $P_a(\cG)$ in these models turn out to be proportional.

Expanding further \eqref{pa3} and using a change of variable
$k\to k-(k_{\max}-q)$, it can be found
\bea
&&
0=2P_a(\cG) =
  (2a-\dim G_D)q
 +\sum_{k \in S'_{k_{\max},q}}\Big[ 2\dim G_D+ (2a-\dim G_D)k\Big]  V_k \crcr
&& 
0=(2a-\dim G_D)q
 +\sum_{k\in S_{q}}  (2a-\dim G_D)(k-q) V_{k+(k_{\max}-q)}
 = (2a-\dim G_D)\Big(q -\sum_{k \in S_{q}} (q-k) V_{k+(k_{\max}-q)} \Big) \label{int}
\eea
where $S_{q}=\{2, \dots, q-2\}$ still includes only even integers. 

Since $2a-\dim G_D <0$, we understand now that solving $P_a(\cG)=0$ turns out to find non trivial partitions of $q/2\geq 0$ (i.e.
excluding $q/2=q/2+0$). Indeed, changing variables
$q\to q'=q/2$, $k \to k' =k/2$, we then rename $k'$ in $k$ and
get from \eqref{int} 
\bea
0 =q' -\sum_{k=1 }^{q'-1} (q'-k) V_{k_{\max}-2(q'-k)} \,.
\label{partion}
\eea
In order to find the set $\{V_{k_{\max}-2(q'-k)}\}_{k=1,\dots, q'-1}$ which fulfills \eqref{partion}, one performs an integer partition $(l_1,\dots, l_\alpha)$ of $q'$, 
namely $\sum_{a}l_a=q'$, $1\leq l_a\leq q'-1$ and then 
count the number of times that $q'-k=l_{a_0}$ appears in the partition. 
Since in a partition, the number of times that each addend appears
uniquely determines the partition, there is a one-to-one correspondence
between a set of solutions $V_{k_{\max}-2(q'-k)}$ of \eqref{partion}
and a given partition of $q'$. It is clearly difficult to work out for an arbitrary $q'$ the possible set of solutions but certainly these solutions are in finite
number and order by order in $q'$ can be read off.
\footnote{
For an illustration, we apply the formalism to the lowest order $k_{\max}=6$, such that $q=2,4\leq 6-2$. We concentrate on the possibility 
$q =4 \geq 4$, providing $q'=2$. A non trivial partition of $q'=2$ is given by $2=1+1$. Then, from \eqref{partion}, we have 
\beq
0=2 - (2-1) V_{6-2(2-1)}= 2 - V_4 \qquad \Leftrightarrow 
\qquad V_4 =2. 
\eeq
$V_4=2$ gives the number of times that 1 appears the above
partition of 2. Requiring $V_{2;1}=0$ and $g_{\tilde \cG}=0$ leads to
a log--divergent amplitude.  We can apply also the formalism to $k_{\max}=8$, for which are relevant $q=4,6$. For $q=6$, one gets $q'=3$, and \eqref{partion} gives
\beq
0= 3 -\sum_{k=1 }^{2} (3-k) V_{8-2(3-k)}
 = 3 - 2V_{4} - V_6 \qquad \Leftrightarrow 
\qquad (V_4,V_6) \in \{ (1,1), (0,3)\}\, 
\eeq
expressing the fact that $V_4$ and $V_6$ are the number
of times that 2 and 1, respectively, should appear in the partitions of  3=2+1=1+1+1. Imposing in addition $V_{2;1}=0$ and $g_{\tilde \cG}=0$, this configuration will be log--divergent. 

For $q=4$, $q'=2$,  \eqref{partion} becomes 
\beq
0 =2 -(2-1) V_{8-2(2-1)} =  2 -V_{6} \;\qquad \Leftrightarrow 
\qquad V_6 = 2\,.
\eeq 
with the same interpretation in terms of the partition of 2.  
Now putting $V_4=0=V_{2;1}$ will lead to a log--divergent
amplitude. }

\medskip

(i) Let us consider the case $P_{a}(\cG)= \frac{1}{2}(2a-\dim G_D)q_1<0$, 
where $q_1$ even and $2\leq q_1 \leq q$. (Note that we must have once again $C_{\bG}=1$ otherwise (i.e. $C_{\bG}>1$), by \eqref{cgk}, $P_a(\cG)>0$.)  One has
\bea
P_a(\cG) &=& \frac{1}{2}(2a-\dim G_D)(q-q_1+q_1)
 +\frac{1}{2} \sum_{k\in S_{q}}  (2a-\dim G_D)(k-q) V_{k+(k_{\max}-q)}
\crcr
&=&
 \frac{1}{2}(2a-\dim G_D)q_1 + \frac{1}{2}(2a-\dim G_D) (q-q_1 -\sum_{k\in S_{q}} (q-k) V_{k+(k_{\max}-q)})\,.
\eea
Hence the hypothesis $P_{a}(\cG)= \frac{1}{2}(2a-\dim G_D)q_1<0$ 
would require
\bea
&&
0= q-q_1 -\sum_{k\in S_{q_1}} (q-k) V_{k+(k_{\max}-q)}
-\sum_{k\in S'_{q_1}} (q-k) V_{k+(k_{\max}-q)} \,,
\eea
where $S_{q_1}=\{2,\dots, q_1-2\}$ contains only even integers
and $S'_{q_1,q} =\{q_1,q_1+2,\dots, q-2\}= S_{q}\setminus S_{q_1}$.

If for any $2\leq k \leq q_1-2$, $V_{k+(k_{\max}-q)} >0$
then $q-q_1- (q-k) V_{k+(k_{\max}-q)}<0$, then $P_a(\cG)\neq
 \frac{1}{2}(2a-\dim G_D)q_1$. Then necessarily, for any $2\leq k < q_1$, $V_{k+(k_{\max}-q)}=0$. We obtain, changing the variable
such that $k \to k -q_1$,
\beq
0= q-q_1 -\sum_{k\in S'_{0,q-q_1}} ((q-q_1)-k) V_{k+(k_{\max}-(q-q_1))} \,,
\eeq
with $S'_{0,q-q_1}=\{0,2,\dots, q-q_1-2\}$ including only even integers,
with $q-q_1\geq 2$. 
Thus this case again reduces again to the search of partitions of $(q-q_1)/2$ including trivial ones. If $q-q_1=0$, there is a unique
possibility given by $ V_{k}=0$, for all $k \in S_q$.

Assuming now that $P_{a}(\cG)= \frac{1}{2}(2a-\dim G_D)q_1<0$, 
where $q_1$ is odd and $2\leq q_1 < q$, one has
\beq
P_a(\cG) = \frac{1}{2}(2a-\dim G_D)q_1 + \frac{1}{2}(2a-\dim G_D) (q-q_1 -\sum_{k\in S_q} (q-k) V_{k+(k_{\max}-q)})\,,
\eeq
and thus the hypothesis requires
\bea
0= q-q_1
-\sum_{k\in S_q} (q-k) V_{k+(k_{\max}-q)} \,.
\eea
Noting that $q-q_1>0$ is odd and that the summation $k$ runs on even indices so that $q-k$ is even, this contradicts the fact that the above expression vanishes. In conclusion, writing $P_{a}(\cG)= \frac{1}{2}(2a-\dim G_D)q_1<0$, $q_1$ should be always even. 

The divergence degree in the present situation is 
bounded by $\omega_d(\cG) \leq \frac12(\dim G_D -2a)q_1$, 
$2\leq q_1 \leq q < k_{\max}$. Consider $g_{\tilde\cG}>0$, one infers
\bea
\omega_d(\cG) &=& -2\dim G_D g_{\tilde \cG} + \frac12(\dim G_D -2a)q_1 \leq -2 \dim G_D  +  \frac12(\dim G_D -2a)q_1\crcr
&\leq & -2\dim G_D  + \frac12(\dim G_D -2a)q < -2 \dim G_D  + (\dim G_D -2a) k_{\max} =0\,.
\eea
Thus whenever $g_{\tilde \cG}>0$, the amplitude is convergent. 
We only have a divergent amplitude for $ g_{\tilde \cG}=0$ such that $\omega_d(\cG) = \frac12(\dim G_D -2a)q_1$. 
\footnote{
As an illustration, we study $k_{\max}=8$, $q=6$ such that $S_6=\{2,4\}$. 
If $q_1=2$, $S'_{0,4}=\{0,2\}$
\beq
P_a(\cG) = \frac12(2a-\dim G_D)2 + \frac12(2a-\dim G_D)\big(4-4V_{4} -2V_{6}\big). 
\eeq
Requiring $P_a(\cG) =\frac12(2a-\dim G_D)2 $, leads to the solutions
of $2-2V_4-V_6=0$, which are $(V_4,V_6)=\{(1,0),(0,2)\}$ related to the partitions of $2=2+0=1+1$. It is simple to fix the rest
of parameters to zero in order to get $\omega_d(\cG)=-P_a(\cG)/a>0$. 

If $q_1=4$, $S'_{0,2}=\{0\}$, we have $P_a(\cG) = \frac12(2a-\dim G_D)4$, if $2- 2V_{6}=0$ corresponding to the partition of $1=1+0$.
We have a divergence degree $\omega_d(\cG)=-P_a(\cG)>0$
in this case too after setting the other parameters
to zero. If $q_1=6$, then $S'_{0,2}=\emptyset$ and
$\omega_d(\cG)=-P_a(\cG)>0$ for all parameters $V_k$
and $g_{\tilde \cG}$ set to 0. }

This achieves the study of divergent contributions in matrix
models presented in \eqref{matrixmod}. 
Appendix \ref{app:diverg}  illustrates the formalism by discussing a nontrivial example of the list of divergent amplitudes for the model $(_{\dim G_D}\Phi^8_2,a)$.

\subsection{Renormalization in matrix models}
\label{subsect:ren}

We address, in this section, the renormalization analysis of the diverging 
$N$-point functions in the matrix models studied in 
previous section. 

\medskip 

\noindent{\bf Renormalization of marginal $k_{\max}$-point functions.}
Consider a marginal $k_{\max}$-point function with 
$k_{\max}$ propagators hooked to it such that the graph $G^i_k$
associated with that amplitude obeys $g_{\tilde G^i_k}=0$, $C_{\partial G^i_k}=1$, $V_k=0$ for $2\leq k < k_{\max}$, and with external data following a cyclic matrix trace invariant pattern. The associated amplitude 
is given by
\beq
A_{k_{\max}}(\{P^{\ext}_f\}) =\sum_{P_f} \int [\prod_{\ell}d\alpha_\ell e^{-\alpha_\ell \mu^2}]
\prod_{f \in \cF_{\ext}}\Big[
e^{-(\sum_{\ell \in f}\alpha_\ell) |P_{f}^{\ext}|^a} \Big]
\prod_{f \in \cF_{\inter}}\Big[\dee_{P_{f}}^2
e^{-(\sum_{\ell \in f}\alpha_\ell) |P_{f}|^a}  \Big],
\label{akmax}
\eeq
Next we perform a Taylor expansion in each external
face amplitude as done in \eqref{tayface} and insert 
the result in \eqref{akmax}. The 0th order term in the
expansion factorizes in a similar way as found in  \eqref{a60} as
\beq
A_{k_{\max}}(\{P^{\ext}_f\};0) =\Big[\int [\prod_{l}d\alpha_l e^{-\alpha_l \mu^2}]
\prod_{f\in \cF_{\ext}}
e^{-(\alpha_l+\alpha_{l'}) |P_{f}^{\ext}|^a} \Big]\sum_{P_f} \int [\prod_{\ell\neq l}d\alpha_\ell e^{-\alpha_\ell \mu^2}]
\prod_{f \in \cF_{\inter}}\Big[\dee_{P_{f}}^2
e^{-(\sum_{\ell \in f}\alpha_\ell) |P_{f}|^a}  \Big].
\label{akmax0}
\eeq
We write $\cF_{\ext}=\{f_1,\dots, f_{k_{\max}}\}$
such that the first factor of $A_{k_{\max}}(\{P^{\ext}_f\};0)$ expands as
\beq\label{vcyc}
\int [\prod_{l=1}^{k_{\max}}d\alpha_l e^{-\alpha_{l} \mu^2}]
\prod_{j=1}^{k_{\max}}e^{-(\alpha_{l_j}+\alpha_{l'_j}) |P_{f_j}^{\ext}|^a}
=
\int [\prod_{l=1}^{k_{\max}}d\alpha_l]
\prod_{l=1}^{k_{\max}}
e^{-\alpha_{l}( \sum_{s=1}^2 |P_{l,s}^{\ext}|^a + \mu^2)}\,,
\eeq
where one identifies the pattern of the cyclic matrix interaction $V_{k_{\max}}$ for external momenta, 
\bea
j=1,\dots, k_{\max} \,, \qquad
P_{j,1}^{\ext} = P^{\ext} _{f_j}\,, \qquad P^{\ext} _{j,2}= P^{\ext} _{f_{j+1}} \qquad 
\text{and} \qquad 
\alpha_{l'_{j}}=\alpha_{l_{{j+1}}}=\alpha_{j}\,. 
\eea
The first factor of \eqref{akmax0} represents the gluing of $k_{\max}$ propagators in the pattern of the $V_{k_{\max}}$ interaction.  
By the power counting theorem this term is log--divergent. 

Studying the remainders, in same anterior notations,  the first order 
can be bounded up to some constant $K$ as 
\bea
|R_{k_{\max}}| \leq K  M^{-2a(i(G^i_k)- e(G^i_k))} M^{\omega_d(G^i_k)=0}
\label{rkmax}\,.
\eea
This shows that the remainder converges and will bring
additional convergence during the final assignment summation.

\medskip 

\noindent{\bf Renormalization of $(k_{\max}-q)$-point functions.}
We treat now the case of an amplitude associated with a graph $G^i_k$ with  $N=k_{\max}-q$ external propagators, $2\leq q \leq k_{\max}-2$.
According to the previous dissection of the type of divergent 
graphs, this case splits in several situations. Positive degree of divergence
may vary as $0 \leq \omega_d(G^i_k) \leq  -P_a(G^i_k)= \frac{q_1}{2}(\dim G_D -2a)$, where $0 \leq  q_1 \leq q$ and $q_1$ is even.  

$\star$ If $q_1=0$, then the amplitude is log--divergent and can be
handled in the same way as in the marginal $N=k_{\max}$-point
functions. The quasi local part of the amplitude  
renormalizes the $(k_{\max}-q)$-valent vertex if the initial external
data configuration of the graph follow the cyclic pattern of
a $V_{k_{\max}-q}$ vertex. All remainders are convergent. 

$\star$ Let us assume now that $q_1 \geq 2$. 
We choose a graph $G^i_k$ such that $g_{\tilde G^i_k}=0$, $C_{\partial G^i_k}=1$, $V_{p+ (k_{\max}-q)}=0$, for $2\leq p \leq q_1$ 
and a partition of $(q-q_1)/2$ as
\bea
&&
\text{if } q-q_1 >0\,, \qquad
0= q-q_1 - \sum_{p =0; p \text{ even}}^{q-q_1-2} (q-q_1-p)V_{p+k_{\max}-(q-q_1)} \,, 
\eea
and if  $q=q_1$ set $V_{p}=0$ for all $2\leq p \leq q-2$.
The amplitude of such a graph $G^i_k$ is given by
\beq
A_{k_{\max-q}}(\{P^{\ext}_f\}) =\Big[\int [\prod_{l}d\alpha_l e^{-\alpha_l \mu^2}]
\prod_{f\in \cF_{\ext}}
e^{-(\alpha_l+\alpha_{l'}) |P_{f}^{\ext}|^a} \Big]\sum_{P_f} \int [\prod_{\ell\neq l}d\alpha_\ell e^{-\alpha_\ell \mu^2}]
\prod_{f \in \cF_{\inter}}\Big[\dee_{P_{f}}^2
e^{-(\sum_{\ell \in f}\alpha_\ell) |P_{f}|^a}  \Big].
\label{akmax0fin}
\eeq

Two main cases occur: 

(A) Assuming $2\leq q < k_{\max}-2$, we perform a Taylor expansion of the face amplitude  at first  order as given by \eqref{tayface}. 
Using the same previous techniques,  
it is direct to show that the 0th order term $A_{k_{\max-q}}(\{P^{\ext}_f\})$ factors and reproduces the gluing of $k_{\max-q}$
propagator according the pattern of a vertex with $k_{\max-q}$
number of legs. Let us discuss the remainder and concentrate
on the first term involving $\sum_{f}R_f$. This term can be bounded as 
\bea
|R_{q_1}| &\leq& K M^{-2a(i(G^i_k)-e(G^i_k))}\sum_{P_f} \int [\prod_{\ell}d\alpha_\ell e^{-\alpha_\ell \mu^2}]\Big[
\prod_{f\in \cF_{\ext}}
e^{-(\alpha_l+\alpha_{l'}) |P_{f}^{\ext}|^a} \Big]
\prod_{f \in \cF_{\inter}}\Big[\dee_{P_{f}}^2
e^{-(\sum_{\ell \in f}\alpha_\ell) |P_{f}|^a}  \Big]\,.
\eea
The integration over $\alpha_{\ell \neq l}$ and summation
over $P_f$ yield the overall divergence degree of the $G^i_k$. 
We have
\bea
|R_{q_1}| &\leq& K M^{-2a(i(G^i_k)-e(G^i_k))} M^{\omega_d(G^i_k)}\,.
\label{rq1}
\eea
We need the following result (in the same previous notations). 
\begin{lemma}\label{lem1}
$\forall q_1$  such that $2 \leq q_1 \leq q \leq k_{\max}-2$, 
\bea
&&
\text{if } \quad q_1 < k_{\max}-2 \,,\qquad   
-2a + \frac{q_1}{2}(\dim G_D -2a) < 0\,,\crcr
&& 
\text{if } \quad q_1 = k_{\max}-2 \,,\qquad   
-2a + \frac{q_1}{2}(\dim G_D -2a) = 0\,.
\eea
\end{lemma}
\proof We have
\bea
 \frac12 \big(-4a + q_1(\dim G_D -2a) \big)
&\leq&  \frac12 \big(-4a + (k_{\max}-2)(\dim G_D -2a) \big) \crcr
&\leq & \frac12 \big(-4a + (\frac{2\dim G_D}{\dim G_D -2a}-2)(\dim G_D -2a) \big)=0
\eea
The inequation saturates only if $q_1= k_{\max}-2$. 

\qed

Using Lemma \ref{lem1}, one shows that the remainder $R_{q_1}$
converges since $-2a + \omega_d(G^i_k)<0$ because $q_1 \leq q <k_{\max}-2$. This means that having a $(k_{\max}-q)$-point
function, where $q<k_{\max}-2$ diverging like $\omega_d(G^i_k)=\frac{q_1}{2}(\dim G_D - 2a)$, for all $2\leq q_1\leq q$, with all graph properties required for being renormalizable, 
then expanding this function gives a unique contribution 
renormalizing the coupling constant of a vertex $V_{k_{\max}-q}$
and all remainders are convergent.

(B) Let us assume that $q= k_{\max}-2$, we are dealing necessarily with a $N_{\ext}=2$-point function. The graph $G^i_k$ has a 
divergence degree of the form $\omega_d(G^i_k)=\frac{q_1}{2}(\dim G_D - 2a)$, for all $2\leq q_1 \leq k_{\max}-2$. 

$\bullet$ If $q_1 < k_{\max}-2$, we perform a Taylor expansion on
external faces as \eqref{tayface} and, just as in the above situation, it is simple to check that the 0th order of the graph amplitude yields a vertex coupling renormalization for $V_{k_{\max}-q}$ whereas all remainders are convergent (by Lemma \ref{lem1}) and obey a bound like \eqref{rq1}.

$\bullet$ If $q_1 = k_{\max}-2$, we use in this case a Taylor expansion 
of the form \eqref{tayface2} for each external face. The procedure is similar to  tensor situation:  the Taylor expansion at 0th order
of the graph amplitude yields a mass renormalization for $V_{2;1}$
the first order remainder containing $\sum_{f}R_f$ provides the log--divergent term embodying the wave function renormalization term. 
This again holds by invoking symmetry arguments on graphs. 
All other remaining terms are simply convergent and 
will bring additional decay useful for the summation over
momentum assignments. 

In conclusion of this part, we realize that all expansions of 
diverging graphs respecting precise renormalizability criteria 
yield diverging local terms which can be recast as term
present in the matrix model action \eqref{matrixmod}. 
The remainders give enough decay allowing the summation
over scale attributions in the last stage and proof of the
finiteness of the Schwinger functions when removing the cut-off
\cite{Rivasseau:1991ub}. Thus, Theorem \ref{theo:renormat} holds.

\section{Super-renormalizable models}
\label{sect:Supren}

Let consider models such that $\dim G_D(d-1)=2a$. We have identified
some situations for which this occurs. For $d=3$, 
the divergence degree of a graph $\cG$ in the model $(_1 \Phi^k_3, a=1)$ assumes the form
\beq
\omega_d(\cG) = - (\omega(\cexG) - \omega(\bG)) - (C_\bG -1) 
+2(1 -V)\,.
\label{degsup}
\eeq
On the other hand, for $d=2$, the divergence degree of a  graph in the models $(_{\dim G_D}\Phi^k_2, a)$, with $(\dim G_D,a)\in \{(1,\frac12), (2,1)\}$, is given by
\beq
\omega_d(\cG) = - 4g_{\tilde\cG} - 2(C_\bG -1) 
+2(1 -V)  \,,  
\label{matsup}
\eeq
For  any rank, \eqref{degsup} and \eqref{matsup} tell us that only graphs with $V = 1$ vertex may diverge. Graphs with $V=1$ are called tadpoles. This situation is typical of super-renormalizable models. 

Given a maximal valence of vertices $k_{\max}$, the number
of melonic or planar interactions which can be built with maximal valence
$k_{\max}$ is certainly a finite number. 
Consider an action including all these  vertices
up to order $k_{\max}$. We do not need to consider any anomalous
term here ($C_\bG >1$  leads to convergence).

For the $(_1 \Phi^{k_{\max}}_3, a=1)$ model, 
introducing a UV cut-off $\Lambda$,   in a similar
way as in \eqref{actphi6}, it can be inferred that 
\beq
S^{\Lambda} = \sum_{k=4}^{k_{\max}} \sum_{i}\frac{\lambda^{\Lambda}_{k;i}}{\sigma_{k;i}} S^{\inter}_{k;i} +  CT^{\Lambda}_{2;1} S_{2;1}
\label{supermod}
\eeq
where $\sigma_{k;i}$ is a symmetry factor of the interaction 
$S^{\inter}_{k;i}$ including all interactions of the melonic form 
of valence $k$ up to some color permutation. 
The index $i$ is
at this point totally formal and parameterize the types of melonic 
interactions which differ up to a color permutation. 

Concerning matrix models $(_{\dim G_D} \Phi^{k_{\max}}_2, a)$, we introduce an interaction 
\beq
S^{\Lambda} = \sum_{k=4}^{k_{\max}} \frac{\lambda^{\Lambda}_{k}}{\sigma_{k}} S^{\inter}_{k} +  CT^{\Lambda}_{2;1} S_{2;1}\,.
\label{supermod2}
\eeq
where $S^{\inter}_{k}$ is the familiar trace invariant of
order $k$ for matrices. 
Note that we did not introduce a wave-function renormalization
counter-term but only a mass counter-term vertex in both 
\eqref{supermod} and \eqref{supermod2}.

The following statement holds. 

\begin{theorem}[Super-renormalizable tensor models]\label{theo:superren}
The models $(_{\dim G_D} \Phi^{k_{\max}}_d, a)$ defined by 
\bea
\forall k \geq 2\,,\qquad && 
(_1\Phi^{2k}_{2}, a=\frac12) \text{ over } G=U(1)\,, 
\qquad 
(_2\Phi^{2k}_{2}, a=1) \text{ over } G=U(1)^2\,, \cr\cr
&& 
(_1\Phi^{2k}_{3}, a=1) \text{ over } G=U(1)\,, 
\eea
with action defined by \eqref{supermod} and \eqref{supermod2} are super-renormalizable. 
\end{theorem}

The multi-scale analysis of a connected graph as performed in anterior sections will lead to power counting governed  by \eqref{degsup} or \eqref{matsup}. 
We can investigate the type of divergences which occur in the model
and prove that they appear  in finite number. Their expansion and 
subtraction  scheme can be done as in  Subsection \ref{subsect:rentensr}
and will lead to finiteness after summing over scale attribution.
Since this last part is completely standard, it will be not addressed here.  

\medskip 

We already know that all possible diverging connected graphs are 
generated by one vertex. 

- Considering $N_{\ext} \geq k_{\max}$ implies that one uses
more than 1 vertex, then it is immediate that the amplitude will 
be convergent. Having a graph such that $ N_{\ext}=k_{\max}$
defined by  a unique vertex necessarily means that the graph (which 
should be connected) is defined to be the open vertex
itself. This also leads to the convergence of the amplitude. 

- Consider now graphs such that $N_{\ext} < k_{\max}$. 
Given any vertex with valence $k \leq k_{\max}$, we can 
only built a finite number of tadpoles out of it. 
Hence tadpole graphs are certainly of finite number.
The number of divergent graphs which should be chosen
among these tadpoles is therefore finite. 

Associated with a tadpole with have a divergence
degree
\bea
\omega_d(\cG) = - (\omega(\cexG) - \omega(\bG)) - (C_\bG -1) 
\eea
which leads at most to a log--divergent amplitude. Thus 
$\omega_d(\cG)=0$ if and only if $\omega(\cexG) =0=\omega(\bG)$
and $C_\bG =1$. 

Performing a Taylor expansion around the local part of a $N_{\ext}$-point amplitude graph such that  $\omega(\cexG) =0=\omega(\bG)$
and $C_\bG =1$ and such that the external momentum data of this graph follows the pattern of a vertex of the theory with valence $N_{\ext}$ can be done in exact conformity with the previous developments. We can show 
that the 0th order term is log--divergent should renormalize a vertex of valence equals $N_{\ext}$.
The remainders are convergent and there is no need to introduce a wave function renormalization. 

Using the  techniques of \cite{Rivasseau:1991ub}, we can 
sum over momentum assignments using the additional decay
of the remainders appearing in the Taylor expansion of tadpoles.
This achieves the proof of Theorem \ref{theo:superren}.

\section{First order $\beta$-functions}
\label{sect:beta}

The renormalizability of the above models 
leads to another important question related to the UV asymptotic behavior
of these models. This section undertakes the computation
of the $\beta$-functions of the $_{\dim G_D}\Phi^{4,6}_{d=2,3,4,5}$ models
at enough number of loops so that we may conclude about
their UV behavior. The generic situation $\Phi^{k\geq 8}$ is intricate
and will have numerous perturbative corrections.
We will not address the study of these coupling here. 

\subsection{Method}
\label{subsect:Gen}

Consider a bare coupling constant $\lambda_{k}$ associated with 
some interaction $S^{\inter}_k$ itself associated with a vertex $V_{2k}$ of
valence $2k\geq 4$, and its renormalized coupling constant $\lambda^{\ren}_{k}$. The first order $\beta$-functions and renormalized coupling constant equations of the models are encoded in the ratio
\bea
\lambda^{\ren}_{k} = -\frac{\Gamma_{2k}(\{0\})}{Z^{k}}\,,
\label{reneq}
\eea
where $\Gamma_{2k}(\{b^{\ext}\})$ is the sum of all amputated one-particle irreducible (1PI) $2k$-point functions satisfying the renormalization criteria in order
to be associated with the coupling $\lambda_{k}$ and 
evaluated at first loops. In particular among these criteria, the external momentum data $\{b^{\ext}\}$ of any graph contributing to this 
this quantity should reproduce the pattern of the $V_{2k}$ vertex and should be planar or melonic. The quantity $Z$ is the so-called wave function renormalization which evaluates from 
\bea
Z = 1 - \partial_{(b^{\ext})^{2a}}\; \Sigma\big|_{b^{\ext}=0}\,,
\label{genwfr}
\eea 
where $\Sigma$ is the self-energy or sum of all amputated 1PI
two-point functions evaluated at the first loop orders.
Note that, according to the maximal valence $k_{\max}$ of the 
 interactions, the number of loops may vary from one
model to another. 
The function $\Sigma(b^{\ext}_1, \dots, b^{\ext}_d)$ is symmetric in its variables $b^{\ext}_s$ where $s$ refers again to strand momentum variables. 

In order to make clear the following developments, let us 
consider a simple coupling formulation of some renormalizable theory. 
In such case, the wave function renormalization $Z$ and $\Gamma_{2k}$ function generally express at first loop expansion in the simple form as
\beq
Z = 1- {\rm a}\lambda + O(\lambda^{2}) \,, 
\qquad 
\Gamma_{2k}= - \lambda + {\rm b}\lambda^2 +
O(\lambda^3) \,,
\eeq
where ${\rm a}\geq 0$ and ${\rm b} \geq 0$ are real numbers involving the different
graph contributions and their combinatorics. Computing now
the ratio \eqref{reneq}, one finds
\beq
\lambda^{\ren} =  \lambda + (k{\rm a}-{\rm b}) \lambda^2 + O(\lambda^3) \,.
\eeq
Then the quantity $k{\rm a}-{\rm b}$ determine the first order
$\beta$-function related to the model and its coupling
constant $\lambda$. If $k{\rm a}-{\rm b}>0$ the model is said asymptotically free,
if $k{\rm a}-{\rm b}<0$, it possesses the so-called Landau ghost with a 
coupling constant blowing in the UV \cite{Rivasseau:1991ub}. 
Meanwhile, in the case $k{\rm a}={\rm b}$ we call the model perturbatively 
safe at one-loop. It is a striking observation that the more $k$ is large
the more the quantity $k{\rm a}-{\rm b}$ is likely positive. But
the quantities ${\rm a}$ and ${\rm b}$ themselves are in fact function
of $k$ and this makes difficult to know a priori the 
sign of $\beta(k)= k{\rm a}(k)-{\rm b}(k)$ as $k$ may vary. 

The goal of this section is to show that for the previous renormalizable
tensor models $\Phi^{4}$, $\beta=k{\rm a}-{\rm b}>0$. Concerning
the renormalizable tensor models $\Phi^6$, the renormalized coupling
equations are much more involved and require further loop
calculations. 
It would be very interesting to show that the asymptotic freedom still holds for all renormalizable tensor models and therefore this feature 
is generic. The tower of potentially renormalizable model in rank $d=3$
prevents us to conclude anything at this stage if we include models with a dynamics which is not Laplacian.

The general procedure that we will use, even though lengthy, turns out to be efficient to get a definite result for the several types of $\beta$-functions for all interactions given by the above renormalizable models.
We use the following method: 

First, we enlarge the space of couplings and consider for each renormalizable
action defined by \eqref{actphi6} and \eqref{actphi4} as a multiple coupling theory. This means that 
we give to each interaction term associated with a certain permutation of colors a different coupling. Doing so, we have
(forgetting wave function and mass vertices)

- For $k_{\max}=6$, $\dim G_D= 1$ and $(d,a)\in \{(3, \frac23),(4,1)\}$, 
\beq
S = \sum_{\rho} \frac{\lambda_{6;1;\rho}}{3} S^{\inter}_{6;1;\rho} + \sum_{\rho\rho'}
 \lambda_{6;2;\rho\rho'}S^{\inter}_{6;2;\rho\rho} + 
\sum_{\rho}\frac{\lambda_{4;1;\rho}}{2} S^{\inter}_{4;\rho} 
 + \frac{\lambda_{4;2}}{2} \delta_{d,4}\,S^{\inter}_{4;{\rm a }}
\,,
\label{mcoup6}
\eeq
where the single label $\rho \in \{1,\dots,d\}$, and the second 
double index $\rho\rho'$ has to be chosen in all symmetric pairs of color such that $\rho\neq \rho'$. Precisely, $d=3$, $\rho\rho'\in\{12,13,23\}$, whereas for $d=4$,  $\rho\rho'\in\{12,13,14,23,24,34\}$;

- For $k_{\max}=4$, $(d, \dim G_D,a)\in \{(3,1,\frac12), (3,2,1), (4,1,\frac34), (4,1,1), (5,1,1)\}$,
\beq
S =\sum_{\rho} \frac{\lambda_{4;\rho}}{2} S^{\inter}_{4;\rho}\,,
\label{mcoup4}
\eeq
with $\rho$ keeping its above meaning. 

- For $d=2$ or matrix models,  the renormalizable actions \eqref{inmat2} are already in the proper ``multi-coupling'' form. 

Then, at the end, we collapse all couplings of all terms
which can be mutually identified up to a permutation of colors to a single value, that is
$\lambda_{6;1;\rho}\to \lambda_{6;1}$, 
$\lambda_{6;2;\rho\rho'}\to \lambda_{6;2}$, and 
$\lambda_{4;1;\rho}\to \lambda_{4;1}$. This will provide us
with the renormalized coupling equation for the models \eqref{actphi6}
and \eqref{actphi4}. Importantly, we will concentrate on the maximal valence interaction coupling in this work. The relevant couplings of the $\Phi^4$ type occurring in the $\Phi^{k_{\max}=6}_{d\geq 3}$ models can be simply inferred from this point whereas the  $\Phi^4$ couplings occurring in the $\Phi^{k_{\max}=6}_{2}$ matrix models might be very involved and will be not addressed.

\subsection{First order $\beta$-functions of tensor models}
\label{subsect:tensbeta}

\subsubsection{One-loop $\beta$-functions of the $\Phi^{4}$ models}
\label{subsub:beta4}

We simplify in the following the notations and use $P$ for $|P|$, for some momentum (multi-)variable $P$. 
Furthermore, in this notation, we recall that $P_s^{\,a}$ expands fully as $\sum_{i=1}^{D} |p_{s,i}|^{2a}$ for the representation of the group $U(1)^D$. We also use the block matrix notation $[1]=1$ and
$\{\check{1}\}=(2,\dots,d)$. The following formal series will be useful: 
\beq
\label{s1tens}
S^1 := \sum_{P_{s}} \frac{1}{ [ P_{\check{1}}^{a}+ \mu^2 ]^2}\,,
\eeq
where $P^a_{\check{1}}:=\sum_{s\in \{\check{1}\}} P^a_s$.
The self-energy  is expressed as
\bea
\Sigma(\{b\}) = \Sigma(b_{1}, \dots, b_{d})=\Sigma(b_{[1]},b_{\{\check{1}\}})
= \bra \bar\varphi_{[1]\{\check{1}\}} \, \varphi_{[1]\{\check{1}\}} \ket^{t}_{1PI}  
\label{sigmad}
\eea
where $b_s=(b_{s,1}, \dots, b_{s,D})$, $D=1,2$,
$s=1,\dots,d$, are external momenta. 

We start by evaluating the self-energy as
\bea
\Sigma(\{b\}) = \sum_{\cG_c} K_{\cG_c}\, \cS_{\cG_c}(\{b\})
\label{selfexp}
\eea
where the sum is over all amputated 1PI 2-point graphs $\cG_c$
that we truncate at one-loop, $K_{\cG_c}$ is a combinatorial 
factor and $\cS_{\cG_c}(\{b\})$ is the amplitude of the graph. 
These graphs should be listed in Table \ref{tab:listprim2} for $N_{\ext}=2$.  

\begin{figure}[t]
 \centering
    \begin{minipage}[t]{.8\textwidth}
\includegraphics[angle=0, width=7cm, height=2cm]{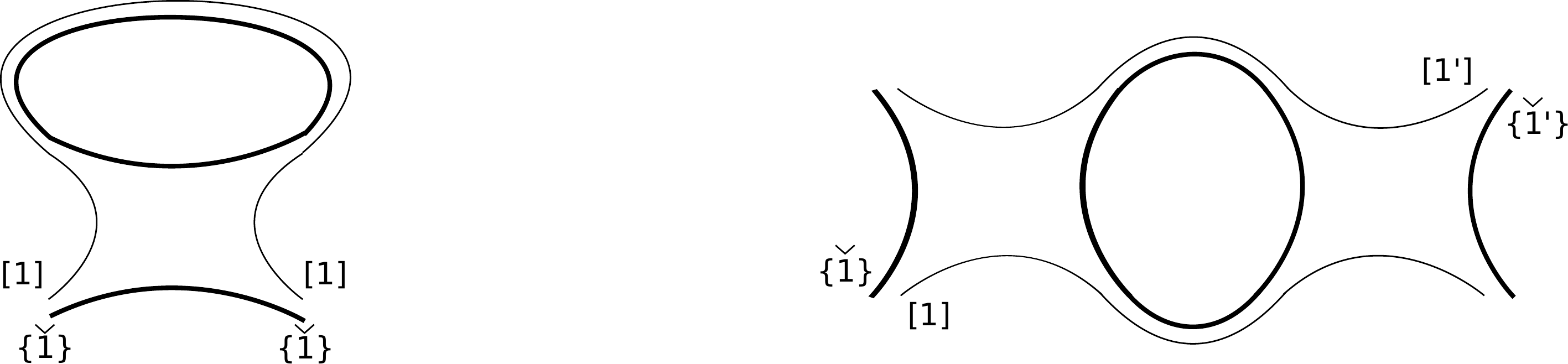}
\vspace{0.3cm}
\caption{ {\small Tadpole graph $T_1$ and 4-point graph $F_1$.}} \label{fig:beta40}
\end{minipage}
\put(-285,-12){$T_1$}
\put(-157,-12){$F_1$}
\end{figure}

Up to color permutation, the graphs which are the most divergent and which will contribute to \eqref{selfexp} are tadpoles of the melonic-type  $T_\rho$ coined by one particular color $\rho=1,\dots, d$ ($T_1$ has been illustrated in Figure \ref{fig:beta40}; $T_{\rho}$ can be obtained by color permutation). 
Let $A_{T_\rho}$ be the amplitude of $T_\rho$ which necessarily depends on $b_{\rho,i}$, $i\in \{1,\dots,D\}$. 
We fix a particular variable $b_{\rho,1}$ and derivate
this amplitude. Note that if $D=1$, there is no other choice
than the one dictated by the color $\rho$. 

The amplitude $A_{T_\rho}$ can be evaluated as
\bea
A_{T_\rho} (b_\rho) =
K_{T_\rho} \left(-\frac{\lambda_{4;\rho}}{2}\right) S_{\rho}(b_{\rho})  \,,
\qquad K_{T_\rho} =2 \,, \qquad 
S_{\rho}(b)   := 
\sum_{P_{s}} \frac{1}{  b^{a} + P_{\check{1}}^{a} +\mu^2 }\,,
\eea 
such that the  wave function renormalization is given by
\beq
Z = 1-\partial_{(b_{\rho,1})^{a}} \Sigma \Big|_{b_{s,i}=0}
= 1-  \partial_{(b_{\rho,1})^{a}} A_{T_\rho} \Big|_{b_{s,i}=0} =  1- \lambda_{4;\rho}S^1 + O(\lambda_{4}^2)\,, 
\eeq 
where $S^1$ is given in \eqref{s1tens} and $O(\lambda_{4}^2)$ involves
all quadratic power of coupling constants.

\medskip 

Next, we must evaluate the $\Gamma_{4;\rho}$ function at 0 external momenta. Formally, 
\bea
\Gamma_4(\{b\}) = \sum_{\cG_c} K_{\cG_c}\, \cS_{\cG_c}(\{b\})\,,
\eea
where the sum runs over all amputated 1PI 4-point graphs
which satisfy the first line of Table \ref{tab:listprim2}, and the external 
momentum data of which should reproduce the pattern
of the vertex having $\lambda_{4;\rho}$ as coupling constant. 

At second order of perturbation, there is a unique way to build these graphs (see Figure \ref{fig:beta40}). Denote $F_\rho$ such a graph where $\rho$ is the external color index used as well in the propagators. 
The amplitude of such graph can be written as
\bea
&&
A_{F_\rho}(\{b_{\rho}\}) = K_{F_\rho} \frac12\left( \frac{-\lambda_{4;\rho}}{2}\right)^2
S'_{\rho}(b_{\rho},b'_{\rho}) \,, \crcr
&&
\qquad  K_{F_\rho} = 2^3 \,\qquad
S'_{\rho}(b,b') = 
\sum_{P_s} \frac{1}{( b^{a} + P_{\check{1}}^{a}+\mu^2 )}\frac{1}{( b'^{a} + P_{\check{1}}^{a}+\mu^2 )}\,.
\eea
Hence,
\beq
\Gamma_{4;\rho}(\{0\})=- \lambda_{4;\rho} + A_{F_\rho}(\{0\})
 =- \lambda_{4;\rho}+ \lambda^2_{4;\rho}S^1 + O(\lambda^3)\,.
\eeq 

We are in position to calculate the renormalized coupling equation of $\lambda_{4}$. We first set $\lambda_{4;\rho}\to \lambda_{4}$ 
in all equations. Using \eqref{reneq}, one then gets 
\beq
\lambda^{\ren}_{4} = -\frac{\Gamma_{4}(\{0\})}{Z^2}
=  \lambda_{4}+ \lambda_{4}^2 S^1 +  O(\lambda_4^3)\,.
\label{ren4bet} 
\eeq
The $\beta$-function at one-loop for all renormalizable $\Phi^4$-models is always fixed and given by 
\bea
\beta = 1 \,.
\eea
The renormalized coupling equation \eqref{ren4bet} also 
exhibits the fact that $\lambda^{\ren}_{4} >\lambda_{4}$, for strictly positive coupling $\lambda_{4}>0$. This means that the 
models are asymptotically free in the UV. The free theory describes
non interacting topological $d$-spheres. Meanwhile going in the other IR direction, the renormalized coupling becomes larger and larger. 
This generally hints at a phase transition. A widely known example of this
kind of theory is certainly QCD where, in the IR, quarks and gluons 
experience a phase transition for making hadrons. We hope that, for large group
distances, the present models may lead to new condensate-type of degrees of freedom (quite different from the basic simplexes used in the initial models) which might be useful to describe interesting geometric properties.

\subsubsection{Two- and four-loop $\beta$-functions  of the $\Phi^{6}$ models}
\label{subsub:2loop6}

The computation of the $\beta$-functions for the $\Phi^6$ models
is inferred from a previous work \cite{BenGeloun:2012yk}. 
The types of graphs relevant for the calculation of
the $\beta$-function for each model 
$_1\Phi^6_{3}$ or $_1\Phi^6_{4}$ have been listed in that 
work. Indeed, the relevant graph construction does not
depend much on the characteristics of the models
($\dim G_D$, $a$) but on the rank $d$, $k_{\max}$ and on the combinatorics of constructing melonic graphs using $\Phi^6$ interaction. 
The reduction from $d=4$ to $d=3$ is quite immediate.

We introduce the block index notation $[1]=1$, and depending
on the model $[2]= (2,3)$ and $[3]=4$ for $_1\Phi^6_4$
or $[2]=2$ and $[3]=3$ for $_1\Phi^6_3$.
The following formal sums
will be also useful (avoiding multiplication of notations we shall
 use again $S^1$):
\bea
S^1 &:=&  \sum_{p_{s},p'_{s}}
\frac{1}{(p_{1}^{2a}+p_{\check{1}}^{2a} +\mu^2)^2}
\frac{1}{(p'^{2a}_{1}+ p'^{2a}_{\check{1}} +\mu^2)} \,,
  \crcr
S^{12} &:=& \sum_{p_{s}, p'_{s}}
\frac{1}{(p_{1}^{2a}+p_{\check{1}}^{2a} +\mu^2)^2}
\frac{1}{(p_{1}^{2a} +p'^{2a}_{\check{1}}+ p'^{2a}_{\check{1}} +\mu^2)}  \,,
\label{s0tens}
\eea
where $p_{1}^{2a}:= |p_1|^{2a}$ and 
$p_{\check{1}}^{2a}:= |p_{2}|^{2a} +\dots + |p_{d-1}|^{2a}$. 

From Lemma 1 in \cite{BenGeloun:2012yk}, 
at two loops, the self-energy $\Sigma$ and wave function renormalization $Z$ compute to 
\bea
\Sigma(b_{[1]},b_{[2]},b_{[3]})&=&  \Sigma^0(b_{[1]},b_{[2]},b_{[3]})+ \Sigma'(b_{[2]},b_{[3]})\, ,
\cr\cr
\Sigma^0(b_{[1]},b_{[2]},b_{[3]})&=&-\lambda_{6;1;1} \tilde{S}^{1}(b_{[1]},b_{[1]})
-\sum_{\rho\in \{[2],[3]\}}\Big[\lambda_{6;2;1\rho} \tilde{S}^{1}(b_{[1]},b_\rho)\Big]
 - \Big[\sum_{\rho\in \{[2],[3]\}}\lambda_{6;2;1\rho} \Big] \tilde{S}^{12}(b_{[1]}) + O(\lambda^2),
\label{sig0}\\
Z &=& 1- \partial_{(b_{[1]})^{2a}} \Sigma^0\Big|_{b_{[s]}=0} = 1 - \Big[2\lambda_{6;1;1} + \sum_{\rho\in \{[2],[3]\}}\lambda_{6;2;1\rho}\Big] S^1 
 - \Big[\sum_{\rho\in \{[2],[3]\}}\lambda_{6;2;1\rho} \Big]S^{12}  + O(\lambda^2),
\label{wfren}\\
\tilde{S}^{1}(b,b') &:=& 
 \sum_{p_{s}, p'_{s}}
\frac{1}{(b^{2a} + p_{1}^{2a}+ p_{\check{1}}^{2a}+ \mu^2)}
\frac{1}{(b'^{2a} + p'^{2a}_{1}+ p'^{2a}_{\check{1}}+\mu^2)} \,,
\label{s1}  \\
 \tilde{S}^{12}(b) &:=&  \sum_{p_{s}, p'_{s}}
\frac{1}{(b^{2a} + p_{1}^{2a}+ p_{\check{1}}^{2a}+\mu^2)}
\frac{1}{(p_{1}^{2a} + p'^{2a}_{1}+ p'^{2a}_{\check{1}}+\mu^2)}  \,,
\nonumber
\eea
where $\Sigma' =\Sigma- \Sigma^0$ consists in the self-energy remaining part which is independent of the variable $b_{[1]}$ 
and $O(\lambda^2)$ denotes a sum of O-functions with 
arguments any quadratic power of the coupling constants
$O(\lambda_{6;1;\bullet}^2) +  O(\lambda_{6;2;\bullet}^2) 
+ O(\lambda_{6;1;\bullet}\lambda_{6;2;\bullet})$.

\begin{figure}[h]
 \centering
    \begin{minipage}[t]{.8\textwidth}
\includegraphics[angle=0, width=3cm, height=1cm]{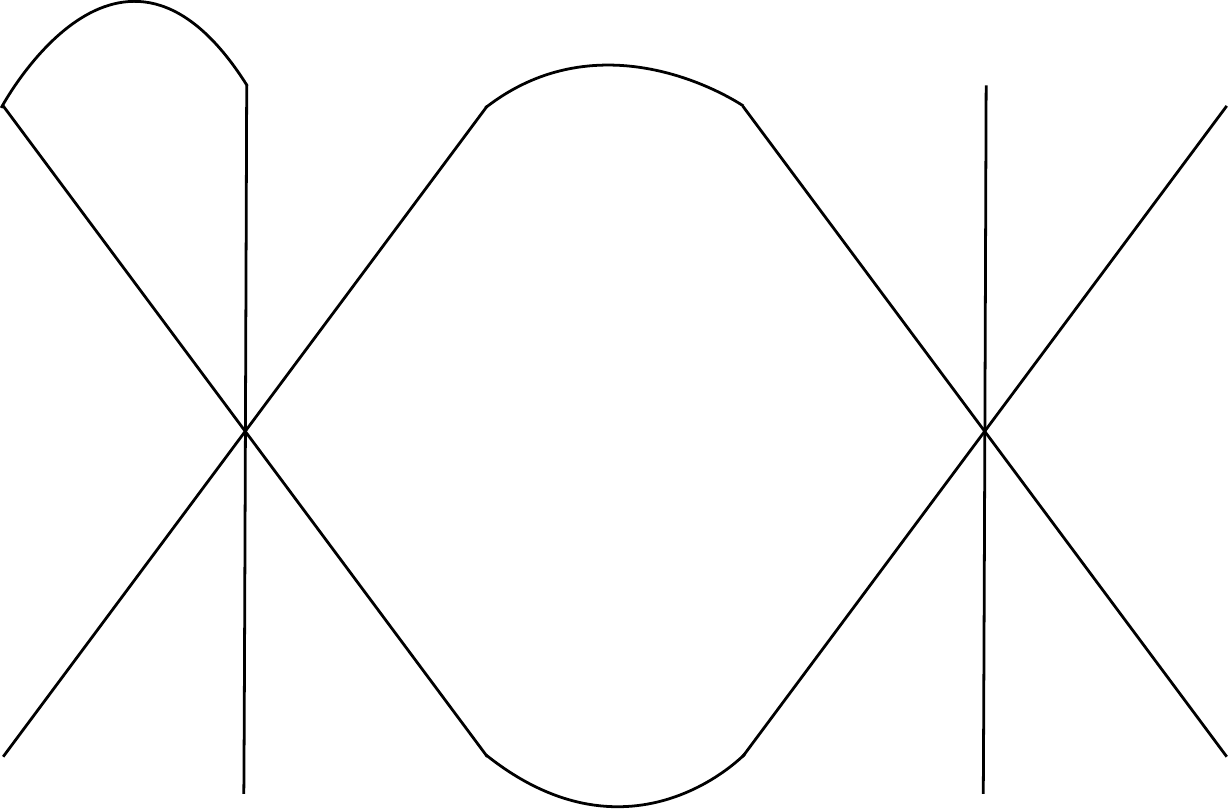}
\caption{ {\small General form of 1PI 6-point graphs.}} \label{fig:Gener6}
\end{minipage}
\end{figure}

1PI amputated 6-point functions truncated at two loops
are of the form  by Figure \ref{fig:Gener6}
(where we use the most simple representation of the 
$\Phi^6$-vertex as a vertex with 6 external legs).
The computations are lengthy but we do not need to 
rederive these.  Thus, we use Lemma 2 in \cite{BenGeloun:2012yk}, 
in order to get, at two loops, the amputated truncated six-point functions at zero external momenta given as, for $\rho\in \{[1],[2],[3]\},$
\beq
\Gamma_{6;1;\rho} (0,\dots,0)=
- \lambda_{6;1;\rho} +  \lambda_{6;1;\rho}
\Bigg[ 6\,\lambda_{6;1;\rho}S^1 \,  + 3\Big[\sum_{\rho' \in \{[1],[2],[3]\} \setminus \{\rho\}}\lambda_{6;2;\rho\rho'} \Big]  [S^1 +  S^{12}] \Bigg] + O(\lambda^3) \,,
\eeq
where $S^1$ and $S^{12}$ are given by \eqref{s0tens} and $O(\lambda^3)$ stands for a sum of $O$-functions of any cubic power in
the coupling constants. On the other hand,  for $\rho' \in \{[1],[2],[3]\} \setminus \{\rho\}$, the second function of interest reads
\bea
&&
\Gamma_{6;2;\rho\rho'}(0,\dots,0) = \crcr
&&
-\lambda_{6;2;\rho\rho'} +  \lambda_{6;2;\rho\rho'} \Bigg[
2[ \lambda_{6;1;\rho} +  \lambda_{6;1;\rho'} ]S^1
+
\Big[ \sum_{\bar\rho \in \{[1],[2],[3]\} \setminus \{\rho\} }\lambda_{6;2;\rho \bar\rho}
 +  \sum_{\bar\rho \in \{[1],[2],[3]\} \setminus \{\rho'\} }\lambda_{6;2;\rho' \bar\rho}\Big]   [S^1 +  S^{12}] \Bigg] +O(\lambda^3)\,. \crcr
&&
\eea
The renormalized coupling equation can be evaluated at two-loops
by equating all coupling constants $\lambda_{6;1;\rho}= \lambda_{6;1}$
and $\lambda_{6;2;\rho\rho'}= \lambda_{6;2}$. After a straightforward evaluation, one obtains
\bea
\lambda^{\ren}_{6;1} =-\frac{\Gamma_{6;1} (\{0\})}{Z^3}
=  \lambda_{6;1} + O(\lambda^3) \,.
\eea
Thus both models $\Phi^6$ are safe at two-loops in 
this sector $\lambda_{6;1}$. In order to know the UV behavior 
of this sector, we need to carry out the computations  of the $\beta$-function
up to  four loops. This will be performed after the next paragraph. 

Inspecting the second coupling, we have
\bea
\lambda^{\ren}_{6;2} = -\frac{\Gamma_{6;2} (\{0\})}{Z^3}
=
\lambda_{6;2}  + (d-1)\lambda^2_{6;2}[S^1 + S^{12}]+ 2 \lambda_{6;1} \lambda_{6;2} S^1 + O(\lambda^3) \,.  
\eea
We find that, for  $\lambda_{6;1}>0$ and $\lambda_{6;2}>0$,
 the $\beta$-function in this sector
splits as
\beq
\beta_{6;2(2)} =  (d-1) \,,\qquad 
\beta_{6;2;(21)} = 2\,.
\eeq
Thus, we still have $\lambda^{\ren}_{6;2}>\lambda_{6;2}$ and, interestingly, this sector is asymptotically free for both models $_1\Phi^{6}_{3,4}$. Note also that the splitting of the $\beta$-function
occurs in other (condensed matter) contexts \cite{Ftrub}.

The behavior of the coupling constant 
$\lambda_{6;1}$ needs still to be investigated. Four-loop calculations are required in this
case. Note that it has been already established that the coupling constant
$\lambda_{6;2}$ tends to 0 in the UV. In order, to determine
the behavior of the coupling constant $\lambda_{6;1}$, 
we can assume that we far enough in the UV such that $\lambda_{6;2}\sim 0$. Under such circumstances, we do not need to involve 
vertices of the form $V_{6;2}$. 

We use Lemma 3 in \cite{BenGeloun:2012yk} and find at four loops the wave function renormalization and $\Gamma_{6;1;\rho}(0,\dots,0)$ function as
\bea
Z = 1 - 2 \lambda_{6;1;1} S^1 +  2 \lambda_{6;1;1}^2  \Big[2 \cS^{1}_{(1)} +3\cS^{1}_{(2)} \Big]
 + 2 \lambda_{6;1;1}\Big(\sum_{\rho \in \{[2],[3]\}}
\lambda_{6;1;\rho}\Big)
\Big[2\cS^{12}_{(1)}+\cS^{12}_{(2)}\Big]  +O(\lambda^3)
\label{wphi61}
\eea
and the sum of truncated amputated six-point functions at four loops 
satisfies, for any $\rho\in \{[1],[2],[3]\}$,
\bea
\Gamma_{6;1;\rho}(0,\dots,0) &=& - \lambda_{6;1;\rho} + 2\cdot 3\, \lambda_{6;1;\rho}^2 S^1
- 2\cdot 3\cdot 5\,\lambda_{6;1;\rho}^3 \,\cS^{1}_{(2)} 
 - 2\cdot 3\,\lambda_{6;1;\rho}^2
\Big[\sum_{\rho' \in \{[1],[2],[3]\}\setminus \{\rho\}}\lambda_{6;1;\rho'}\Big]  \cS^{12}_{(2)} \crcr
&& - 2^2\cdot 5\,\lambda_{6;1;\rho}^3 \,\cS^{1}_{(1)}
 - 2^2\cdot 3\,\lambda_{6;1;\rho}^2
\Big[\sum_{\rho' \in \{[1],[2],[3]\}\setminus \{\rho\}}\lambda_{6;1;\rho'}\Big]  \cS^{12}_{(1)} 
+O(\lambda^4)
\label{gam61pur}
\eea
where $O(\lambda^4)$ stands for a function involving a quartic number of couplings
and where
\bea
\cS^{1}_{(1)}  &:=&  \sum_{p_s,p'_{s},p''_{s},p'''_{s}} \Big[ 
\frac{1}{(p_{1}^{2a}+ p_{\check{1}}^{2a}+\mu^2)^3}
\frac{1}{(p'^{2a}_{1}+ p'^{2a}_{\check{1}}+\mu^2)}
\frac{1}{(p''^{2a}_{1}+ p''^{2a}_{\check{1}}+\mu^2)} 
\frac{1}{(p'''^{2a}_{1}+ p'''^{2a}_{\check{1}}+\mu^2)} \Big]  
\crcr
\cS^{1}_{(2)}  &:=& \sum_{p_s,p'_{s},p''_{s},p'''_{s}} \Big[ 
\frac{1}{(p_{1}^{2a}+ p_{\check{1}}^{2a}+\mu^2)^2}
\frac{1}{(p'^{2a}_{1}+ p'^{2a}_{\check{1}}+\mu^2)^2}
\frac{1}{(p''^{2a}_{1}+ p''^{2a}_{\check{1}}+\mu^2)} 
\frac{1}{(p'''^{2a}_{1}+ p'''^{2a}_{\check{1}}+\mu^2)} \Big]  \crcr
\cS^{12}_{(1)}  &:=&  \sum_{p_s,p'_{s},p''_{s},p'''_{s}} \Big[ 
\frac{1}{(p_{1}^{2a}+ p_{\check{1}}^{2a}+\mu^2)^3}
\frac{1}{(p_{1}^{2a}+ p'^{2a}_{1}+ p'^{2a}_{\check{1}}+\mu^2)}
\frac{1}{(p_{1}^{2a}+ p''^{2a}_{1}+ p''^{2a}_{\check{1}}+\mu^2)}\times  \crcr
&& 
\frac{1}{(p'''^{2a}_{1}+ p'''^{2a}_{\check{1}}+\mu^2)} \Big]  \crcr
\cS^{12}_{(2)}  &:=&  \sum_{p_s,p'_{s},p''_{s},p'''_{s}} \Big[ 
\frac{1}{(p_{1}^{2a}+ p_{\check{1}}^{2a}+\mu^2)^2}
\frac{1}{(p_{1}^{2a}+ p'^{2a}_{1}+ p'^{2a}_{\check{1}}+\mu^2)}
\frac{1}{(p_{1}^{2a}+ p''^{2a}_{1}+ p''^{2a}_{\check{1}}+\mu^2)} \times\crcr
&&
\frac{1}{(p'''^{2a}_{1}+ p'''^{2a}_{\check{1}}+\mu^2)^2} \Big] .
\eea
After identifying all couplings, the renormalized coupling equation 
at four loops becomes
\bea
\lambda^{\ren}_{6;1} = - \frac{\Gamma_{6;1}(\{0\})}{Z^3} 
=  \lambda_{6;1} 
+ 8\lambda_{6;1}^3\cS^{1}_{(1)}  +O(\lambda^4) \,,
\eea
where we used $(S^1)^2 = \cS^{1}_{(2)}$. Hence,
the $\beta$-function at this order of perturbation reads 
\bea
\beta_{6;1} = 8\,.
\eea
Thus the models are asymptotically free in the UV. 
Similar remarks as in the previous section about the meaning
of such free theory hold in the present situation as well. 
It is also remarkable that, for the $\Phi^6$ tensor models, 
the maximally divergent graphs with $N_{\ext}=4$
are graphs without $\Phi^4$ vertices
($V_{4}=0=V_{4;{\rm a}}= V_{2}$) but only with $\Phi^6$
interaction terms. This immediately implies that at the
UV limit, since both $\lambda^{\Lambda}_{6;1}$ and
$\lambda^{\Lambda}_{6;2}$  are vanishing, then 
the renormalized coupling equation for the $\lambda^{\ren}_{4;1}$
reads 
\bea
\lambda^{\ren}_{4} = \lambda_{4} \,,
\eea
which means that this sector is always safe at all loops. 
The last sector $\lambda_{4;{\rm a}}$ is slightly more subtle as it turns out to be disconnected and can generate divergent amplitudes with only $V_{4;{\rm a}}$ vertices \cite{BenGeloun:2012yk}. In all situation, it means that we have for both models a UV fixed manifold
determined by 
\bea
\lambda^{UV}_{6;1} = 0 =\lambda^{UV}_{6;2}\,, 
\qquad
\lambda^{UV}_{4} = k\, ,\qquad 
\lambda^{UV}_{4;{\rm a}} =0 \,,
\eea
for some arbitrary $k$. Adding small perturbations around
this line implies that  the coupling constants $\lambda^{\ren}_{6;1}$
and $\lambda^{UV}_{6;2}$  grow in the IR.

\subsection{First order $\beta$-functions of matrix models $\Phi^{4,6}_2$}
\label{subsect:matbeta}

We now turn our attention to the renormalizable matrix models and their $\beta$-function at small number of loops.  
Consider then the $_{\dim G_D}\Phi^{4,6}_2$ models with their list of
all divergent graphs. We will focus on the main 
interactions with maximal valence $k_{\max}=4$ and 6. 

\subsubsection{One-loop $\beta$-function of $\Phi^4_2$ models}
\label{subsub:matphi4}

We discuss here the $\Phi^4$ models such that 
\beq
(\,  _1 \Phi^{4}_2\,, a=\frac{1}{4}) \,, \quad (\,   _2 \Phi^{4}_2 , a=\frac{1}{2})\,, \quad
(\,  _3 \Phi^{4}_2\,, a=\frac34) \,, \quad (\,   _4 \Phi^{4}_2 , a=1)
\label{matrixmodbeta}
\eeq
with ribbon-like propagator and 4-valent vertex represented as in Figure \ref{fig:ribbon} and aim at computing the renormalized coupling equation
\bea
\lambda^{\ren}_{4} = -\frac{\Gamma_4(\{0\})}{Z^2}\,. 
\eea
We will establish that the $\beta$-functions at one-loop of the
models \eqref{matrixmodbeta} are all vanishing. This is strongly related 
with the same property of the GW model which holds at all orders. 

Let us introduce the formal sum $S^{1}$ with now a different content 
given by 
\bea\label{s1mat}
S^1 = \sum_{P} \frac{d_f(P)}{(P^{\,a} + \mu^2)^2}\,, 
\eea
where according to the group nature, $d_f(P)$ is defined such that:

- if $G_D=U(1)^D$ then $d_f(P)=1$;

- if $G=SU(2)$ then $d_f(P)=2P+1$, $P\in \frac12\N$ (this can happen only for the
model $(\,  _3 \Phi^{4}_2\,, a=\frac34)$ for the choice $G_D=SU(2)$).

\begin{figure}[t]
 \centering
    \begin{minipage}[t]{.8\textwidth}
\includegraphics[angle=0, width=7cm, height=1.5cm]{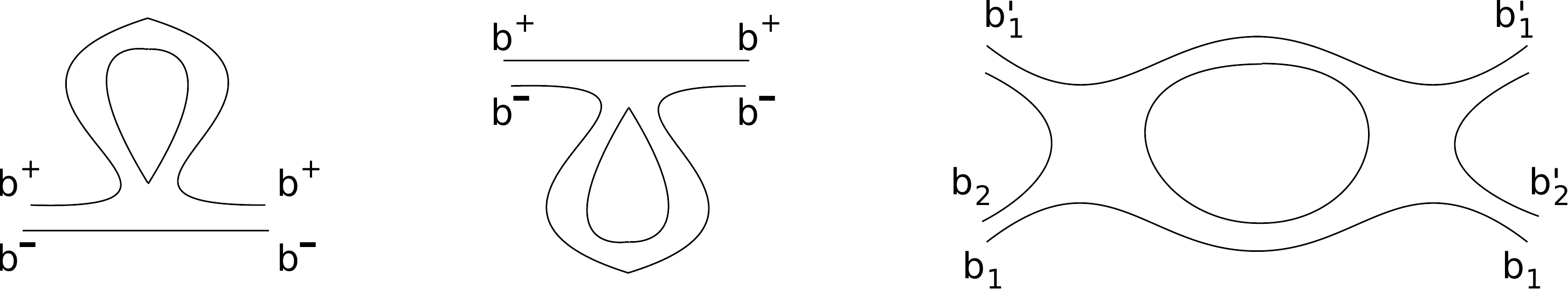}
\vspace{0.3cm}
\caption{ {\small Tadpole graphs $T^\pm$ and a 4-point graph $G_4$.}} \label{fig:beta4mat}
\end{minipage}
\put(-290,-12){$T^+$}
\put(-228,-12){$T^-$}
\put(-148,-12){$G_4$}
\end{figure}
\medskip 

Evaluating the self-energy at one-loop, the tadpole graphs in Figure \ref{fig:beta4mat},
should contribute. These are generally referred to as the tadpole up (+) and down (-). 
We have at one-loop, for external momenta $b_{\epsilon=\pm}$,
\bea
\Sigma(b_+,b_-) =\sum_{\epsilon=\pm} A_{T^\epsilon}(b_\epsilon)\,,
\qquad 
A_{T^\epsilon}(b_{\epsilon}) = \frac{-\lambda_4}{2} K_{T^{\epsilon}} 
\tilde S^{1}(b_{\epsilon})\,,\qquad 
\tilde S^{1}(b) := \sum_{P} \frac{d_f(P)}{b^{a} + P^a + \mu^2}\,,
\eea
with $K_{T^{\epsilon}}=2$. One therefore infers the wave function renormalization   as
\bea
Z= 1- \partial_{b_+^{a}}\Sigma|_{b_\epsilon=0}
 = 1 - \lambda_4 S^1 + O(\lambda_4^2)\,.
\eea
where $S^1$ is given by \eqref{s1mat}. Next, we focus on 1PI amputated 4-point functions and evaluate $\Gamma_4$ at low
external momenta. 
There is a single planar connected graph with one connected component
of the boundary, it is given by $G_4$ in Figure \ref{fig:beta4mat}. Computing the amplitude
associated with this graph and inserting the result in $\Gamma_4$ yields 
\bea
\Gamma_4(\{0\}) = -\lambda_4 + \frac{1}{2!}\big(\frac{-\lambda_4}{2}\big)^2
K_{G_4} S^1+ O(\lambda_4^3) = -\lambda_4 + 2\lambda_4^2 S^1 + O(\lambda_4^3)\,, 
\eea
where we use the fact that the combinatorial factor associated with that graph is $K_{G_4}=2^4$. The renormalized coupling constant equation is straightforward
and given by 
\bea
\lambda^{\ren}_4 = \lambda_4\,.
\eea
Thus, the $\beta$-function is vanishing at one-loop
\bea
\beta_4 =0 \,.
\eea
In fact,  we can see that the matrix models \eqref{matrixmodbeta}
reproduce the same features as the complex GW model. 
At small number of loops, in the derivations of the wave function renormalization, the graph amplitude which may very well vary from
one model to the other, keeps at least its overall form. A similar fact happens in the calculation of the $4$-point functions. All graphs which should be involved in calculation of $\beta$-function of the GW model should appear in the $\beta$-function of the present class of models with the same combinatorial factor. 

We must emphasize that this vanishing $\beta$-function should be strongly 
correlated with a recent breakthrough \cite{Grosse:2012uv},
that it is worth to quickly review. Consider real matrices $M_{ab}$, where $a$ and $b$ belong to a set $I$ of discrete indices (though the following is valid for 
continuous indices, we will only discuss the discrete case). We can define a product
on these $(MN)_{ab}=\sum_{c\in I}\mu_c M_{ac}N_{cb}$, for a constant weight $\mu_c$. We can also introduce a trace ${\rm tr}M=\sum_{a}\mu_a M_{aa}$ 
with a quartic interaction of the form $S =V{\rm tr}( MEM + (M)^4)$,
with $V$ a volume factor, $E$ kinetic term which is not the identity operator but an unbounded self-adjoint operator on an Hilbert space with compact resolvent
so that $MEM$ is traceclass. Hence, one must restrict the set of matrices $M$. Theorem 3.2 of \cite{Grosse:2012uv} states that the model defined by $S$ has a vanishing $\beta$-function. This result holds at the non perturbative level. 

The model $(_1\Phi^4_2, a = \frac14)$
fits in the above category of models for a suitable set of matrices $\varphi_{mn}$. The fact that $\beta_4=0$ at all orders for this case is a simple corollary of Theorem 3.2 of \cite{Grosse:2012uv}. For the other models written in \eqref{matrixmodbeta}, the group
dimension is greater than 2, and the fields $\varphi$ are not really matrices
but implicitly tensors (see Remark \ref{rem}). But we have also seen that
the GW model in 4D may be written in terms of tensors, so this might not
be a great issue for applying that theorem to the rest of these models. 
However, using the representation of the group $SU(2)$,
we obtain face amplitude contributions of the form $d_f(P)$ which modifies the overall amplitude of any $N$-point function. One must carefully check if these features
may or not affect that theorem. 

In any case, we conjecture that the models have all a vanishing $\beta$-function at all orders or perturbation. At the perturbative level, this could be achieved using the same techniques developed in \cite{Disertori:2006nq}\cite{Geloun:2008zk}.  If this statement is true, all these models might be asymptotically safe which means that they have a non trivial fixed point in the UV and Theorem 3.2 of \cite{Grosse:2012uv} would be valid on a larger domain from matrices to tensors.

\subsubsection{Two-loop $\beta$-function of the $\Phi^6_2$ models}
\label{subsub:mbeta6}

\begin{figure}[h]
 \centering
    \begin{minipage}[t]{.8\textwidth}
\includegraphics[angle=0, width=5.5cm, height=1.5cm]{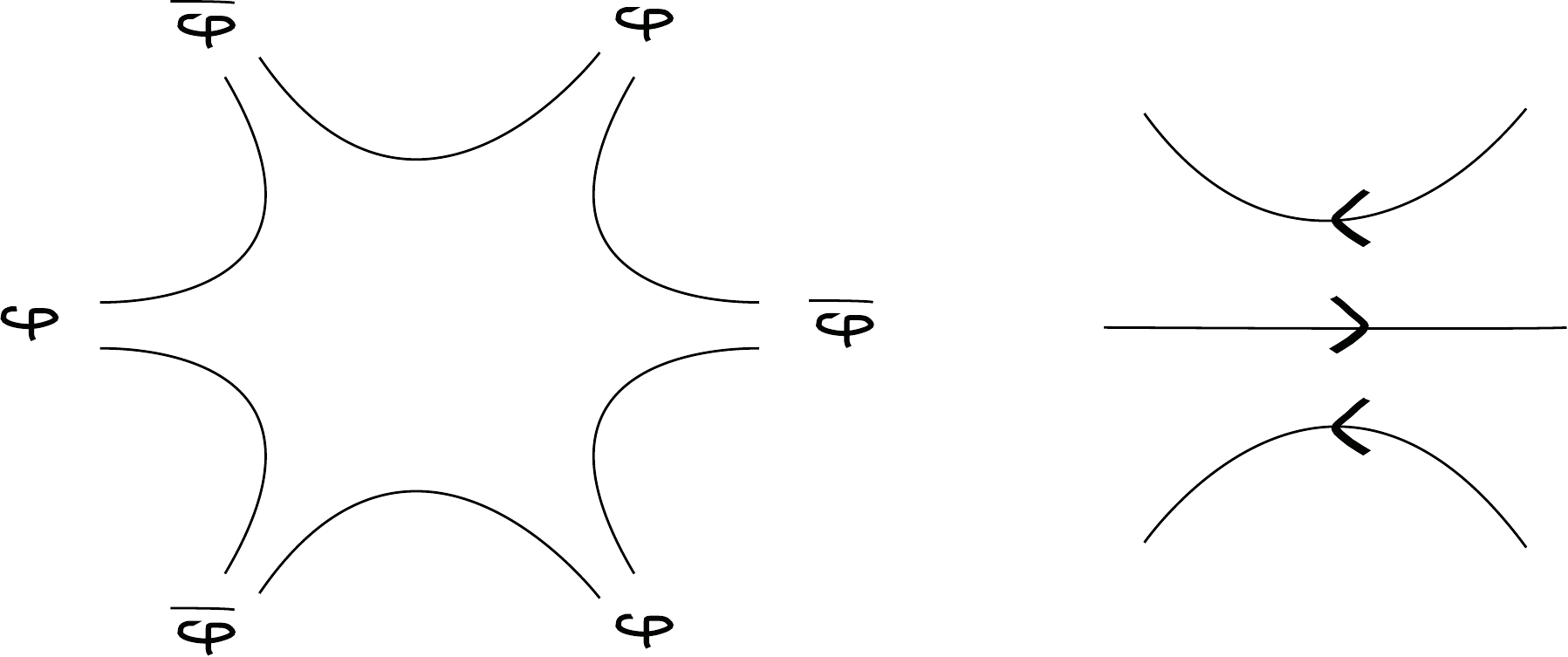}
\caption{ {\small The $\Phi^6$ vertex and its simplified representation.}} \label{fig:spimphi6}
\end{minipage}
\end{figure}

We are now interested in the UV behavior of the models
\beq
(\,  _1 \Phi^{6}_2\,, a=\frac{1}{3}) \,, \quad (\,   _2 \Phi^{6}_2 , a=\frac{2}{3})\,, \quad
(\,  _3 \Phi^{6}_2\,, a=1)\,.
\label{matrixmodbeta6}
\eeq
The vertices of these models are 6-valent and we will use instead a 
simplified representation for these as given in Figure \ref{fig:spimphi6}. 
One must pay attention to the fact that, although this simplified vertex notation
does not seem to be cyclic, there is no way to distinguish 
pairs of external lines obtained from one another after a cyclic permutation.
In fact, this simplified notation is not canonical in the sense that it can be found (easily) a distinct simplified graph encoding the same vertex. However, introducing this notation will be enough for capturing the essential properties that we want to discuss.    
\medskip 

Our goal is to compute at two loops the renormalized coupling equation
\bea
\lambda^{\ren}_{6} = -\frac{\Gamma_6(\{0\})}{Z^3}\,. 
\eea
We will use also the same anterior compact notation for the momentum $P^a$ and 
denote $S^1$ and $S^{12}$ now as the formal sums 
\beq\label{sum6mat}
S^1 = \sum_{P_1,P_2} \frac{d_f(P_1)d_f(P_2)}{(P_1^{\,a} 
+ \mu^2)^2(P_2^{\,a} + \mu^2)}\,,
\qquad 
S^{12} = \sum_{P_1,P_2} \frac{d_f(P_1)d_f(P_2)}{(P_1^{\,a} + \mu^2)^2(P_1^{a} + P_2^{\,a} + \mu^2)}\,,
\eeq
where as usual $d_f(P_s)$ depends on the group manifold. 

\begin{figure}[h]
 \centering
    \begin{minipage}[t]{.8\textwidth}
\includegraphics[angle=0, width=5.5cm, height=1.5cm]{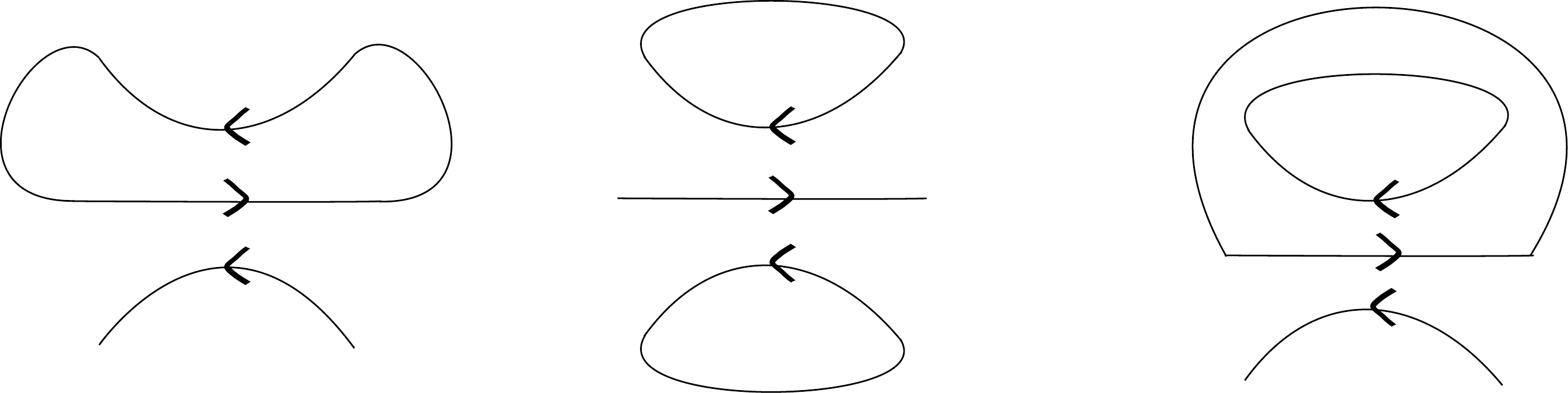}
\vspace{0.3cm}
\caption{ {\small Tadpoles graphs.}} \label{fig:tadmat6}
\end{minipage}
\put(-265,-10){$T^+_1$}
\put(-210,-10){$T_2$}
\put(-150,-10){$T_3^+$}
\end{figure}

At two loops, the wave function renormalization is evaluated 
from the self-energy which includes the 
amplitudes associated with the tadpole graphs
$\{T_1^{\pm}, T_2, T_{3}^{\pm}\}$; $T_1^+, T_2$
and $T_3^+$ appear in Figure \ref{fig:tadmat6}, 
and $T_1^-$ and $T^-_3$ are obtained either by flipping (top-down) the graphs
$T_1^+$ and $T^+_3$, respectively, or by conserving the same graphs and switching the orientation of the arrows.

The self-energy $\Sigma(b_+,b_-)$ splits in two sums: one including 
the external variable $b_+$ which is $\Sigma^0(b_+,b_-)$ and a remainder. 
We have 
\beq
\Sigma^0(b_+,b_-) = A_{T_1^+} (b_+,b_-)+ A_{T_2} (b_+,b_-)
+A_{T_3^+} (b_+,b_-) \,,
\eeq
where $A_{\cG}$ is the graph amplitude associated with 
the graph $\cG$. By direct evaluation, using the so far routine,  
we arrive at 
\bea
Z = 1- \partial_{b^a_1}\Sigma^0|_{b_s=0} 
 = 1 -\lambda_6 (3S^1 + S^{12}) + O(\lambda^2)\,, 
\eea
where $S^1$ and $S^{12}$ are given by \eqref{sum6mat}. 

1PI amputated 6-point functions are once again of the rough form  given by Figure \ref{fig:Gener6}.
\begin{figure}[h]
     \begin{minipage}[t]{.8\textwidth}
\includegraphics[angle=0, width=10cm, height=1.5cm]{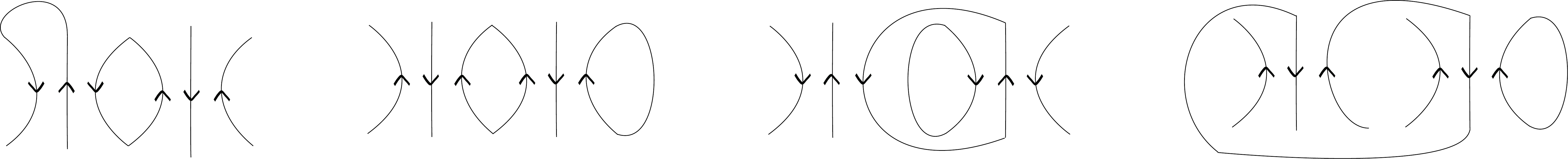}
\vspace{0.3cm}
\caption{ {\small 1PI 6-point graphs contributing to $\Gamma_6$.}} \label{fig:Gamat6}
\end{minipage}
\put(-328,-12){$F^+$}
\put(-260,-12){$G^+$}
\put(-188,-12){$I_1^+$}
\put(-105,-12){$I_2^+$}
\end{figure}
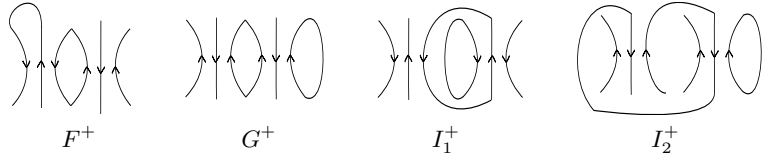
The amplitudes contributing to $\Gamma_6$ are associated to the graphs $\{F^\pm, G^\pm, I_{1,2}^{\pm}\}$ of the form  are listed in Figure \ref{fig:Gamat6} (note that $F^-, G^-$ and  $I_{1,2}^{-}$ are obtained by reversing the orientations of their (+)-partner). At this order of perturbation, the amplitude of any graph $\cG$ at low external momenta is
\beq
A_{\cG}(\{0\}) = \frac{1}{2!} \left(\frac{-\lambda_6}{3}\right)^{2} K_{\cG}
\widetilde{S}_{\cG}(\{0\}) \,,
\eeq
where $\widetilde{S}_{\cG}(\{b\})$ stands for a formal sum and
$K_{\cG}$ is the combinatorial coefficient associated with  each of the significant 
graphs. It can be shown that 
\beq
K_{F^\pm} = 3^3 \cdot 2^2 \,, \;\; \widetilde{S}_{F^\pm}(\{0\})=S^1\,;\qquad
K_{G^\pm} = 3^3 \cdot 2 \,, \;\; \widetilde{S}_{G^\pm}(\{0\})=S^1\,;\qquad
K_{I_{1,2}^\pm} = 3^3 \cdot 2 \,, \;\; \widetilde{S}_{G^\pm}(\{0\})=S^{12}\,,
\eeq
where $S^1$ and $S^{12}$ are still found in \eqref{sum6mat}. A  straightforward evaluation yields
 \beq
\Gamma_6(\{0\}) = -\lambda_6+\sum_{\cG\in \{F^\pm, G^\pm, I_{1,2}^{\pm}\}} A_{\cG}(\{0\}) + O(\lambda^3)
=  -\lambda_6 + 6\lambda_6^2 (3 S^1 + 2 S^{12})+ O(\lambda^3)\,. 
\eeq
We are in position to evaluate  the first order renormalized coupling equation.
One has
\bea
\lambda_{6}^\ren 
= \lambda_6  - 9 \lambda_6^2 (S^1 + S^{12})  +  O(\lambda^3)\,.
\label{lam6mat}
 \eea
The two-loop $\beta$-function of this coupling constant is 
\bea
\beta_6 = -9\,. 
\eea
Clearly, from \eqref{lam6mat} and considering positive coupling constant $\lambda_{6}> 0$,
the models possess a Landau ghost. In other words, the coupling constant
blows in the UV. This matrix models have the same behavior as the
ordinary scalar $\Phi^4$ theory  in 4D. 

\section{Concluding remarks}
\label{concl}

We have investigated the renormalization analysis of field theories
defined with rank $d\geq 2$ tensors defined on $G_D \in \{U(1)^D, SU(2)^D\}$. The actions of the models considered are defined with a general kinetic term written in momentum space involving propagator of the form $\frac{1}{p^{2a}+ \mu^2}$, where $p$ is the eigenvalue of the Laplacian operator 
acting on the group background $G_D$, $a$ a parameter
 free to take any value in $(0,1]$ and $\mu$ is a mass. The limiting case $a=1$ yields the ordinary 
Laplacian dynamics. Our fields are simply random tensors not subjected
to any condition but integrability. This is in contrast with another
type of TGFTs enforcing the so-called gauge invariance  on tensors \cite{oriti}
that we did not consider in this work. 
Within the present framework, we find that there are several just-renormalizable 
models in any rank. In particular, under the above
conditions,  we successfully prove that
 
(A) For the rank $d\geq 3$ case:

- there are 6 just-renormalizable models with rank $d\geq 3$,
$_1 \Phi^6_{3,4} \,, \, _2\Phi^4_3 \,, \, _1 \Phi^4_{3,4,5}$, 
the maximal valence of the vertex is 6; 

- there is no just-renormalizable tensor model with rank $d \geq 6$,

- there is no just-renormalizable tensor model defined with a group dimension
$\dim G_D \geq 3$; in particular there is no just-renormalizable model 
defined on $G=SU(2)$,

- there is a tower of  tensor models in rank $d=3$ with 
group $G=U(1)$ which can be potentially just-renormalizable;
a model in the tower is determined by the maximal valence $k_{\max} \geq 4$ of its vertices,

- all proved just-renormalizable models are so far asymptotically free in the UV,

- the tower $(_1\Phi^k_3, a=1)$, for all $k\geq 4$, of defines super-renormalizable tensor models;

(B) For the $d=2$ or matrix models case:

- there are 6 plus two towers of just-renormalizable models, 

- there is no just-renormalizable model defined on a group with dimension
$\dim G_D \geq 7$,

- all $\Phi^4$ models have a vanishing $\beta$-function at one-loop
which is strongly reminiscent of the vanishing $\beta$-function at all 
orders of the GW model in 4D; we conjecture
that, indeed, these models are asymptotically safe at all loops in the UV,

- all $\Phi^6$ models have a Landau pole in the UV, 

- the tower $(_{\dim G_D}\Phi^k_2, a)$, with 
$(\dim G_D,a)\in \{(1,\frac12), (2,1)\}$, for all $k\geq 4$, of defines super-renormalizable matrix models. 

\medskip 

We can update Table \ref{table:mod} as

\begin{table}[h]
\begin{center}
\begin{tabular}{lccccccccccc|cc|}
\hline
TGFT (type)&& $G_D$ && $\Phi^{k_{\max}}$  &&  $d$ && $a$ && Renormalizability &&   UV behavior \\
\hline
 && $U(1)$ && $\Phi^{4}$ && 4 && 1 && Just- &&  AF \\
 && $U(1)$ && $\Phi^{3}$  && 3 && $\frac12$ && Just- && AF\\
&& $U(1)$ && $\Phi^{6}$ && 3 &&  $\frac23$ && Just- &&  AF \\
&& $U(1)$ && $\Phi^{4}$ && 4 && $\frac34$ && Just- &&  AF \\
&& $U(1)$ && $\Phi^{4}$ && 5 &&  1 && Just- &&  AF \\
&& $U(1)^2$ && $\Phi^{4}$ && 4 && 1 && Just- &&  AF \\
&& $U(1)$ && $\Phi^{2k}$ && 3 && 1 && Super- &&  - \\
 \hline 
 gi-&&  $U(1)$ && $\Phi^{4}$  && 6 && 1 && Just- &&  AF  \\
 gi-&&  $U(1)$ && $\Phi^{6}$  && 5 && 1 && Just- && AF \\
 gi- && $SU(2)^3$ && $\Phi^{6}$  && 3 &&  1 && Just- && ?\\
 gi-&& $U(1)$ && $\Phi^{2k}$  && 4 && 1 &&Super- &&  -\\
 gi-&&  $U(1)$ && $\Phi^{4}$  && 5 && 1 &&  Super- &&-\\
\hline
Matrix&& $U(1)$ && $\Phi^{2k}$ && 2 && $\frac12(1-\frac1k)$ && Just- &&    ($k=2$, AS$^{(\infty)}$); ($k=3$, LG)\\
Matrix&& $U(1)^2$ && $\Phi^{2k}$ && 2 && $1-\frac1k$ && Just- &&   ($k=2$, AS$^{(1)}$); ($k=3$, LG)\\
Matrix&& $U(1)^3$ or $SU(2)$ && $\Phi^{6}$ && 2 && 1 && Just- &&    LG\\
Matrix&& $U(1)^3$ or $SU(2)$ && $\Phi^{4}$ && 2 && $\frac34$ && Just- &&    AS$^{(1)}$\\
Matrix&& $U(1)^4$ && $\Phi^{4}$ && 2 && 1 && Just- &&   AS$^{(1)}$\\
Matrix&& $U(1)$ && $\Phi^{2k}$ && 2 && $\frac12$ && Super- &&   -\\
Matrix&& $U(1)^2$ && $\Phi^{2k}$ && 2 && 1 && Super-  &&   -\\
\end{tabular}\end{center}
\caption{Updated list of renormalizable models and their features
(AF $\equiv$ asymptotically free; LG $\equiv$ existence
of a Landau ghost; AS$^{(\ell)}$ asymptotically safe at
$\ell$-loops).}
\label{table:modfin}
\end{table}

The tower of rank $d=3$ models which might be just-renormalizable
addresses a new combinatoric issue which is the classification of all
melonic interactions of this rank according to some criteria. This problem can
be addressed in any rank $d\geq 3$ of course for its own combinatoric purpose. 
This deserves to be understood in order to complete the list
of just-renormalizable models in rank $d=3$ as well as to check 
whether or not asymptotic freedom is a genuine feature of tensor models
in rank $d\geq 3$.
Finally, the present investigation pertains to the ``discrete to continuum'' approach 
for quantum gravity. To that extent, one might scrutinize all UV asymptotically free theories issued from this work as a potential interesting candidates for describing
new degrees of freedom after a likely phase transition in the IR.

\section*{Acknowledgements}
Research at Perimeter Institute is supported by the Government of Canada through Industry
Canada and by the Province of Ontario through the Ministry of Research and Innovation.

\section*{ Appendix}
\label{app}

\appendix

\renewcommand{\theequation}{\Alph{section}.\arabic{equation}}
\setcounter{equation}{0}

\section{Face amplitude expansion and the Euler Maclaurin formula}
\label{app:euler}

We provide in this appendix further details on the face amplitude expansions \eqref{facea1} 
and \eqref{facea2} for arbitrary $a \in (0,1]$ and small $A \sim M^{-i_\ell}$, 
$A >0$.

 Let us first consider $h_n(x)=x^ne^{- A x^a}$, with $x \geq 0$, $n\in \N$, and the sum $\sum_{p=0}^{\infty}h(p)$. Using the Euler-Maclaurin formula, 
we have, for a finite integer $q\geq 1$,
\bea
\sum_{p =1}^q h(p) 
= \int_{1}^{q}h(p) dp  
 + R(q) \,,\qquad 
R(q) = - B_1(h(1)+  h(q) ) 
 + \sum_{k=1}^\infty \frac{B_{2k}}{(2k!)} (h^{(2k-1)}(q) -h^{(2k-1)}(1) )\,,
\eea
where $B_{k}$ are Bernoulli numbers. A rapid checking shows that
\bea
h'(x) = (x^ne^{- A x^a})' = x^{-1+n}(n- A a x^{a}) h(x)\,, \dots\,, \,
h^{(m)}(x) = x^{-m+n}\,( nF_n(m)+Aa G_{n,m}(a,Ax^a)) h(x)\,,
\eea
where $F_n(m)=\prod_{l=1}^{m-1}(n-l)$ and $G_{n,m}$ is polynomial in the variable $Ax^{a}$ so that the remainder $R$ is of the form
\bea
&&
R(q)=-B_1(e^{-A}+ q^ne^{- A q^a} ) \crcr
&&+  \sum_{k=1}^\infty \frac{B_{2k}}{(2k!)} 
[q^{-(2k-1)+n}\,( nF_n(2k-1)+Aa G_{n,2k-1}(a,Ax^a)) h(q)
 - ( nF_n(2k-1)+Aa G_{n,2k-1}(a,A)) h(1)]\,.
\eea
At the limit $q\to \infty$, $h^{(n)}(q)$ is clearly exponentially suppressed by presence of $h(q)\to 0$, we obtain
\bea
\lim_{q\to \infty} R(q) &=& -B_1(1+ O(A)) 
  - \sum_{k=1}^\infty \frac{B_{2k}}{(2k!)} 
( nF_n(2k-1)+Aa G_{n,2k-1}(a,A))h(1)  \crcr
&& -B_1(1+ O(A)) 
  - \sum_{k=1}^\infty \frac{B_{2k}}{(2k)!} ((2k-1)!\big(^n_{2k-1}\big)+Aa G_{n,2k-1}(a,A))h(1)
= -B_1 -\sum_{k=1}^\infty \frac{B_{2k}}{2k}\big(^n_{2k-1}\big) + O(A)\,.\crcr
&&
\eea
On the other hand, for any $A$, the following integral is exact:
\beq
\lim_{q\to \infty} \int_{1}^{q}h_{n}(p) dp   = \frac1a  \,\Gamma[\frac{1+n}{a},A] \,A^{-\frac{1+n}{a}}
= \frac{1}{a}A^{-\frac{1+n}{a}}\Gamma[\frac{1+n}{a}] -\frac{1}{1+n} + O(A)\,,
\eeq
where $\Gamma[\cdot,\cdot]$ denotes the incomplete Gamma function
and $\Gamma[\cdot]$ stands for the Euler gamma function. Finally, one 
obtains
\bea
\sum_{p=1}^\infty h_n(p) = \lim_{q\to \infty}\sum_{p =1}^q h_n(p) =
\frac{1}{a}A^{-\frac{1+n}{a}}\Gamma[\frac{1+n}{a}] -\frac{1}{1+n}   - \tilde B_n+  O(A) = c_{a,n} A^{-\frac{1+n}{a}}(1+ O(A^{\frac{1+n}{a}}))\,,
\label{facfin}
\eea
with some constant $c_{a,n}=\Gamma[(1+n)/a]/a $.

We are now in position to specifically address \eqref{facea1} 
and \eqref{facea2}. Equation \eqref{facfin} implies at $n=0$ the following relation
\bea
\sum_{p=0}^\infty h_0(p) = 1+ \sum_{p=1}^\infty e^{-A p^a} = 
\frac{1}{a}A^{-\frac{1}{a}}\Gamma[\frac{1}{a}] - \tilde B_0 +  O(A) = c_{a,0} A^{-\frac{1}{a}}(1+ O(A^{\frac{1}{a}}))\,,
\eea
which implies \eqref{facea1}. 

Second, consider the following sum $\sum_{p \in \frac12 N} (2p+1)^2 e^{-A' p^a}$ in relation with \eqref{facea2} and that expands as:
\bea
\sum_{p \in \frac12 N} (2p+1)^2 e^{-A' p^a} =  \sum_{p=1}^{\infty} p^2  e^{-A p^a} + 2  \sum_{p =1}^{\infty} p  e^{-A p^a}  
+\sum_{p=0}^{\infty} e^{-A p^a}\,,
\label{jface}
\eea
where $A=A'/2^a$.  For each resulting sum, we use \eqref{facfin} at $n=2$, $n=1$ and $n=0$, respectively, and get
\bea
\sum_{p \in \frac12 N} (2p+1)^2 e^{-A' p^a} &=& (\frac{1}{a}A^{-\frac{3}{a}}\Gamma[\frac{3}{a}] -\frac{1}{3}   - \tilde B_2) + 2 (\frac{1}{a}A^{-\frac{2}{a}}\Gamma[\frac{2}{a}] -\frac{1}{2}   - \tilde B_1)
+ \frac{1}{a}A^{-\frac1a}\Gamma[\frac{1}{a}]  - \tilde B_0 + O(A) \crcr
&=& c_{a,3}A^{-\frac{3}{a}} (1+ O(A^{\frac1a}))\,,
\label{jface2}
\eea
which implies \eqref{facea2}.

\section{On potential renormalizable real matrix models}
\label{app:real}

We make in this section additional comments on real matrix models which were not analyzed in the Section \ref{sect:matrix}. Due to the occurrence of an odd valence of interactions, these models could be defined via 
real matrix fields. These are
\bea
&& 
(\,  _1 \Phi^{2+\gamma>2}_2\,, a=\frac{\gamma}{2(2+\gamma)}\leq \frac12) \,, \quad (\,   _2 \Phi^{2+\gamma>4}_2 , a=\frac{\gamma}{2+\gamma}\geq \frac12)\,,\quad 
(\,   _2 \Phi^{3}_2 , a=\frac13) \,,
\crcr
&&
(\,  _3 \Phi^{3,5}_2\,, a=\frac12,\frac{9}{10}) 
\,, \quad (\,   _4 \Phi^{3}_2 , a=\frac23)\,,
 \quad (\,   _5 \Phi^{3}_2 , a=\frac56)\,,
 \quad (\,   _6 \Phi^{3}_2 , a=1)\,,
\label{Rmatrixmod}
\eea
where $2+\gamma$ should be an odd integer.

The interaction for these models are of the form 
\beq
 S^{\inter}_k= \sum_{P_{[I]}} {\rm tr} \Big[ (\varphi_{[I]})^{k}\Big]= \sum_{P_{[I]}} 
\varphi_{12}\,\varphi_{1'2}\,\varphi_{1'2'}\,\varphi_{1''2'}\dots
\varphi_{1'''2'''}\,\varphi_{12'''} \,.
\eeq 
where, this time, we now allow $k$ to take odd integer values greater than 2. Given a real matrix model $_{\dim G_D}\Phi^{k_{\max}}_2$, a cut-off 
$\Lambda$ in momentum space, the total interaction may be written 
\beq
S^{\Lambda} = \sum_{k=3}^{k_{\max}}  \frac{\lambda_k^{\Lambda}}{k} S^{\inter}_k  + CT^{\Lambda}_{2;1} + CT^{\Lambda}_{2;2} S_{2;2} \,.
\eeq
Note that $S^{\Lambda}$ includes even and odd valence interaction terms. 
 
The next stage is to list all primitively divergent graph. For this purpose, 
we adopt the same method of Section \ref{sect:matrix} and write the divergence degree of a connected graph $\cG$, with $N_{\ext} \geq 1$ 
external leg(s), $C_{\bG}\geq 1$ and $V_2=V_{2;1}$ number of mass vertices as
\bea
 \omega_d(\cG) &=& - 2\dim G_D g_{\tilde\cG} - P_a(\cG) \,,
\crcr
P_a(\cG) &=&
\dim G_D(C_\bG -1) 
+  \frac12\Big[(\dim G_D -2a) N_{\ext} - 2\dim G_D\Big] +\frac12\sum_{k=2}^{k_{\max}-1}\Big[ 2\dim G_D+ (2a-\dim G_D)k\Big]  V_k \,, \nonumber 
\eea
where the sum $\sum_{k=2}^{k_{\max}-1}$ is performed over
even and odd integers. This is in contrast with complex case where only 
even integers were considered in this sum. 

In the same vein, $N_{\ext} > k_{\max}$ will give $ \omega_d(\cG)<0$
and $N_{\ext} = k_{\max}$ will give $ \omega_d(\cG)=0$ 
if and only if $g_{\tilde\cG}=0$, $C_{\bG}= 1$,  
$V_k=0$, for all $k$.

- If $N_{\ext} = k_{\max}-q$, where $1\leq q \leq k_{\max}-2$, 
one gets
\bea
P_a(\cG) &=&
\dim G_D(C_\bG -1) 
+  \frac12\Big[(\dim G_D -2a) ( k_{\max}-q) - 2\dim G_D\Big] +\frac12\sum_{k=2}^{k_{\max}-1}\Big[ 2\dim G_D+ (2a-\dim G_D)k\Big]  V_k \,, \crcr
&=& 
\dim G_D(C_\bG -1) 
-  \frac12(\dim G_D -2a)\left(q -\sum_{k=1}^{k_{\max}-2}k V_{k_{\max}-k} \right)\,.
\eea
Using the same arguments as in Subsection \ref{subsect:just2}, 
we must have $C_{\bG}=1$ and $g_{\tilde \cG}=0$ in all cases. 
Then the analysis of divergent graphs with $ \omega_d(\cG) =
-P_{a}(\cG)  \geq 0$ can be also recast in terms of partition of $q$ and $q-q_1$ for an integer $q_1 \leq q$. We clearly see that
the number of primitively divergent configurations can be listed
according to these partitions. 

- If $N_{\ext} = 1$, 
\bea
P_a(\cG) &=&
\dim G_D(C_\bG -1) 
-  \frac12(\dim G_D -2a)\left(k_{\max}-1 -\sum_{k=1}^{k_{\max}-2}k V_{k_{\max}-k} \right)\,.
\eea
Then, it may exist some configurations such that 
$V_{k_{\max}-k}=0$ for all $k=1,\dots, k_{\max}-2$. 
Having $N_{\ext} = 1$ then $C_\bG =1$ so that 
\beq
\omega_d(\cG) = - 2\dim G_D g_{\tilde\cG} +  \frac12(\dim G_D -2a)(k_{\max}-1) 
\label{om1pt}
\eeq
Therefore it may exist 1-point function configurations which 
are divergent. Indeed, take the planar tadpole of the $(_1\Phi^{3}_2, a= \frac{1}{6})$ model. It diverges as $\Lambda^{\frac{2}{3}}$. Thus 
\eqref{om1pt} could generate a new type of anomalous terms
of the vector form. An non-invariant interaction vertex which could be introduced is of the form $S^{\inter}_1 ={\rm tr}(\varphi)=\sum_{a}\varphi_{aa}$. Another way to proceed is 
to combine both vector and matrix fields in the initial action. 
One then has to consider this situation with all the care needed
by performing the multi-scale analysis from the beginning
for this new class of mixed rank models ((vector+matrix)-models). 

\section{Primitively divergent graphs for  the $(_{\dim G_D}\Phi^{8}_{2},a)$ model.}
\label{app:diverg}

We now provide a complete application of the method of finding
primitively divergent graphs for the nontrivial order $k_{\max}=8$
in the matrix model $(_{\dim G_D}\Phi^{8}_{2},a)$.

It can be simply proved that $N_{\ext}>8$ yields a convergent amplitude. 
$N_{\ext}=8$ leads to a log--divergent amplitude 
if and only if $g_{\tilde \cG}=0$, and $V_{6}=V_4=V_{2;1}=0$
and $C_{\bG}=1$.

- For $N_{\ext}=6$, we can write
\bea
P_{a}(\cG) &=& \dim G_D(C_\bG -1) 
+  \frac12\Big[ 2\dim G_D+(2a-\dim G_D) (8-2)\Big](V_{6}-1) \crcr
&+&  \frac12\Big[ 2\dim G_D+(2a-\dim G_D) (8-4)\Big]V_{4} +
\frac12\Big[ 2\dim G_D+(2a-\dim G_D) (8-6)\Big]  V_{2;1} \,, \crcr
&=& \dim G_D(C_\bG -1) 
-\frac12(\dim G_D-2a)\Big(  -2(V_{6}-1) -  4V_{4} -  6V_{2;1}\Big) \,,
\eea

\noindent (8j) Seeking solution of $P_a(\cG)=0$, we have
 $V_6=1$, $V_4=0=V_{2;1}$ and $C_\bG=1$, giving
log--divergent graphs if $g_{\tilde \cG}=0$. 

\medskip
 
\noindent (8l) Solutions of $P_a(\cG)=(2a-\dim G_D)$ are given 
by $C_\bG=1$, $V_6=0$, $V_4=0=V_{2;1}$. This case gives 
divergent graphs with $\omega_d(\cG)=\dim G_D-2a$, 
if $g_{\tilde \cG}=0$. 

\medskip 

- For $N_{\ext}=4$, we have
\bea
P_{a}(\cG) &=& \dim G_D(C_\bG -1) 
+  \frac12\Big[ 2\dim G_D+(2a-\dim G_D) (8-2)\Big]V_{6} \crcr
&+&  \frac12\Big[ 2\dim G_D+(2a-\dim G_D) (8-4)\Big](V_{4}-1) +
\frac12\Big[ 2\dim G_D+(2a-\dim G_D) (8-6)\Big]  V_{2;1} \,, \crcr
&=& \dim G_D(C_\bG -1) 
-\frac12(\dim G_D-2a)\Big(  -2V_{6}-  4(V_{4}-1)  -  6V_{2;1}\Big) \,,
\eea

\noindent (8m) Solving $P_a(\cG)=0$ yields a log--divergent amplitude if $g_{\tilde\cG}=0$ and if 

(8m1) $V_{4}=1$, $V_{2;1}=0$, $V_{6}=0$, $C_\bG=1$;

(8m2) $V_{4}=0$, $V_{2;1}=0$,  $V_{6}=2$, $C_\bG=1$
(corresponding to the trivial partition of 4/2=2=2+0);

\noindent (8n) Solving $P_a(\cG)=(2a-\dim G_D)$ yields a
divergent amplitude with $\omega_d(\cG)=\dim G_D-2a$ if $g_{\tilde\cG}=0$ and if 
$V_{4}=0$, $V_{2;1}=0$, $C_\bG=1$, and $V_{6}=1$
(corresponding to the trivial partition of 1=1+0);

\noindent (8o) Solving $P_a(\cG)=2(2a-\dim G_D)$ yields a
divergent amplitude with $\omega_d(\cG)=2(\dim G_D-2a)$ if $g_{\tilde\cG}=0$ and if 
$V_{4}=0$, $V_{2;1}=0$, $C_\bG=1$, and $V_{6}=0$.

\medskip

- For $N_{\ext}=2$, it can be written 
\bea
P_{a}(\cG) = \dim G_D(C_\bG -1) 
-\frac12(\dim G_D-2a)\Big(  -2V_{6}-  4V_4 -  6(V_{2;1}-1)\Big) \,,
\eea

\noindent (8p) Solving $P_a(\cG)=0$ yields a log--divergent amplitude if $g_{\tilde\cG}=0$ and if 

(8p1) $V_{2;1}=1$, $V_{6}=0$, $V_{4}=0$, $C_\bG=1$;  

(8p2) $V_{2;1}=0$, $V_{6}=3$, $V_{4}=0$, $C_\bG=1$
(corresponding to the partition of 3=1+1+1);

(8p3) $V_{2;1}=0$, $V_{6}=1$, $V_{4}=1$, $C_\bG=1$
(corresponding to the partition of 3=1+2);

\noindent (8q) Solving $P_a(\cG)=(2a-\dim G_D)$ yields a
divergent amplitude with $\omega_d(\cG)=\dim G_D-2a$ if $g_{\tilde\cG}=0$ and if 

(8q1) $V_{2;1}=0$, $V_{6}=2$, $V_{4}=0$, $C_\bG=1$
(corresponding to the partition of 2=1+1); 

(8q2) $V_{2;1}=0$, $V_{6}=0$, $V_{4}=1$, $C_\bG=1$
(corresponding to the trivial partition of 2=2+0);

\noindent (8r) Solving $P_a(\cG)=(2a-\dim G_D)2$ yields a
divergent amplitude with $\omega_d(\cG)=2(\dim G_D-2a)$ if $g_{\tilde\cG}=0$ and if 
$V_{2;1}=0$, $V_{6}=1$, $V_{4}=0$, $C_\bG=1$
(corresponding to the partition of 1=1+0); 
 
\noindent (8r) Solving $P_a(\cG)=(2a-\dim G_D)3$ yields a
divergent amplitude with $\omega_d(\cG)=3(\dim G_D-2a)$ if $g_{\tilde\cG}=0$ and if 
$V_{2;1}=0$, $V_{6}=0$, $V_{4}=0$, $C_\bG=1$.

Table \ref{tab:8} gives the list of primitively divergent
graph for the $_{\dim G_D} \Phi^8_{2}$ models. 

 \begin{table}[h]
\begin{center}
\begin{tabular}{lcccccc cccccc c|}
\hline
$N_{\ext}$ && $V_{2;1}$   && $V_4$ && $V_{6}$ &&$C_{\bG}-1$   &&$g_{\tilde \cG}$ && $\omega_d(\cG)$  \\
\hline
8 && 0 && 0 && 0 && 0 && 0 && 0\\
6 && 0 && 0 && 0 && 0 && 0 &&  $\dim G_D -2a$ \\
6 && 0 && 0 && 1 && 0 && 0 && 0\\
4 && 0  && 0 && 0 && 0 && 0 &&  $2(\dim G_D -2a)$\\
4 && 0  && 0 && 1 && 0 && 0 &&  $\dim G_D -2a$\\
4 && 0  && 1 && 0 && 0 && 0 &&  0\\
4 && 0  && 0 && 2 && 0 && 0 &&  0\\
2 && 0  && 0 && 0 && 0 && 0 &&  $3(\dim G_D -2a)$\\
2 && 0  && 0 && 1 && 0 && 0 &&  $2(\dim G_D -2a)$\\
2 && 0  && 1 && 0 && 0 && 0 &&  $\dim G_D -2a$\\
2 && 0  && 0 && 2 && 0 && 0 && $\dim G_D -2a$\\
2 && 1 &&  0 && 0 && 0 && 0 && 0\\
2 && 0  && 1 && 1 && 0 && 0 && 0\\
2 && 0  && 0 && 3 && 0 && 0 && 0\\
\hline
\end{tabular}
\caption{ List of primitively divergent graphs of matrix 
models $_{\dim G_D}\Phi^{8}_{2}$. \label{tab:8}}
\end{center}
\end{table}

\end{document}